\documentclass[]{aa} 
 
\usepackage{graphicx, txfonts}

\newcommand{\rsol}{\mbox{R$_{\odot}$}}

\newcommand{\ks}{km s$^{-1}$}

\begin{document} 

%%%%% REFEREE VERSION %%%%%%%
%\renewcommand{\baselinestretch}{1.2}
%%%%% REFEREE VERSION %%%%%%%
 
\title{
Baade-Wesselink distances to Galactic and Magellanic Cloud Cepheids and the effect of metallicity
%\thanks{Table~3 is only 
%available in electronic form at the CDS via 
%anonymous ftp to cdsarc.u-strasbg.fr (130.79.128.5) or via 
%http://cdsweb.u-strasbg.fr/cgi-bin/qcat?J/A+A/. 
%Figure~*** is available in the on-line edition of A\&A. 
%}
%\fnmsep
%\thanks{
%Based on observations 
%} 
}  
 
\author{ 
M.A.T.~Groenewegen 
%\inst{1}  
%\and 
%P.~Harmanec 
%\inst{4,5}  
}

\institute{ 
Koninklijke Sterrenwacht van Belgi\"e, Ringlaan 3, B--1180 Brussels, Belgium \\ \email{marting@oma.be}
} 
 
\date{received: 2012,  accepted:  2012} 
 
\offprints{Martin Groenewegen} 
 
\authorrunning{M.~Groenewegen} 
\titlerunning{Baade-Wesselink distances and the effect of metallicity in Cepheids} 
 
\abstract{The metallicity dependence of the Cepheid period-luminosity (PL) relation is
  of importance in establishing the extra-galactic distance scale.}
{The aim of this paper is to investigate the metallicity dependence of
  the PL relation in $V$ and $K$, based on a sample of 128 Galactic, 36 Large Magellanic Cloud (LMC), and 6 Small Magellanic Cloud (SMC)
  Cepheids with individual Baade-Wesselink (BW) distances (some of the
  stars also have an \it Hubble Space Telescope \rm (HST) based and \it Hipparcos \rm parallax or are in clusters) and individually determined
  metallicities from high-resolution spectroscopy.}
{Literature values of the $V$-band, $K$-band, and radial velocity data
  were collected for the sample of Cepheids. Based on a $(V-K)$ surface-brightness relation and a
  projection factor, distances were derived from a BW analysis.}
{ 
The  $p$-relation finally adopted is $1.50 -0.24 \log P$. The slope of this relation is based 
on the condition that the distance to the LMC does not depend on period or $(V-K)$ colour 
and that the slope of the PL relation based on the BW distances agrees with that based on apparent magnitude.
The zero point of the relation is tight to the Cepheids with HST and revised Hipparcos parallaxes as well as to Cepheids in clusters.
The slope of the Galactic and LMC $K$-band relation formally agrees within the errors, and combining all Cepheids (including the SMC) 
results in a negligible metallicity dependence and a relation of $M_{\rm K} = (-2.50 \pm 0.08) + (-3.06 \pm 0.06) \log P$. 
A similar conclusion is found for the reddening-free Wesenheit relation ($W(VK)= K - 0.13\;(V-K)$), with 
$M_{\rm WVK} = (-2.68 \pm 0.08) + (-3.12 \pm 0.06) \log P$.
In the $V$-band the situation is more complex. The slope of the LMC and the Galactic PL relation differ at the 3$\sigma$ level.
Combining the sample nevertheless results in a metallicity term significant at the 2$\sigma$ level: 
$M_{\rm V} = (-1.55 \pm 0.09) + (-2.33 \pm 0.07) \log P + (+0.23 \pm 0.11)$[Fe/H]. 
Taking only the Galactic Cepheids, the metallicity term is no longer significant, namely $(+0.17 \pm 0.25)$.
Compared to the recent works by Storm et al. (2011a, b), there is both agreement and disagreement.
A similar dependence of the $p$-factor on period is found, but the zero point found here implies a shorter distance scale.
The distance modulus (DM) to the LMC and SMC found here are 18.29 $\pm$ 0.02 and 18.73 $\pm$ 0.06 (statistical error on the mean), respectively.
Systematic differences in reddening could have an effect of order +0.05 in DM. 
The details of the comparison of BW-based distances and Cepheids with HST and revised Hipparcos parallaxes also play a role.
The method used by Storm et al. would lead to larger DM of 18.37 and 18.81 for the LMC and SMC, respectively.
The LMC DM is shorter than the currently accepted value, which is in the range 18.42 to 18.55 (Walker 2012), 
and it is speculated that the $p$-factor may depend on metallicity. 
This is not predicted by theoretical investigations, but these same investigations do not predict a steep dependence on period either, 
indicating that additional theoretical work is warranted.
}
{}

\keywords{Stars: distances - Cepheids - distance scale}

\maketitle

\section{Introduction} 
 
Cepheids are considered an important standard candle because they are
bright and thus the link between the distance scale in the nearby
universe and that further out via those galaxies that contain both Cepheids and SNIa (e.g. Riess et al. 2009)

Distances to local Cepheids may be obtained in several ways, e.g. through main-sequence fitting 
for Cepheids in clusters (e.g. Feast 1999, Turner 2010) 
% distances to clusters An et al. 2007 ApJ 671, 1649
%
%
or via determination of the parallax.  
Benedict et al. (2007) published absolute trigonometric parallaxes for ten Galactic Cepheids
using the {\it Fine Guidance Sensor} on board the {\it Hubble Space Telescope}, and revised Hipparcos parallaxes have
also become available (van Leeuwen et al. 2007).

In addition, distances to Cepheids can be obtained from the
Baade-Wesselink (BW) method. This method relies on the availability of
surface-brightness (SB) relations to link variations in colour to variations
in angular diameters and an understanding of the projection ($p$-)
factor, which links radial velocity to pulsational velocity variations.

A lot has been published on the subject over the past decade by the group of Storm/Gieren/Fouqu\'e and coworkers  
(Storm et al. 2004, % 34 galactic + 5 SMC ; FV = 3.947 − 0.131(V − K)0 ; discuss influence of metallicity ; p = 1.39 − 0.03 log P
Gieren et al. 2005, % 13 LMC Cepheids + revision of Storm et al. ; p = 1.58 +-0.02 −0.15+-0.05 log P
Fouqu\'e et al. 2007), % FG SB-relation ; p = 1.366−0.075 logP
which culminated in the recent works by
Storm et al. (2011a, b). % SB Kervella et al. ; p = 1.550−0.186 log(P) ; 70 Gal + 36 LMC +5 SMC Cepheids
In those two papers, the authors analysed 70 Galactic and 41 Magellanic Cloud (MC) Cepheids. They found that, 
(a) the $p$-factor depends quite steeply on period, confirming Gieren et al. (2005), based on the requirement 
that the distance to the (barycenter of the) Large Magellanic Cloud (LMC) should not depend on period, 
(b) the $K$-band period-luminosity (PL) relation is universal, $M_{\rm K}= -3.30 \,(\pm 0.06)\, (\log P - 1) - 5.65 \,(\pm 0.02)$,
(c) the $K$-band PL relation is insensitive to metallicity, and 
(d) a distance modulus to the barycenter of the LMC of 18.45 $\pm$ 0.04 for a $p$-factor relation 
$1.550 - 0.186 \log P$, where the zero point was calibrated on the Cepheids with  {\it Hubble Space Telescope} (HST) parallaxes (Benedict et al. 2007).

Independently, Groenewegen (2007, hereafter G07) investigated the SB relation (finding excellent agreement with the relation by Kervella et al. 2004a) 
and the $p$-factor, based on six Cepheids with interferometrically measured angular-diameter variations and known distances.
Groenewegen (2008, hereafter G08) presented BW distances to 68 Galactic Cepheids with individually determined metallicities 
from high-resolution spectroscopy.

The main aim of that particular work was to address the metallicity dependence of the PL relation, which remains a matter of debate.
Observations seem to consistently indicate that metal-rich Cepheids
are brighter, and various estimates have been given in the literature,
$-0.88 \pm 0.16$ mag/dex ($BRI$ bands, Gould 1994),
$-0.44^{+0.1}_{-0.2}$ mag/dex ($VR$ bands, Sasselov et al. 1997),
$-0.24 \pm 0.16$ mag/dex ($VI$ bands, Kochanek 1997), 
$-0.14 \pm 0.14$ mag/dex ($VI$ bands, Kennicutt et al. 1998), 
$-0.21 \pm 0.19$ in $V$, $-0.29 \pm 0.19$ in $W$, $-0.23 \pm 0.19$ in $I$, 
$-0.21 \pm 0.19$ mag/dex in $K$ (Storm et al. 2004), 
$-0.29 \pm 0.10$ mag/dex ($BVI$ bands, Macri et al. 2006).
G08 found values of
$+0.27 \pm 0.30$ mag/dex ($V$-band) and
$-0.11 \pm 0.24$ mag/dex ($K$-band).

Since then, values have been reported of
$-0.29 \pm 0.11$ mag/dex ($W_{\rm VI}$, Scowcroft et al. 2009), and a very steep value of % 2 fields in M33
$\sim$$-0.8 \, \pm$ $\sim$0.2 ($W_{\rm VI}$, Shappee \& Stanek 2011). % M101 
Finally, Storm et al. (2011) quote slopes of
$-0.23 \pm 0.10$ mag/dex ($W_{\rm VI}$),
$+0.09 \pm 0.10$ mag/dex ($V$-band), and
$-0.10 \pm 0.10$ mag/dex ($K$-band).

With the exception of Groenewegen (2008), no individual abundance
determinations of individual Cepheids are used in these studies.
Instead, abundances of nearby H{\sc ii} regions or even a mean
abundance of the entire galaxy are used.
Even in the recent work by Storm et al. all Galactic, Small Magellanic Cloud (SMC) and LMC Cepheids 
were assigned a  metallicity typical of that galaxy.

The present paper is an update of Groenewegen (2007, 2008) and takes into account the latest available data, 
in terms of interferometrically determined angular diameters, photometric and radial velocity data, and individually determined metallicities.
In this paper, we revisit BW distances to Cepheids with metallicity
determinations, increasing the sample to over 120 Galactic stars, adding 6 SMC and 42 LMC objects, 
and using updated $p$-factor and SB relations.
Section~2 describes the selection of the photometric and radial velocity data. 
Section~3 outlines how the data are modelled. 
In Sect.~4, the surface-brightness relation is discussed.
Section~5 describes how the binary Cepheids are treated, and new and updated orbital elements are presented. 
Section~6 describes the results regarding the period-luminosity(-metallicity) (PL(Z)) relation.
Section~7 presents conclusions.

\section{The sample}

Table~\ref{TAB-DATA} lists the 128 Galactic classical Cepheids used in this study. 
This is a significant increase with respect to the 68 stars considered in G08.
This sample represents essentially all Galactic Cepheids with accurate individually determined metallicities 
(mostly from Luck \& Lambert 2011 and Luck et al. 2011)
that have sufficient optical, $K$-band, and radial velocity data available for a 
BW analysis\footnote{Some Cepheids have considered, but the datasets were deemed insufficient for an accurate 
BW analysis: BK Aur and GH Cyg (too poor radial velocity data), EV Sct and X Sct (too poor NIR data).}.
The single largest new dataset that made this increase in sample size
possible is the recent publication of near-infrared light curves of 131
northern Cepheids by Monson \& Pierce (2011).

Contrary to G08, 42 Magellanic Cloud (MC) Cepheids are also considered here, and Table~\ref{TAB-DATAMC} lists that sample.
The radial velocity data presented by Storm et al. (2011b) contribute significantly to the fact that a 
BW analysis is now feasible for a sizeable sample of MC Cepheids.

Tables~\ref{TAB-DATA} and \ref{TAB-DATAMC} list the sources of the $V$-band, $K$-band, and radial velocity data, 
as well as data which were not considered in the 
BW analysis\footnote{When deriving the binary orbits, all radial velocity (RV) data were used (see Sect~\ref{sect-bin}).}.
Sometimes a certain range in Julian date is excluded, mostly to exclude some of the older datasets, which are less accurate. 
This is also done because there is a clear change in pulsation period, and these cases are marked by $\dot{P}$.

Since 2008, additional radial velocity data has been obtained using the 1.2m Euler telescope located at 
the La Silla observatory (see G08 for a description of the data taking and data reduction). 
The new Radial Velocity (RV) data are presented in Table~\ref{TAB-RV}.

\begin{table*} 

\caption{Sources of $V$-, $K$-band and RV data for the Galactic Cepheids. } 
\begin{tabular}{rrrrl} \hline \hline 

Name       & $V$                      & $K$             & RV                                           &  Data not considered\tablefootmark{a} \\  
\hline 

   AK Cep &                         3 &             111 &                                            33 & \\                               
   AN Aur &                         3 &             111 &                                         33,39 & -,-,39 \\                            
   AQ Pup &           1,2,3,4,5,84,99 &           6,7,8 &                                 1,4,39,44,113 & 3,-,- JD $<$ 44100, JD $>$ 48500, $\dot{P}$ \\       
   AV Sgr &                  3,16,112 &             112 &                                  16,26,41,112 & JD $<$ 45000 \\                      
   AW Per &          2,3,13,27,30,116 &             111 &                        9,28,33,39,117,118,119 & -,-,9 JD $<$ 40000 \\  % RV=binary             
   BB Sgr &              2,3,16,23,63 &            6,24 &                            20,33,39,44,47,113 & -,-,44 JD $<$ 44400 \\                   
   BE Mon &                     3,125 &             111 &                                             4 & \\                               
 beta Dor &                3,11,16,35 &            6,18 &                             11,19,20,21,22,47 & -,-,20 JD $<$ 40000 \\                   
   BF Oph &              2,3,16,23,63 &         6,18,24 &                                10,20,39,44,76 & JD $<$ 36000 \\                      
   BG Lac &                 2,3,12,13 &          12,111 &                                   14,15,39,44 & JD $<$ 33000 \\                      
   BM Per &                     3,120 &             111 &                                        33,121 & \\                               
   BN Pup &                  1,3,5,16 &             6,7 &                                          1,17 & JD $<$ 33000 \\  % RV                
   BZ Cyg &                  3,64.120 &             111 &                                            33 & \\                               
   CD Cyg &                    2,3,64 &          24,111 &                                         33,44 & \\                               
   CF Cas &                       2,3 &             111 &                                   4,17,33,105 & \\                               
   CK Sct &                      3,16 &             111 &                                     17,26,105 & \\                               
   CP Cep &                         3 &             111 &                                     17,33,105 & \\                               
   CR Cep &                       2,3 &             111 &                                      4,44,105 & -,-,44 \\                            
   CR Ser &                         3 &             111 &                                            33 & \\                               
   CS Vel &                     3,120 &          6,7,24 &                                          4,26 & -,24,- \\      
   CV Mon &              2,3,13,16,23 &        6,24,111 &                                      25,26,44 & \\                               
   DD Cas &                    2,3,13 &             111 &                                       4,33,44 & \\                               
delta Cep &           2,3,11,12,13,27 &           12,98 &                         4,9,14,25,28,29,37,39 & -,-,9 JD $<$ 43500 \\  % K-mag             
   DL Cas &                 2,3,11,13 &             111 &                               4,33,44,105,122 & 11,-,- JD $<$ 40000 \\                   
   DT Cyg &            2,3,4,27,30,31 &           24,32 &                               4,9,28,29,33,79 & -,32,9+28 JD $<$ 33400 \\  % K-mag             
  eta Aql &     2,3,12,13,16,23,27,35 &           12,24 &                        9,14,20,25,28,29,36,37 & -,-,9 \\                            
   FF Aql &     2,3,11,16,23,27,30,31 &        18,24,32 &                              9,20,28,29,33,38 & 3,32,- JD $<$ 44400 \\  % K, RV=binary          
   FM Aql &           2,3,12,13,16,23 &       12,24,111 &                                   14,33,39,44 & JD $<$ 30000 \\                      
   FM Cas &                    2,3,13 &             111 &                                       4,33,44 & -,-,44 \\                            
   FN Aql &           2,3,12,13,16,23 &       12,24,111 &                                      14,33,39 & -,-,39 \\                            
   GH Lup &                 1,3,16,23 &               6 &                                             1 & JD $<$ 43000 \\                      
   GY Sge &                         3 &           6,111 &                                        33,105 & JD $<$ 45500, JD $>$ 48200, $\dot{P}$ \\          
   KN Cen &         1,3,5,11,16,40,84 &               6 &                                       1,17,41 & -,-,41 JD $<$ 40000, JD $>$ 47000 \\  % RV    
   KQ Sco &                    1,3,11 &               6 &                                      1,42,114 & \\  % RV                         
    l Car &              3,5,11,16,35 &               6 &                             11,19,20,23,43,47 & 5,-,23 JD $<$ 46500 \\                   
   LS Pup &                    1,3,11 &               6 &                                     1,113,114 & -,-,1 \\  % RV                      
   MW Cyg &                    2,3,13 &             111 &                                      33,44,80 & -,-,44 \\                            
   QZ Nor &                    1,3,11 &               6 &                                 26,81,114,127 & -,-,81 \\                            
   RR Lac &                2,3,13,120 &             111 &                                       4,33,44 & -,-,44 \\                            
   RS Cas &                         3 &             111 &                                         33,39 & -,-,39 \\                            
   RS Ori &              2,3,13,16,30 &             111 &                                            33 & \\                               
   RS Pup &            2,3,5,11,16,84 &          6,8,24 &                                   11,19,25,44 & -,-,44 JD $<$ 42000, JD $>$ 50000, $\dot{P}$ \\       
   RT Aur &             2,12,27,31,45 &       12,32,111 &                              9,28,29,33,37,46 & -,32,9 JD $<$ 40000 \\                   
   RU Sct &                    2,3,16 &           6,111 &                                  26,33,39,105 & -,-,39 JD $<$ 25000 \\                            
   RW Cam &                 2,3,30,64 &             111 &                                            44 & \\                               
   RW Cas &                    2,3,64 &             111 &                                         33,44 & \\                               
   RX Aur &                    2,3,64 &             111 &                                      28,33,44 & \\                               
   RX Cam &                    2,3,13 &             111 &                                      33,44,80 & JD $<$ 40000 \\                      

\hline 
\end{tabular} 
\end{table*}

\setcounter{table}{0}
\begin{table*} 

\caption{Continued } 
\begin{tabular}{rrrrl} \hline \hline 

Name       & $V$                      & $K$             & RV                                            & Data not considered\tablefootmark{a}  \\  
\hline 

   RY Cas &                      3,30 &             111 &                                            33 & \\                               
   RY CMa &                    2,3,16 &             111 &                                         33,44 & \\                               
   RY Sco &             1,2,3,5,11,16 &               6 &                                       1,20,44 & \\                               
   RY Vel &               1,3,5,11,16 &               6 &                                       1,11,17 & JD $<$ 44000, JD $>$ 50000, $\dot{P}$ \\          
   RZ CMa &                    2,3,16 &             111 &                                         17,44 & \\                               
   RZ Gem &                    2,3,13 &             111 &                                         33,44 & \\                                      % RV
   RZ Vel &               1,3,5,11,16 &               6 &                                    1,11,20,47 & \\                               
    S Mus &                3,11,23,40 &            6,24 &                    10,11,20,22,41,47,48,49,50 & -,-,41 JD $<$ 30000 \\                % RV=binary    
    S Nor &                 3,4,23,40 &            6,24 &                     4,10,20,41,47,49,51,52,53 & -,-,41 JD $<$ 30000 \\                % RV=binary  
    S Sge &  2,3,11,12,13,23,27,30,40 &        12,24,32 &                     9,14,28,29,33,37,54,55,56 & -,32,9+28+54 \\                       % RV=binary   
   SS Sct &              2,3,16,23,63 &          24,111 &                               39,42,44,63,114 & JD $<$ 30000 \\                      
   ST Tau &              2,3,12,31,90 &             111 &                                       4,33,44 & 90,-,- \\                            
   SU Cas &               2,3,9,27,31 &           12,32 &             4,9,25,28,29,33,57,58,59,60,61,62 & -,32,9+28+59+61+62 JD $<$ 43000 \\    % RV=binary 
   SU Cyg &              2,3,27,30,31 &          24,111 &                       9,28,33,106,107,108,109 & -,-,9 JD $<$ 35000 \\                 % RV=binary   
   SV Mon &               1,2,3,64,90 &             111 &                                       1,33,44 & \\                               
   SV Per &                 2,3,64,90 &             111 &                                     33,44,121 & -,-,44 \\                            
    S Vul &                      3,30 &           6,111 &                                        33,110 & JD $<$ 45400, JD $>$ 49200, $\dot{P}$ \\          
   SV Vul &             2,3,4,9,27,64 &     6,12,24,111 &                   4,9,14,15,25,28,29,33,39,65 & JD $<$ 45500, JD $>$ 48600, $\dot{P}$ \\          
   SW Cas &                    2,3,13 &             111 &                                            33 & \\                               
   SW Vel &            1,3,5,11,16,84 &               6 &                                          1,11 & JD $>$ 48000, $\dot{P}$ \\                   
   SX Vel &                   3,23,40 &               6 &                                            10 & \\ %RV                            
   SY Cas &                      3,31 &             111 &                                          4,33 & \\                               
   SZ Aql &        2,3,11,12,16,23,64 &     6,12,24,111 &                                      11,14,23 & \\                               
   SZ Cyg &                    2,3,64 &             111 &                                     39,44,123 & JD $<$ 30000 \\                      
   SZ Tau &            2,3,4,12,27,31 &         6,12,32 &                                  4,9,28,29,33 & -,32,9 JD $<$ 43000, JD $>$ 48000, $\dot{P}$ \\       
    T Mon &         1,2,3,11,16,35,64 &     6,24,32,111 & 4,9,25,29,33,37,48,65,66,67,68,69,70,71,72,73 & -,32,70 JD $<$ 44500 \\               % RV=binary  
   TT Aql &     2,3,11,12,16,27,64,74 &       12,24,111 &               9,11,14,15,22,25,28,33,37,39,74 & -,-,74 JD $<$ 47000 \\                   
    T Vel &                   3,16,75 &            6,18 &                                      10,20,75 & -,18,- \\                            
    T Vul &            2,3,4,12,27,31 &     12,24,32,98 &                               4,9,14,28,29,37 & -,32,9 JD $<$ 43600 \\                   
   TW Nor &                3,5,16,120 &            6,24 &                                          4,26 & \\                               
   TY Sct &                      3,16 &             111 &                                     26,39,105 & -,-,39 \\                            
   TZ Mon &                  3,16,121 &             111 &                                     17,39,121 & JD $<$ 40000 \\                      
    U Aql &                 2,3,16,27 &             111 &                                9,11,20,28,124 & -,-,44 JD $<$ 42000 \\                   
    U Car &            1,3,5,11,16,35 &            6,24 &                                       1,11,20 & \\                               
    U Nor &                 1,5,11,16 &               6 &                                    1,11,17,41 & \\                               
    U Sgr &         2,3,4,16,23,35,63 & 6,24,32,111,115 &       4,9,10,11,20,25,28,33,39,47,51,76,77,78 & -,32,9 JD $<$ 37000 \\                   
   UU Mus &              1,5,11,40,84 &               6 &                                          1,11 & \\                               
    U Vul &           2,3,12,13,23,27 &          12,111 &                           4,14,33,37,44,79,80 & -,-,44 \\  % RV=binary                      
   UZ Sct &                3,5,16,112 &         111,112 &                              17,26,39,105,112 & JD $<$ 47500 \\                      
V1162 Aql &                         3 &             111 &                                        33,114 & \\                               
 V340 Ara &                    3,5,23 &             112 &                                     17,23,112 & JD $<$ 50000 \\                      
 V340 Nor &                    3,4,82 &               6 &                                      4,26,114 & \\  % RV                         
 V350 Sgr &              2,3,23,63,99 &           18,24 &                      20,22,33,39,44,76,83,114 & -,18,76 JD $<$ 42000 \\                 % RV=binary  
 V386 Cyg &                    2,3,13 &             111 &                                     33,44,105 & \\                               
 V402 Cyg &                 2,3,31,90 &             111 &                                        33,105 & 90,-,- \\                            
 V459 Cyg &                       2,3 &             111 &                                     17,33,105 & \\                               
 V495 Cyg &                     3,125 &             111 &                                        33,105 & \\                               
 V496 Aql &                 2,3,16,63 &              24 &                         9,10,20,33,76,113,114 & -,-,9 JD $<$ 40500 \\                   % K, RV=binary   
 V538 Cyg &                         3 &             111 &                                            33 & \\                               
 V600 Aql &                 2,3,13,16 &             111 &                                     17,33,105 & \\                               
 V916 Aql &                         3 &             111 &                                            33 & \\                               

\hline 
\end{tabular} 
\end{table*}

\setcounter{table}{0}
\begin{table*} 

\caption{Continued } 
\begin{tabular}{rrrrl} \hline \hline 

Name      & $V$                       & $K$             & RV                                           & Data not considered\tablefootmark{a}  \\  
\hline 

    V Car &                   3,16,23 &               6 &                                      10,20,47 & \\  % RV                         
    V Cen &             3,11,16,23,63 &            6,24 &                                10,11,20,47,76 & -,24,- JD $<$ 40000 \\                   
   VW Cen &                 1,3,16,84 &               6 &                                      1,41,114 & -,-,41 \\  % RV                      
   VX Cyg &                  3,64,120 &             111 &                                         33,39 & JD $<$ 30000 \\                      
   VX Per &             2,3,30,84,120 &             111 &                                     33,44,105 & 30,-,- \\                            
   VY Car &     1,3,5,11,16,84,99,100 &            6,24 &                                    1,11,20,41 & 84,-,- JD $<$ 42000, JD $>$ 50490, $\dot{P}$ \\       
   VY Cyg &                       2,3 &             111 &                                     39,44,123 & JD $<$ 40000 \\                      
   VY Sgr &                  3,16,112 &             112 &                                  17,26,39,112 & JD $<$ 47000 \\                      
   VZ Cyg &                 2,3,12,31 &       12,24,111 &                             4,14,33,39,44,113 & JD $<$ 40000 \\  % RV=binary                
   VZ Pup &               1,3,5,11,16 &           6,7,8 &                                      1,11,113 & \\                               
    W Gem &              2,3,13,16,27 &             111 &                                       9,28,33 & \\                               
    W Sgr &           2,3,23,35,40,85 &        24,32,86 &             4,9,10,20,22,28,47,48,85,87,88,89 & -,-,9+28 JD $<$ 39000,\\  % K VERY POOR, RV=binary   
   WZ Car &               1,3,5,11,16 &               6 &                                          1,11 & \\                               
   WZ Sgr &             1,2,3,5,11,40 &           6,111 &                                    1,11,33,39 & JD $<$ 40000 \\                      
    X Cyg &               2,3,4,27,64 &     12,24,32,98 &                              4,14,25,29,33,37 & -,32,- \\                            
    X Lac &                 2,3,12,13 &          12,111 &                                    4,14,33,44 & -,12,44 \\                            
    X Pup &            2,3,5,11,16,23 &               6 &                                  11,23,44,113 & 5,-,- \\                            
%    X Sct &                      3,16 &             111 &                                    26,105,114 & \\                               
    X Sgr &        2,3,11,16,23,35,99 &       24,32,104 &           9,10,11,20,28,47,52,101,102,103,113 & -,32,9 JD $<$ 48800 \\  % RV=binary             
    X Vul &                    2,3,13 &             111 &                                       4,33,44 & -,-,44 \\                            
   XX Cen &                3,11,40,74 &            6,24 &                             10,11,20,41,52,74 & -,24,41 \\  % RV=binary                      
   XX Sgr &                    2,3,23 &              24 &                                       113,114 & JD $<$ 44000 \\                      
    Y Lac &              2,3,12,31,90 &          12,111 &                                14,15,33,39,44 & -,111,- JD $<$ 40000 \\                      
    Y Oph &                  1,2,3,16 &            6,24 &                      1,9,19,20,28,33,34,91,92 & -,-,9 JD $<$ 40000 \\                   
    Y Sct &                    2,3,16 &             111 &                                         11,44 & \\                               
    Y Sgr &           2,3,11,16,23,35 &           24,32 &             9,11,19,20,28,46,49,93,94,113,114 & -,32,9 JD $<$ 28500 \\  % RV             
   YZ Aur &                      3,64 &             111 &                                      3,39,126 & JD $<$ 45000 \\                      
   YZ Sgr &                 2,3,23,40 &          24,111 &                                     20,44,113 & JD $<$ 44900 \\                      
 zeta Gem &            2,3,4,27,35,64 &          32,104 &                        4,9,19,28,29,33,37,113 & -,32,9 JD $<$ 23500 \\                   
    Z Lac &                 2,3,12,64 &          12,111 &                       14,33,39,44,80,95,96,97 & JD $<$ 40500 \\  % RV=binary                

\hline 
\end{tabular} 
\tablefoot{
\tablefoottext{a}{The number indicates the dataset not used in $V$, $K$, RV, respectively.}
}

\bf References. \rm
1= Coulson \& Caldwell (1985a);   % SAAO Circulars, 9, 5
2= Moffett \& Barnes (1984); 
3= Berdnikov et al. (2000), a datafile named ``cepheids-16-03-2006'' was retrieved from the ftp address listed in that paper;
4= Bersier et al. (1994); 
5= Madore (1975); 
6= Laney \& Stobie (1992); 
7= Schechter et al. (1992);
8= Welch (1985);  
9= Barnes et al. (1987), points with uncertainty flag ``:'' were removed; % ApJS, 65, 307
10= Stibbs (1955);  % MN 115, 963
11= Bersier (2002), data points with weight 0 and 1 in the Geneva photometry were removed;  % ApJS, 140, 465
12= Barnes et al. (1997);  %PASP 109, 645
13= Szabados (1980);   %Konkoly No.76  
14= Barnes et al. (2005); 
15= Imbert (1999);
16= Pel (1976);     %  A&AS, 105, 165 
17= Pont et al. (1994);
18= Lloyd Evans (1980a); %SAAO Circulars, 1, 163
19= Nardetto et al. (2006); 
20= Lloyd Evans (1980b);  %SAAO Circulars, 1, 257 
21= Taylor \& Booth (1998); 
22= Petterson et al. (2005); % MN362 1167
23= Caldwell et al. (2001);  % JAR 7,4
24= Welch et al. (1984);  %ApJS, 54, 547
25= Storm et al. (2004); 
26= Metzger et al. (1992); % AJ, 103, 529
27= Kiss (1998); 
28= Wilson et al. (1989); 
29= Kiss (2000); 
30= Szabados (1991);   % Commun. Konkoly. , No.96 
31= Szabados (1977);  %  Mitt. Sternw. ung. Akad. Wiss., No.70 
32= Wisniewski \& Johnson (1968); 
33= Gorynya et al. (1998, VizieR On-line Data Catalog: III/229); 
34= Sanford (1935); 
35= Shobbrook (1992);  % MN255 486
36= Jacobsen \& Wallerstein (1981); 
37= Evans (1976);  % ApJS 32 399
38= Evans et al. (1990);  
39= Joy (1937);  % ApJ 86, 363
40= Walraven et al. (1964); 
41= Grayzeck (1978);  % PASP 90, 742
42= Groenewegen (2008);
43= Taylor et al. (1997); 
44= Barnes et al. (1988), points with uncertainty flag ``:'' were removed;  % ApJS 66, 43
45= Turner et al. (2007); 
46= Duncan (1908); 
47= Lloyd Evans (1968);  % MN 141, 109
48= Petterson et al. (2004); % MN250   
49= Campbell \& Moore (1928); 
50= B\"ohm-Vitense et al. (1990); 
51= Mermilliod et al. (1987);
52= Feast (1967); 
53= Breger (1970); 
54= Evans et al. (1993);
55= Herbig \& Moore (1952);
56= Breitfellner \& Gillet (1993);
57= Adams \& Shapley (1918); 
58= Abt (1959);
59= Niva \& Schmidt (1979);
60= Gieren (1976);
61= H\"aupl (1988);
62= Beavers \& Eitter (1986);
63= Gieren (1981b);
64= Szabados (1981);  %Comm 77
65= Sanford (1956);
66= Evans et al. (1999); 
67= Frost (1906);
68= Sanford (1927);
69= Wallerstein (1972); 
70= Coulson (1983); 
71= Evans \& Lyons (1994);
72= Gieren (1989); 
73= Harper (1934); 
74= Coulson et al. (1985); 
75= Gieren (1985); 
76= Gieren (1981a); 
77= Jacobsen (1970);
78= Breger (1967);
79= Sanford (1951); % ApJ, 114, 331
80= Imbert (1996); 
81= Coulson \& Caldwell (1985b); 
82= Eggen (1983); 
83= Evans \& Sugars (1997); 
84= Hipparcos Epoch Photometry;
85= Babel et al. (1989); 
86= Kimeswenger et al. (2004); 
87= Albrow \& Cottrell (1996); 
88= Jacobsen et al. (1984); 
89= Jacobsen (1974); 
90= Henden (1980); 
91= Evans \& Lyons (1992); 
92= Abt \& Levy (1978); 
93= Duncan (1922);
94= ten Bruggencate (1930); 
95= Evans \& Welch (1993); 
96= Gieren (1989);
97= Sugars \& Evans (1996); 
98= Fernley et al. (1989);
99= Dean (1977); 
100= Dean (1981); 
101= Duncan (1932); 
102= Slipher (1904);
103= Moore (1909); 
104= Feast et al. (2008);
105= Metzger et al. (1993);  % ApJS, 76, 803
106= Evans (1988);
107= Maddrill (1906);
108= Abt (1973);         % ApJS 26, 365 
109= Imbert (1985);
110= Joy (1952);

111= Monson \& Pierce (2011);
112= Pedicelli et al. (2010);
113= Storm et al. (2011a);  % AA 534 A94
114= This paper (Table~\ref{TAB-RV}); 
115= McGonegal et al. (1983); % ApJ269, 641
116= Vink\'o (1993);  % MN260, 273
117= Evans (1983);
118= Evans et al. (2000); % AJ120, 407
119= Welch \& Evans (1989);
120= Harris (1980);
121= Pont et al. (1997); % AA318, 416
122= Gieren et al. (1994); % Astron. J. 107, 2093
123= Struve (1945);
124= Welch et al. (1987);
125= Schmidt \& Seth (1996)  % AJ, 112, 2769
126= Szabados \& Pont (1998);
127= Kienzle et al. (1999).

\label{TAB-DATA} 
\end{table*} 
%%%%%%%%%%%%%%%%%%%%%%%%%%%%%%%%%%%%%%%%%%%%%%%%%%%%%%%%%%%%%%%%%%%%%%%%%%%%%%%%%%%%%%%%%%%%%%%%%%%%%%%%
%\setcounter{table}{0}
\begin{table*} 

\caption{Sources of $V$-, $K$-band and RV data for the MC Cepheids.} 
\begin{tabular}{rrrrl} \hline \hline 

Name\tablefootmark{*} & $V$               & $K$            & RV             & Data not considered\tablefootmark{a}  \\  
\hline 

HV 822\tablefootmark{*}  & 1001,1002,1003                & 1001           & 1001 & 1002,-,- \\
HV 837\tablefootmark{*}  & 1002,1003,1004,1005,1006,1007 & 1008,1009      & 1010 &          \\
HV 873   & 1002,1004                     & 1009,1011      & 1012 & -,1009,- \\
HV 876   & 1004                          & 1011           & 1012 &          \\
HV 877   & 1002,1003,1004,1005,1006      & 1008,1009,1011 & 1012 &          \\
HV 878   & 1002,1004                     & 1009,1011      & 1012 &          \\
HV 879   & 1004,1006,1007                & 1008,1009,1011 & 1010 & -,1009,- \\
HV 881   & 1002,1004,1005                & 1009,1011      & 1012 &          \\
HV 899   & 1002,1004,1006,1007           & 1009,1011      & 1010 & 1006,1009,- \\
HV 900   & 1002,1003,1004,1005,1007      & 1009,1011      & 1010,1012 & -,-,- JD $<$ 44600 JD $<$ 51000 \\
HV 909   & 1002,1003,1006,1007           & 1009,1011      & 1010 & -,1009,-  \\
HV 914   & 1004                          & 1009,1011      & 1012 & -,1009,-  \\
HV 1005  & 1004,1006                     & 1011           & 1012 &           \\
HV 1006  & 1004                          & 1011           & 1012 &           \\
HV 1023  & 1004,1006                     & 1011,1008      & 1012 &           \\
HV 1328\tablefootmark{*} & 1001,1003,1004                & 1001,1009      & 1001 &  -,-,- JD $<$ 47000  \\
HV 1333\tablefootmark{*} & 1001,1004                     & 1001           & 1001 &           \\
HV 1335\tablefootmark{*} & 1001,1004                     & 1001           & 1001 &           \\
HV 1345\tablefootmark{*} & 1001,1004                     & 1001           & 1001 &           \\
HV 2257  & 1002,1004,1005,1007           & 1009,1011      & 1010 & -,1009,- JD $<$ 44500, JD $>$ 50500, $\dot{P}$ \\
HV 2282  & 1004                          & 1011           & 1012 &           \\
HV 2338  & 1002,1005,1006,1007           & 1009,1011      & 1010 & -,1009,-  \\
HV 2369  & 1002,1003,1004,1005,1006,1013 & 1009,1011      & 1012 &          JD $<$ 44300, $\dot{P}$ \\
HV 2405  & 1004                          & 1009,1011      & 1012 &           \\
HV 2527  & 1004,1006                     & 1009,1011      & 1012 &           \\
HV 2538  & 1004                          & 1011           & 1012 &           \\
HV 2549  & 1003,1004,1006                & 1009,1011      & 1012 &           \\
HV 2827  & 1003,1006,1007                & 1008,1009,1011 & 1010 & -,1008,-  \\
HV 5655  & 1004                          & 1011           & 1012 &           \\
HV 6093  & 1004                          & 1011           & 1012 &           \\

HV 12197 & 1014,1015,1016,1017           & 1018           & 1015,1018,1022 & JD $<$ 37000 \\
HV 12198 & 1014,1015,1016,1017           & 1018           & 1001,1016,1018 & JD $<$ 37000 \\
HV 12199 & 1014,1015,1016,1017           & 1018           & 1015,1018,1022 & JD $<$ 37000 \\
HV 12202 & 1014,1015,1016,1017           & 1018,1019      & 1015,1018 & -,1019,- JD $<$ 37000 \\
HV 12203 & 1014,1015,1016,1017           & 1018           & 1015,1018 & JD $<$ 37000 \\
HV 12204 & 1014,1015,1016                & 1018           & 1015,1018 & JD $<$ 37000 \\
HV 12452 & 1004                          & 1011           & 1012      &              \\
HV 12505 & 1004                          & 1011           & 1012      &              \\
HV 12717 & 1004                          & 1011           & 1012      &              \\
HV 12815 & 1003,1006,1007,1020           & 1008,1009,1011 & 1020      & -,1009,-     \\
HV 12816 & 1006,1007,1020                & 1008,1009,1011 & 1020,1021 & -,-,1020     \\

U 1      & 1004                          & 1011           & 1018      &              \\

\hline 
\end{tabular} 
\tablefoot{
\tablefoottext{a}{The number indicates the dataset not used in $V$, $K$, RV, respectively.}
\tablefoottext{*}{Objects marked by a $\star$ are located in the SMC.}
}

{\bf References.}
1001= Storm et al. (2004);  % A&A 415, 521
1002= Madore (1975);     %ApJS 29,219
1003= Van Genderen (1983);  %A&AS 52
1004= OGLE-III data (Soszynski et al., 2008, 2010);
1005= Eggen (1977); %ApJs 34,33
1006= Martin \& Warren (1979); %SAOC4,98
1007= Moffett et al. (1998); % ApJS 117
1008= Laney \& Stobie (1986);  % SAAO C 10, 51 
1009= Welch et al. (1987);  %ApJ 321, 162
1010= Imbert (1989);  % A&AS 81
1011= Persson et al. (2004);  %AJ 128
1012= Storm et al. (2011b); % AA 534, 95
1013= Freedman et al. (1985); % APJS 59,331
1014= Gieren et al. (2000);
1015= Welch et al. (1991);  % AJ, 101, 490
1016= Walker (1987a); % MNRAS, 225, 627
1017= Arp \& Thackeray (1967); 
1018= Storm et al. (2005); % AA 440,487
1019= Testa et al. (2007); 
1020= Caldwell et al. (1986); % MN220,671
1021= Gieren et al. (2005); % 05ApJ627_224
1022= Molinaro et al. (2012).

\label{TAB-DATAMC} 
\end{table*} 
%%%%%%%%%%%%%%%%%%%%%%%%%%%%%%%%%%%%%%%%%%%%%%%%%%%%%%%%%%%%%%%%%%%%%%%%%%%%%%%%%%%%%%%

\begin{table*} 

\caption{New radial velocity data. } 
\begin{tabular}{crcrcrcr} \hline \hline 

JD          & RV  (\ks) & JD          & RV  (\ks)  & JD          & RV  (\ks)   & JD          & RV  (\ks)    \\  
\hline 

\multicolumn{2}{c}{KQ Sco} & \multicolumn{2}{c}{V340 Nor} & \multicolumn{2}{c}{VW Cen} & \multicolumn{2}{c}{Y Sgr}   \\
2454625.7165 &  -43.06     & 2454626.5964  & -47.27  & 2454626.5170 & -63.59   & 2454302.5565 &   -9.03 \\
2454626.8125 &  -47.14     & 2454626.6375  & -47.17  & 2454626.5863 & -63.60   & 2454303.7440 &    2.45 \\
2454627.5483 &  -46.52     & 2454627.5619  & -45.44  & 2454627.5270 & -60.25   & 2454304.6563 &   10.42 \\
2454628.6980 &  -45.57     & 2454628.6236  & -40.48  & 2454627.6088 & -59.83   & 2454305.7003 &   14.94 \\
2454631.5682 &  -32.87     & 2454631.5810  & -29.51  & 2454631.5496 & -33.87   & 2454306.7067 &  -20.91 \\
2454632.7716 &  -28.10     & 2454633.6032  & -40.26  & 2454633.5906 & -21.36   & 2454307.6489 &  -15.31 \\
2454633.6530 &  -25.24     & 2454635.5536  & -45.34  & \multicolumn{2}{c}{X Sct} & 2454308.6302 & -6.21 \\
2454953.6445 &  -12.65 & \multicolumn{2}{c}{V350 Sgr} & 2454302.7712 & 18.62   & 2454309.5865 &    2.96 \\
\multicolumn{2}{c}{LS Pup} & 2454302.5747  &  12.56  & 2454304.7842 &  -6.32   & 2454310.7130 &   14.48 \\
2454627.4580 &   102.98    & 2454303.7480  &  25.54  & 2454305.7605 &   0.02   & 2454311.7705 &   -0.97 \\
2454631.4584 &    66.99    & 2454304.7665  &   6.63  & 2454306.6959 &  13.83   & 2454312.7778 &  -20.24 \\
2454635.4577 &    69.31    & 2454305.7794  &  -7.51  & 2454307.7367 &  31.34   & 2454331.6950 &   -6.40 \\
\multicolumn{2}{c}{QZ Nor} & 2454306.8017  &   2.13  & 2454309.7736 &  -2.81   & 2454332.5063 &    1.38 \\
2454303.5850 &  -31.84     & 2454307.7016  &  12.21  & 2454310.7698 &  12.04   & 2454334.5199 &   14.87 \\
2454307.6148 &  -30.76     & 2454308.6363  &  21.25  & 2454312.7944 &  17.84   & 2454339.5444 &   13.17 \\
2454311.6411 &  -30.72     & 2454309.6259  &  24.82  & 2454332.6218 &  27.46   & 2454341.5468 &  -21.03 \\
2454334.5106 &  -32.10     & 2454310.7948  &  -8.45  & 2454335.6077 &   6.87   & 2454625.8749 &   -8.96   \\ 
2454341.5283 &  -31.64     & 2454311.6175  &  -1.54  & 2454341.6291 &  31.67   & 2454626.8939 &    1.02   \\
2454625.5461 &  -30.94     & 2454312.8122  &  11.59  & 2454626.6591 &  29.84   & 2454627.6650 &    7.31   \\
2454633.6613 &  -32.27     & 2454331.7205  &  -6.69  & 2454627.8317 &   2.80   & 2454630.8372 &  -16.37   \\
\multicolumn{2}{c}{SS Sct} & 2454332.7166  &   3.26  & 2454630.8799 &  29.94   & 2454631.8182 &   -7.29   \\
2454625.8712 &    7.07     & 2454334.6189  &  23.99  & 2454633.6960 &   7.21   & 2454632.8076 &    2.19   \\
2454627.6691 &  -13.88     & 2454335.5375  &  15.52  & 2454953.7207 &  23.38   & 2454633.6681 &    9.60   \\
2454630.8872 &  -21.58     & 2454339.6487  &  20.88  & \multicolumn{2}{c}{XX Sgr} & 2454634.8012 & 14.13   \\
2454631.8791 &   -5.11     & 2454341.6415  &  -9.50  & 2454302.5656 &  23.35   & 2454953.7064 &  -19.59   \\
2454632.8718 &    7.04     & 2454625.8994  &  -8.11  & 2454303.7571 &  35.71  \\ 
2454633.6781 &   -8.44     & 2454626.6336  &  -0.31  & 2454304.6605 &  -4.33  \\ 
2454634.8687 &  -16.28     & 2454627.7019  &  10.97  & 2454305.7946 &  -2.23  \\ 
\multicolumn{2}{c}{V1162 Aql} & 2454628.6315 & 20.74 & 2454306.7194 &   4.97  \\ 
2454302.7986 &    2.26     & 2454630.8919  &  -9.61  & 2454307.6692 &  12.70  \\ 
2454304.8177 &   19.95     & 2454632.8342  &  10.67  & 2454308.6166 &  18.08  \\ 
2454305.8033 &   30.86     & 2454633.6831  &  19.28  & 2454309.7481 &  36.45  \\ 
2454306.8129 &    9.31 & \multicolumn{2}{c}{V496 Aql} & 2454310.7181 &   9.30  \\ 
2454307.7181 &   -0.40     & 2454304.7716  &  10.86  &  2454311.7645 &  -4.85  \\ 
2454308.6830 &    6.64     & 2454302.5863  &  -0.34  &  2454312.7287 &   1.62  \\ 
2454309.5984 &   15.19     & 2454302.5914  &  -0.32  &  2454331.6906 &  -0.65  \\ 
2454310.8143 &   26.18     & 2454303.7521  &   4.29  &  2454332.5108 &   5.84  \\ 
2454311.7788 &   25.46     & 2454305.7241  &  14.40  &  2454334.5262 &  20.30  \\ 
2454312.8241 &   -0.59     & 2454306.7725  &   1.55  &  2454339.5496 &  10.37  \\ 
2454331.7254 &   19.79     & 2454307.7225  &  -5.59  &  2454341.5520 &  33.12  \\ 
2454332.7340 &   30.87     & 2454308.6425  &  -4.00  &  2454625.8782 &  -4.40   \\ 
2454333.6771 &    9.67     & 2454309.6342  &   0.67  &  2454626.8872 &  -3.10   \\ 
2454335.6354 &    7.36     & 2454310.7901  &   5.43  &  2454627.6480 &   2.69   \\ 
2454339.7009 &   -0.89     & 2454311.6308  &  11.40  &  2454630.8323 &  35.04   \\ 
2454341.6982 &   13.71     & 2454312.8179  &  12.64  &  2454631.8732 &  12.36   \\ 
2454625.9016 &    6.83     & 2454332.7026  &  14.41  &  2454632.8667 &  -5.53   \\ 
2454626.8971 &   16.15     & 2454335.5413  &  -5.49  &  2454633.6727 &  -0.62   \\ 
2454627.9075 &   24.67     & 2454336.6798  &  -0.42  &  2454634.8377 &   8.53   \\ 
2454631.7591 &   11.39     & 2454341.6457  &  -5.96  &  2454953.7113 &  -6.04   \\ 
2454632.8787 &   20.74     & 2454626.7744  &  -1.12  &               &          \\
2454633.8050 &   29.90     & 2454627.7452  &  -7.57  &               &         \\ 
             &             & 2454630.8954  &   4.48  &               &          \\
             &             & 2454633.6873  &  -2.61  &               &         \\

\hline 
\end{tabular} 
\label{TAB-RV}
\end{table*}

Tables~\ref{Tab-metal} and \ref{Tab-metalMC} list the published iron abundances for the Galactic Cepheids 
and for 15 MC Cepheids (only one of those, HV 837, is located in the SMC). Most Galactic [Fe/H] values come from 
Luck \& Lambert (2011) and Luck et al. (2011), while the majority of MC Cepheid determinations are from Romaniello et al. (2008).
To put all metallicities on the scale of Luck \& Lambert (2011), differences  were determined between the other references and them.
Following the reference numbers in Table~\ref{Tab-metal}, the differences are 
2-1 = $-0.07 \pm 0.09$ (N= 184 objects in common),
3-1 = $-0.11 \pm 0.11$ (N= 25),
4-1 = $-0.18 \pm 0.08$ (N= 11), and 
5-1 = $-0.09 \pm 0.07$ (N= 47).
The [Fe/H] values used in this paper are the published values corrected for these offsets.

For the five SMC Cepheids without metallicity determination the value in Storm et al. (2011b) is adopted: [Fe/H]= $-0.68$.
For the four Cepheids in the LMC cluster NGC 1866 without a metallicity
determination a value of $-0.39$ is adopted. This is the median of the values of
the three Cepheids in the cluster that do have a metallicity determination (HV 12197, HV 12199, and We2 (Molinaro et al. 2012)) 
and is in agreement with the average metallicity of $-0.43$ determined by Mucciarelli et al. (2011).
For the other LMC Cepheids without metallicity determination, the value used in Storm et al., $-0.34$ is used. This is very close
to the median value of $-0.36$ of the 12 LMC Cepheids not in NGC 1866 listed in Table~\ref{Tab-metalMC}.

Reddening values with errors were preferentially taken from Fouqu\'e et al. (2007, for 105 objects) 
and otherwise from van Leeuwen et al. (2007, for 16 objects, the value listed as ET in their Table~A1), 
Andrievsky et al. (2002, four objects), Luck et al. (2011, one object), and Tammann et al. (2004, two objects).
For the MC Cepheids, the values in Storm et al. (2011b) were adopted with an error in $E(B-V)$ of 0.005.
The $E(B-V)$ values and errors are listed in column~2 of Table~\ref{Tab-dist}.

The following Galactic Cepheids are considered to be first overtone pulsators 
(e.g. Klagyivik \& Szabados 2009): FF Aql, SZ Tau, SU Cas, QZ Nor, GH Lup, DT Cyg.
When relations are plotted against period, their periods are "fundamentalised" following Alcock et al. (1995).
None of the LMC and SMC Cepheids in the sample are believed to be overtone pulsators.

\begin{table}[!hp]

\caption{Published [Fe/H] values for the Galactic Cepheids.} 

%%%%% REFEREE VERSION %%%%%%%
%\tiny
%%%%% REFEREE VERSION %%%%%%%

{
\renewcommand{\arraystretch}{0.98}

\begin{tabular}{rrrr} \hline \hline 

Name     & [Fe/H]  & Name       & [Fe/H] \\  
\hline 

AK Cep & +0.05 (1) & S Vul  & +0.12 (1) \\
AN Aur & -0.10 (1) & SV Vul & +0.05 (2) \\
AQ Pup & +0.04 (1) & SW Cas & +0.13 (2) \\
AV Sgr & +0.34 (2) & SW Vel & +0.00 (1) \\
AW Per & +0.04 (1) & SX Vel & +0.06 (1) \\
BB Sgr & +0.08 (2) & SY Cas & +0.04 (2) \\
BE Mon &  0.08 (1) & SZ Aql & +0.17 (2) \\
beta Dor & -0.01 (3) & SZ Cyg & +0.09 (2) \\
BF Oph & +0.14 (1) & SZ Tau & +0.07 (2) \\
BG Lac & +0.07 (1) & T Mon  & +0.23 (1) \\
BM Per & +0.23 (1) & TT Aql & +0.22 (1) \\
BN Pup & +0.11 (1) & T Vel  & +0.04 (1) \\
BN Pup & +0.11 (1) & T Vul  & +0.01 (2) \\
BZ Cyg & +0.19 (2) & TW Nor & +0.33 (1) \\
CD Cyg & +0.15 (1) & TY Sct & +0.37 (1) \\
CF Cas & +0.02 (1) & TZ Mon & +0.01 (1) \\
CK Sct & +0.21 (1) & U Aql  & +0.17 (1) \\
CP Cep & -0.01 (2) & U Car  & +0.04 (1) \\
CR Cep & -0.06 (2) & U Nor  & +0.19 (1) \\
CS Vel & +0.12 (1) & U Sgr  & +0.08 (2) \\
CV Mon & +0.01 (1) & UU Mus & +0.19 (1) \\
DD Cas & +0.10 (2) & U Vul  & +0.19 (1) \\
delta Cep & +0.12 (1) & UZ Sct & +0.33 (2) \\
DL Cas & -0.01 (2) & V1162 Aql & +0.01 (2) \\
DT Cyg & +0.10 (2) & V340 Ara & +0.31 (2) \\
eta Aql & +0.08 (2) & V340 Nor & +0.16 (1) \\
FF Aql & +0.04 (2) & V350 Sgr & +0.18 (2) \\
FM Aql & +0.24 (1) & V386 Cyg & +0.11 (2) \\
FM Cas & -0.09 (2) & V402 Cyg & +0.02 (2) \\
FN Aql & -0.06 (1) & V459 Cyg & +0.09 (1) \\
GH Lup & +0.13 (1) & V495 Cyg & +0.24 (1) \\
GY Sge & +0.29 (1) & V496 Aql & +0.05 (2) \\
KN Cen & +0.41 (1) & V538 Cyg & +0.05 (1) \\
KQ Sco & +0.16 (2) & V600 Aql & +0.03 (2) \\
l Car  & +0.13 (1) & V916 Aql & +0.39 (1) \\
LS Pup & -0.16 (3) & V Car  & +0.04 (1) \\
MW Cyg & +0.09 (2) & V Cen  & +0.03 (1) \\
QZ Nor & +0.06 (4) & VW Cen & -0.02 (3) \\
RR Lac & +0.04 (1) & VX Cyg & +0.09 (2) \\
RS Cas & +0.18 (1) & VX Per & +0.06 (1) \\
RS Ori & -0.10 (2) & VY Car & +0.02 (1) \\
RS Pup & +0.22 (1) & VY Cyg & +0.00 (2) \\
RT Aur & +0.13 (1) & VY Sgr & +0.26 (2) \\
RU Sct & +0.11 (1) & VZ Cyg & +0.05 (2) \\
RW Cam & +0.11 (1) & VZ Pup & -0.11 (1) \\
RW Cas & +0.22 (2) & W Gem  & +0.02 (1) \\
RX Aur & +0.10 (1) & W Sgr  & +0.02 (2) \\
RX Cam & +0.11 (1) & WZ Car & +0.05 (1) \\
RY Cas & +0.26 (2) & WZ Sgr & +0.19 (2) \\
RY CMa & +0.02 (1) & X Cyg  & +0.10 (2) \\
RY Sco & +0.09 (2) & X Lac  & +0.08 (1) \\
RY Vel & +0.09 (1) & X Pup  & +0.08 (1) \\
RZ CMa & -0.03 (1) & X Sgr  & -0.29 (5) \\ 
RZ Gem & -0.17 (1) & X Vul  & +0.07 (2) \\
RZ Vel & +0.04 (1) & XX Cen & +0.18 (1) \\
S Mus  & +0.07 (1) & XX Sgr & +0.10 (2) \\
S Nor  & +0.13 (1) & Y Lac  & +0.03 (1) \\
S Sge  & +0.08 (2) & Y Oph  & +0.06 (2) \\
SS Sct & +0.14 (1) & Y Sct  & +0.23 (1) \\
ST Tau & +0.00 (1) & Y Sgr  & +0.05 (2) \\
SU Cas & +0.06 (2) & YZ Aur & -0.30 (1) \\
SU Cyg & -0.03 (2) & YZ Sgr & +0.06 (2) \\
SV Mon & -0.03 (2) & zeta Gem & +0.10 (1) \\
SV Per & +0.06 (1) & Z Lac  & +0.10 (1) \\

\hline 
\end{tabular} 
}

{\bf References.}
1= Luck \& Lambert (2011);
2= Luck et al. (2011); 
3= Romaniello et al. (2008);
4= Fry \& Carney (1997);
5= Andrievsky et al. (2003). % AN 324, 532

\label{Tab-metal}
\end{table}

\begin{table}
\caption{Published [Fe/H] values for the MC Cepheids.} 
\begin{tabular}{rrrr} \hline \hline 
Name     & [Fe/H]               & Name       & [Fe/H] \\  
\hline 

HV 837   & $-0.83 \pm 0.10$ (1) & HV 877   & $-0.44 \pm 0.10$ (1) \\
HV 879   & $-0.14 \pm 0.10$ (1) & HV 1023  & $-0.28 \pm 0.10$ (1) \\
HV 2369  & $-0.62 \pm 0.10$ (1) & HV 2405  & $-0.27 \pm 0.10$ (1) \\
HV 2827  & $-0.38 \pm 0.10$ (1) & HV 6093  & $-0.60 \pm 0.10$ (1) \\
HV 12452 & $-0.35 \pm 0.10$ (1) & \\ 
HV 12197 & $-0.39 \pm 0.05$ (2) & HV 12199 & $-0.38 \pm 0.06$ (2) \\
HV 900   & $-0.38 \pm 0.10$ (3) & HV 909   & $-0.28 \pm 0.10$ (3) \\
HV 2257  & $-0.34 \pm 0.10$ (3) & HV 2338  & $-0.44 \pm 0.10$ (3) \\

\hline
\end{tabular} 

{\bf References.}
1= Romaniello et al. (2008);
2= Molinaro et al. (2012);
3= Luck \& Lambert (1998).

\label{Tab-metalMC}
\end{table}

\section{The model}

The model is outlined in G07 but will be briefly repeated here. 
The $V$-, $K$- and RV data with error bars are fitted with a function of the form
\begin{equation}
F(t) = F_0 + \sum_{i=1}^{i=N} 
\left( A_{\rm i} \sin (2 \pi \; t \; e^{i f}) + 
       B_{\rm i} \cos (2 \pi \; t \; e^{i f}) \right),
\label{eq-f}
\end{equation}
where $P = e^{-f}$ is the period (in days). The period is
determined from the fit to the available optical photometry as this
dataset is usually the most extensive. The period is then fixed when
fitting Eq.~1 to the $K$-band and RV data.
 
The determination of the parameters is done using the {\sc mrqmin}
routine (following the Levenberg-Marquardt method) from Press et al. (1992), which minimises
\begin{equation} 
  \chi^2= \sum_{i=1}^{i=n} (F_i - F(t_i))^2 / (\sigma_{\rm F_i})^2,
\end{equation} 
with $F_i$ the measurement at time $t_{\rm i}$, which has an error bar $\sigma_{\rm F_i}$.
Also the reduced $\chi^2$ is defined as
\begin{equation} 
  \chi_{\rm r}^2 = \frac{\chi^2}{ (n-p)},
\end{equation} 
and the quantity BIC as
\begin{equation} 
  {\rm BIC} = \chi^2 + (p + 1)\; \ln (n), 
\end{equation} 
where $p = 2 N + 2$ is the number of free parameters ($p = 2 N + 1$
when fitting the RV and $K$ light curve).  As the number $N$ of
harmonics to be fitted to the data is a priori not known, one could
obtain ever better fits (lower $\chi^2$) by increasing $N$.  The
Bayesian information criterion (Schwarz 1978) is a formalism that
penalises this, and $N$ (for the $V$, $K$ and RV curve independently)
is chosen such that BIC reaches a minimum. The number of harmonics
used varies between 3 and 10 in the optical, 1 and 5 in the NIR, and 2
and 8 for the RV curves.

Given the analytical form of Eq.~1, the RV curve can be
exactly integrated to obtain the variation in radius as a function of
time (phase):
\begin{equation} 
\Delta R (t, \delta \theta)  = -p \; \int_{t_0}^{t + P \delta \theta} (v_{\rm R} - \gamma) \, d t,
\label{eq-fit}
\end{equation} 
where $\gamma$ is the systemic velocity, $v_{\rm R}$ the radial
velocity, $p$ the projection factor and $\delta \theta$ allows for a
phase shift between the RV curve and the angular-diameter variations determined via the SB relation.

Then, the equation
\begin{equation} 
  \theta (t) = 9.3038\, {\rm mas} \; \left( \frac{ R_0 + \Delta R (t, \delta \theta) }{ d } \right)
\label{eq-fit1}
\end{equation} 
is fitted with $\theta$ the angular diameter in mas, $R_0$ the stellar 
radius in solar radii and $d$ the distance in pc.

Compared to G08, the fitting procedure was changed. In G08, the fitting was done
implementing the linear bisector (using the code SIXLIN from Isobe et al. 1990) 
as used and preferred by e.g. Storm et al. (2004), Barnes et al. (2005), Gieren et al. (2005), and Storm et al. (2011a, b). 
The bisector is still the preferred method, but errors in both $\Delta R$ and $\theta$ are now taken into account 
using the Bivariate Correlated Errors and intrinsic Scatter (BCES) method
(Akritas \& Bershady 1996)\footnote{http://www.astro.wisc.edu/$^\sim$mab/archive/stats/stats.html}.
The error in $\theta(t)$ includes the error in $V$ and $K$ magnitude and the error in $E(B-V)$, 
while the error in $\Delta R$ includes the error in the Fourier coefficients, see Eq.~\ref{eq-f}.
An alternative method is also considered based on a non-linear least-squares fit with four parameters ($p$, $d$, $R_0$, and $\delta \theta$).
When the distance is known (the Cepheids with HST parallaxes, see Sect.~4), one solves for $p$ ($R_0$ and $\delta \theta$). 
In most cases (Sect.~6), one solves for $d$ ($R_0$ and $\delta \theta$) 
for a given $p$. In G08, this method was already implemented in order to derive $\delta \theta$. Then the bisector method was used 
with this $\delta \theta$ in order to derive the distance. However, the value of $\delta \theta$ that best fits the data from the 
non-linear fit does not necessarily best fit the data using the bisector. In the present paper, the non-linear fit is used to 
derive $\delta \theta$ and its error, and the BCES method is used for 21 values of $\delta \theta$ within $\pm 4\sigma$ 
of its best-fit value to find the best-fitting distance.
In some cases, a phase range around 0.85-0.95 is excluded from the fit (likely related to shocks in the stellar atmosphere close to minimum radius).

\section{The $p$-factor and surface-brightness relation}

One way of deriving the $p$-factor (and its dependence on period) is to use interferometrically 
determined angular diameters for a sample of stars with known distances, e.g. M\'erand et al. (2005) and G07.
The conclusion in G07 was that, statistically, there was no need to include a period dependence 
and that a constant value of $p$ = 1.27 $\pm$ 0.05 fitted the available data at that time, based on six stars.

The theoretical investigation by Nardetto et al. (2007) suggested that there is a difference between the $p$-factor 
to be used with wide-band interferometry (like in G07) and with RV data (when applying the SB technique as in G08 and the present study). 
For $\delta$ Cep, this difference is of the order of 0.06 (Nardetto et al. 2007), in the sense that in SB studies 
the $p$-factor is slightly larger and $p$ = 1.33 was used in G08.

New interferometric angular-diameter determinations have become available since 2008, and the $p$-factor dependence on period is investigated first.
Table~\ref{TAB-PFAC} lists the stars with an independent distance estimate and/or interferometrically determined (limb-darkened) angular diameters.
Columns 2 and 4 give the period and $E(B-V)$ value.

The distances come primarily from van Leeuwen et al. (2007), who
took the weighted average of the HST determined parallax (Benedict et al. 2007) 
and the revised \it Hipparcos \rm parallax given in the same paper (the exception was Y Sgr for which the HST value was adopted).
The Lutz-Kelker (LK) correction (Lutz \& Kelker 1973) in Col.~4 is taken from van Leeuwen et al. (2007) and included in determining 
the most probable distance.
For $\delta$ Cep and $\zeta$ Gem, the recent distance determinations 
to the host clusters (Majaess et al. 2012a, b) were also considered in the distances finally adopted, which are listed in Col.~6.
Column~7 lists the references to the interferometric observations.

Columns 8 and 9 list the derived $p$-factor and mean radius. The first error bar is
the error in the fit, while the second is the error due to the error in the distance.

\begin{figure} 
\includegraphics[width=85mm]{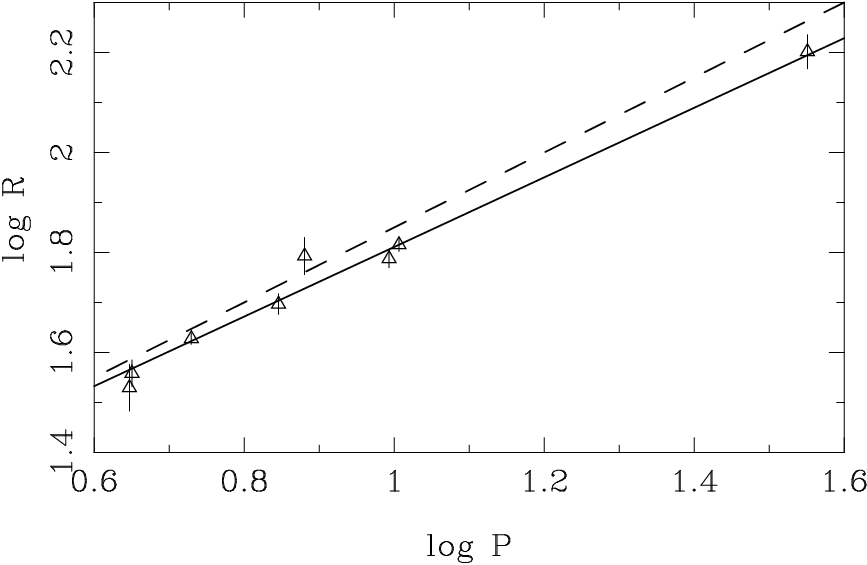} 
\caption[]{ 
The PR relation derived in the present paper (the solid line,
which is the fit to the data points with error bars) compared to the
PR relation of Molinaro et al. (2011), represented by the dashed line.
} 
\label{Fig-RAD} 
\end{figure}

The analysis of these eight stars allows one to derive a period-radius (PR) relation, which is shown in Fig.~\ref{Fig-RAD}.  
A linear weighted least-squares fit results in
\begin{equation} 
\log R = (0.696 \pm  0.033) \log P + (1.115 \pm 0.030), \sigma= 0.022,
\label{Eq-RAD} 
\end{equation} 
where the two error bars for $R$ listed in Table~\ref{TAB-PFAC} have been added quadratically.
Figure~\ref{Fig-RAD} shows the recent PR relation from Molinaro et al. (2011),
$\log R =(0.75 \pm 0.03) \log P + (1.10 \pm 0.03)$, for comparison.

Using the PR relation of Eq.~\ref{Eq-RAD}, the radii of $\eta$ Aql and Y Oph were estimated with their error bar. 
For this radius, the distance and $p$-factor were determined with their error bars 
(the internal error is listed first, and then the error due to the uncertainty in $R$).

\begin{figure} 
\includegraphics[width=85mm]{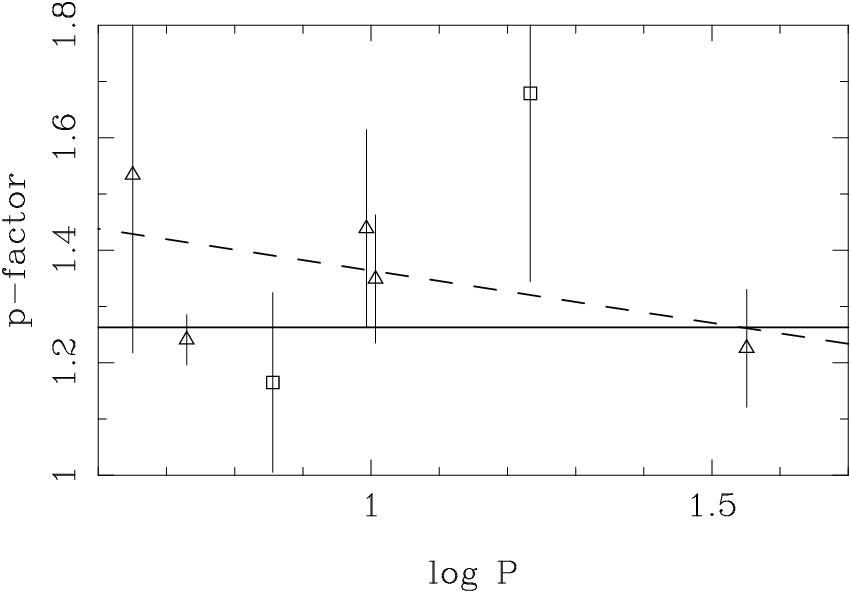} 
\caption[]{ 
The $p$-factor plotted versus $\log$ period for the seven stars that have an error in $p$ smaller than 0.4.
The open squares indicate the two stars for which the radius was fixed from the PR relation.  
Also plotted is a line indicating the best constant value of $p$= 1.264, while the dashed lines represent 
the $p$-factor relation proposed by Storm et al. (2011a), $p$= 1.550 -0.186 $\log P$.
} 
\label{Fig-PF} 
\end{figure}

Figure~\ref{Fig-PF} plots the relation between the derived $p$-factor and $\log P$ for the seven stars with an error 
(the quoted error bars on $p$ were added quadrature) less than 0.4.
A weighted least-squares fit is made to find that there is no evidence from this data alone 
of a dependence on period: $p = (1.24 \pm 0.12) + (+0.03 \pm 0.13) \log P$, nor on period and metallicity:
$p = (1.75 \pm 0.40) + (+0.06 \pm 0.13) \log P + (-4.5 \pm 3.4)$ [Fe/H].
The best-fitting constant value is $p$ = 1.264 $\pm$ 0.036, similar to what was found in G07.

\begin{table*} 

 \caption{Stars with parallaxes and/or interferometrically determined angular diameters.} 

\setlength{\tabcolsep}{1.2mm}

\begin{tabular}{rrccccccccccc} \hline \hline 

Name      &  Period   & $E(B-V)$ &   LK  & Ref &       $d$               & Ref.   &    $p$                           & $R$                             &  $p$   \\  
          &  (days)   &          & (mag) &     &       (pc)              &        &                                  &  (\rsol)                        &        \\
\hline
T Vul     &  4.435421 & 0.064 & -0.09 & 1   & 506   $\pm$ 57             & 4     & 1.781 $\pm$ 0.421 $\pm$ 0.201 &  33.85 $\pm$ 0.38  $\pm$  3.81 & 1.32 $\pm$ 0.16 \\
FF Aql    &  4.470896 & 0.196 & -0.03 & 1   & 384   $\pm$ 24             & 4     & 1.534 $\pm$ 0.301 $\pm$ 0.096 &  36.17 $\pm$ 0.15  $\pm$  2.26 & 0.77 $\pm$ 0.15 \\
delta Cep &  5.366250 & 0.075 & -0.01 & 1,2 & 272.4 $\pm$ 7.3            & 5,6   & 1.241 $\pm$ 0.030 $\pm$ 0.033 &  42.43 $\pm$ 0.038 $\pm$  1.13 & 1.44 $\pm$ 0.12 \\ 
X Sgr     &  7.012745 & 0.237 & -0.02 & 1   & 318   $\pm$ 14             & 7     & 1.025 $\pm$ 0.955 $\pm$ 0.045 &  49.75 $\pm$ 0.98  $\pm$  2.19 & 1.300 $\pm$ 0.087 \\
W Sgr     &  7.594925 & 0.108 & -0.06 & 1   & 447   $\pm$ 38             & 7     & 1.676 $\pm$ 0.871 $\pm$ 0.143 &  62.1  $\pm$ 1.6   $\pm$  5.3  & 2.83 $\pm$ 0.46 \\
beta Dor  &  9.842554 & 0.052 & -0.02 & 1   & 310   $\pm$ 13             & 7,8   & 1.439 $\pm$ 0.165 $\pm$ 0.060 &  61.33 $\pm$ 0.34  $\pm$  2.57 & 1.188 $\pm$ 0.062  \\
zeta Gem  & 10.149922 & 0.014 & -0.02 & 1,3 & 361   $\pm$ 11             & 6,7   & 1.349 $\pm$ 0.106 $\pm$ 0.041 &  65.39 $\pm$ 0.20  $\pm$  2.00 & 1.265 $\pm$ 0.061  \\
l Car     & 35.557209 & 0.147 & -0.05 & 1   & 504   $\pm$ 41             & 9,10  & 1.226 $\pm$ 0.030 $\pm$ 0.100 & 159.06 $\pm$ 0.27  $\pm$ 12.9  & 1.30 $\pm$ 0.11  \\
&         & &&        & & & \\                
eta Aql   &  7.176814 & 0.130 & -     & -   & 268   $\pm$ 1.4 $\pm$ 13.9 & 6,7,8 & 1.165 $\pm$ 0.148 $\pm$ 0.060 &  51.3 $\pm$ 2.7                & - \\  
Y Oph     & 17.126144 & 0.645 & -     & -   & 679   $\pm$ 5.1 $\pm$ 36   & 7     & 1.679 $\pm$ 0.323 $\pm$ 0.087 &  93.9 $\pm$ 4.9                & - \\  
\\                            
RT Aur    & 5.4820695 & 0.059 & -0.06 & 1   & 445   $\pm$ 38             & -     &                               &                                & 1.28 $\pm$ 0.14  \\    
Y Sgr     & 5.7644143 & 0.191 & -0.15 & 1   & 503   $\pm$ 75             & -     &                               &                                & 1.60 $\pm$ 0.37 \\    

\hline 
\end{tabular} 
\tablebib{
(1) van Leeuwen et al. (2007); (2) Majaess et al. (2012a); (3) Majaess et al. (2012b);
(4) Gallenne et al. (2012);   % T Vul, FFAql
(5) Merand et al. (2005);     %             delCep
(6) Nordgren et al. (2002);   %             delta Cep,  etaAql, zeta gem
(7) Kervella et al. (2004c);  % XSgr, WSgr,             etaAql, zeta Gem, betaDor, YOph ,(lCar)
(8) Jacob (2008);             %                         etaAql            betaDor, 
(9) Davis et al. (2009);      %                                                          lCar
(10) Kervella et al. (2004b). % (apj 604)                                                lCar
}
\label{TAB-PFAC}
\end{table*}

Independently of the derivation of the PR relation or the $p$-factor, 
the available interferometric, optical, and infrared data can be used
to calibrate the SB relation for Cepheids, very much in line with Kervella et al. (2004a) and G07. 
An SB relation can be defined as follows (see van Belle 1999):
\begin{equation}
\theta_o = \theta \cdot 10^{(m_1/5)},
\end{equation}
where $\theta$ is the LD angular diameter (in mas) and
$m_1$ a de-reddened magnitude (for example, $V$). The logarithm of
this quantity (the zero magnitude angular diameter) is plotted against a de-reddened colour (for example, $(V-K)_0$), 
\begin{equation}
\log \theta_0 = a \cdot (m_2 - m_3) + b.
\label{EQ-SB}
\end{equation}
The aim is to determine the coefficients $a$ and $b$.

\begin{figure} 
\includegraphics[width=85mm]{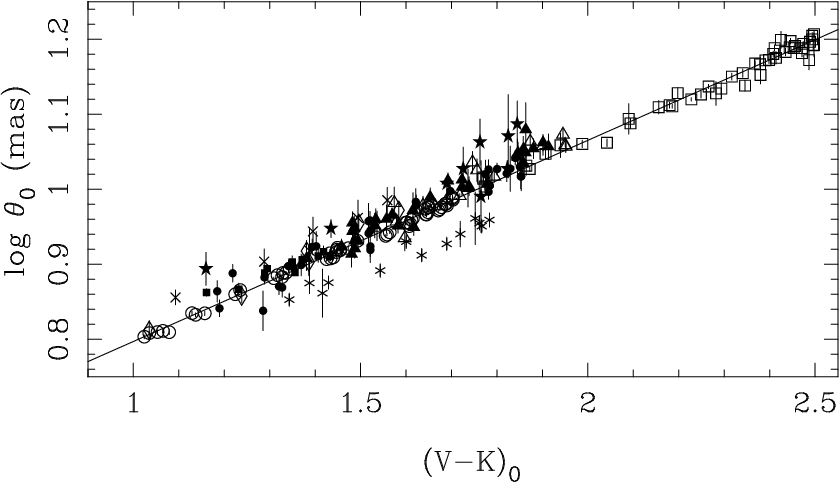} 
\caption[]{ 
Log $\theta_o$ versus de-reddened $(V-K)$ colour for Cepheids.
Different symbols indicate the different stars.
($\delta$ Cep= $\circ$, $l$ Car= $\square$, $\zeta$ Gem= $\triangle$,  $\eta$ Aql= $\bullet$,  $\beta$ Dor=  $\blacktriangle$, 
FF Aql= $\blacksquare$, T Vul= $\Diamond$, W Sgr= $\bigstar$, X Sgr= x, Y Oph= $\ast$)

%
% order 1=delCep 2=lCar 3=zetaGem 4=etaAql 5=betaDor 6=FFAql 7=TVul 8=WSgr 9=XSgr 10=YOph
% pgplot  4        6      7         17       13        16      11     18     5       3
% 
} 
\label{Fig-SB1} 
\end{figure}

Table~\ref{TAB-SB} list the various results.
Figure~\ref{Fig-SB1} shows the results, including all objects and where the error in the angular-diameter determination 
is less than 0.2\arcsec\ to exclude some extreme outliers.
Different stars are marked by different symbols as given in the caption. 
The relation is extremely well defined but as noted in G07 Y Oph (the $\ast$ -symbol) clearly lies below the relation, 
and there are still some data points with large error bars (e.g. W Sgr, indicated by the $\bigstar$ symbol).
The determination of the SB relation is affected by reddening, and Y Oph has by far the largest reddening of the stars 
under consideration here. A reddening of $E(B-V) \sim 1$ would bring it onto the relation.

\begin{figure} 
\includegraphics[width=85mm]{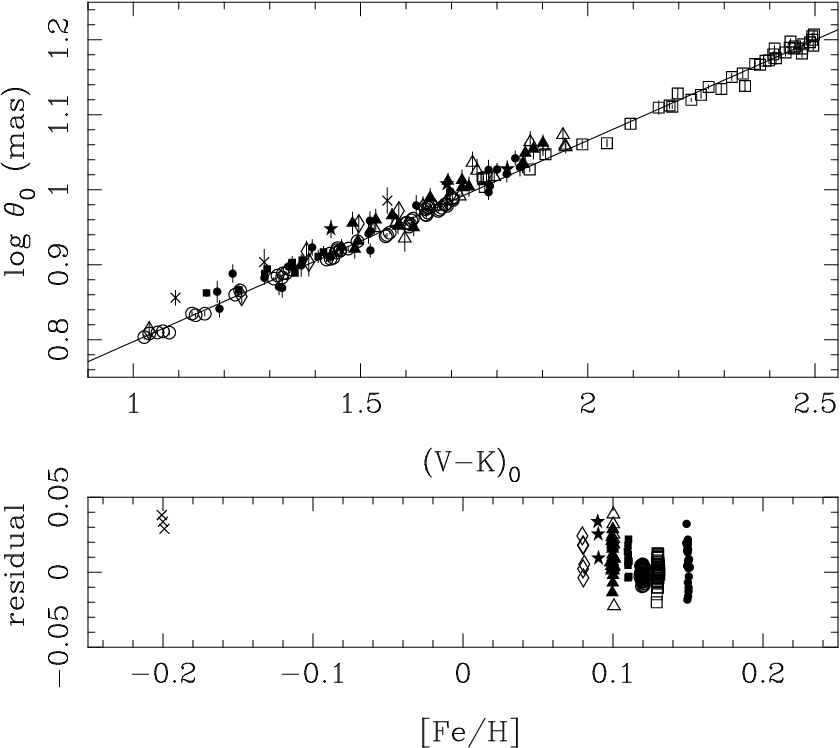} 
\caption[]{ 
Log $\theta_o$  versus  $(V-K)_0$ colour for Cepheids, excluding Y Oph and with errors on the angular diameters $<$0.065\arcsec.
The bottom panel shows the residuals plotted against [Fe/H].
Symbols as Fig.~\ref{Fig-SB1}.
} 
\label{Fig-SB2} 
\end{figure}

Figure~\ref{Fig-SB2} shows the results ignoring Y Oph and selecting only angular diameters determined with an error $<$0.065\arcsec.
More complicated relationships than Eq.~\ref{EQ-SB} were also investigated, in particular 
adding a quadratic term in $(V-K)_0^2$, a linear term on period, and a linear term in [Fe/H].
The quadratic term in colour and a linear term on period do not result in coefficients that are significant.
However, a linear term on metallicity (i.e. $\log \theta_0 = a \, (V - K)_0 + b + c$ [Fe/H] appears significant, 
as illustrated in the bottom panel of Fig.~\ref{Fig-SB2}.

\begin{figure} 
\includegraphics[width=85mm]{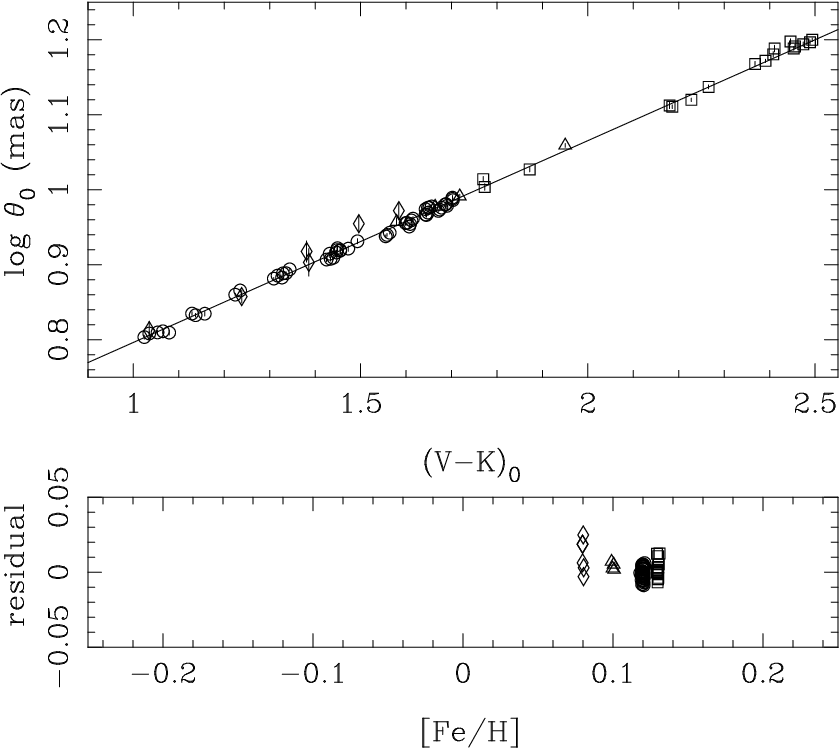} 
\caption[]{ 
Log $\theta_o$ versus $(V-K)_0$ colour for Cepheids, excluding Y Oph, X Sgr and FF Aql and with errors on the angular diameters $<$0.05\arcsec.
The bottom panel shows the residuals plotted against [Fe/H].
Symbols as Fig.~\ref{Fig-SB1}.
} 
\label{Fig-SB3} 
\end{figure}

This result does not depend on X Sgr at [Fe/H]= $-0.20$. Excluding it still results in a coefficient determined at the 3$\sigma$ level.
Uncertainties in reddening could introduce a metallicity dependence. 
%Ignoring FF Aql, X Sgr, W Sgr, $l$ Car and eta Aql, and restricting the sample to four stars with $E(B-V)$ less than 0.075.
When FF Aql is ignored as well, thereby restricting the sample to the stars with $E(B-V)$ less than 0.15, 
and only angular diameters determined with an error $<$0.05\arcsec\ are selected, 
the coefficient is determined at the the 2$\sigma$ level (see Fig.~\ref{Fig-SB3}).

Excluding X Sgr, the spread in [Fe/H] over the sample is small and comparable to the measurement error. 
To test a possible metallicity dependence of the SB relation, a Monte Carlo simulation was performed, 
assuming a Gaussian error of 0.05 dex in [Fe/H] and Gaussian errors in $E(B-V)$, $V$, and $K$.
The resulting SB relations are reported in Cols.~8-12 of Table~\ref{TAB-SB}.
The dependence on metallicity is weaker now, at the 1$\sigma$ level or less.

The SB relation adopted in this paper is based on the second solution from Table~\ref{TAB-SB}, $\log \theta_0 = 0.2674 \, (V - K)_0 + 0.5327$. 
Table~\ref{TAB-SB} also includes other SB relations, including the one by Kervella et al. (2004a), 
which has also been used by Storm et al. (2011a, b)\footnote{Note that in their 2011a paper near their Eq.~2, it is stated that 
the SB relation used is from Fouqu\'e \& Gieren (1997) but this is not the case. The 2011b paper has the correct reference.}.
The agreement is excellent.

\begin{figure} 

\begin{minipage}{0.49\textwidth}
\resizebox{\hsize}{!}{\includegraphics{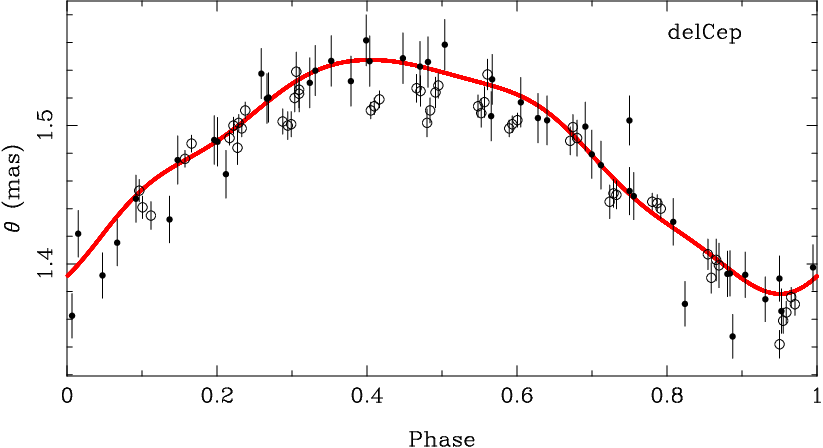}}
\end{minipage}
\begin{minipage}{0.49\textwidth}
\resizebox{\hsize}{!}{\includegraphics{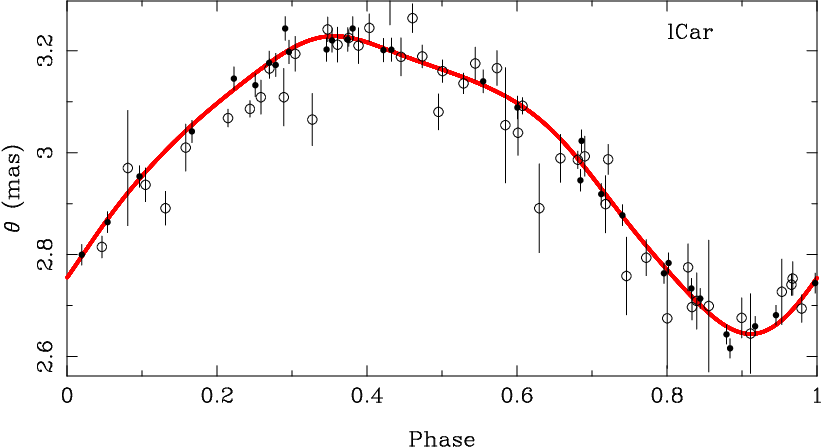}}
\end{minipage}

\caption[]{ 
Comparison between the angular diameters determined from the SB relation (filled circles), 
and from interferometry (open circles) from $\delta$ Cep and $l$ Car. 
} 
\label{Fig-CEPCAR} 
\end{figure}

Figure~\ref{Fig-CEPCAR} compares the angular diameters determined from interferometry with those calculated from photometry and the
SB relation for $\delta$ Cep and $l$ Car, the two Cepheids with the best and most extensive set of interferometrically determined angular diameters.
The agreement is very good and illustrates the power of SB relations.

\begin{table*} 
\setlength{\tabcolsep}{1.7mm}

 \caption{Coefficients of the SB relation.} 

\begin{tabular}{rrccccccccccc} \hline \hline 
Condition                        &  N  &  $b$     &  $a$     &  $b$     &  $a$     &  $c$  &  $b$     &  $a$     &  $b$     &  $a$     &  $c$    \\
\hline

$\sigma_{\theta} < 0.2$           & 226 &  0.5322  &  0.2672 & 0.5435 &  0.2689 &  -0.119 &    0.5335  &  0.2663 & 0.5382 &  0.2667 &  -0.056\\ %& 0.26822 & 0.5288 & 0.2702 & 0.5293  \\
                                 &     &  0.0032  &  0.0019 & 0.0039 &  0.0019 &   0.022 &    0.0029  &  0.0017 & 0.0078 &  0.0039 &   0.048\\ %& 0.00068 & 0.0012 & 0.0026 & 0.0049  \\

$\sigma_{\theta} < 0.065$         & 170 &  {\bf 0.5327}  &  {\bf 0.2674} & 0.5435 &  0.2688 &  -0.111 &    0.5338  &  0.2667 & 0.5384 &  0.2664 &  -0.048 \\ %& 0.26825 & 0.5291 & 0.2667 & 0.5365  \\ 
Y Oph excluded                   &     &  0.0033  &  0.0020 & 0.0043 &  0.0020 &   0.028 &    0.0028  &  0.0016 & 0.0073 &  0.0032 &   0.043\\ %& 0.00069 & 0.0012 & 0.0018 & 0.0034  \\

%$\sigma_{\theta} < 0.05$        & 133 &  0.5311  &  0.2679 & 0.5428 &  0.2694 &  -0.118 & 0.26840 & 0.5287 & 0.2673 & 0.5334  \\
%Y Oph excluded                  &     &  0.0034  &  0.0020 & 0.0050 &  0.0020 &   0.038 & 0.00069 & 0.0012 & 0.0016 & 0.0030  \\

$\sigma_{\theta} < 0.065$         & 167 &  0.5317  &  0.2679 & 0.5479 &  0.2691 &  -0.151 &    0.5329  &  0.2672 & 0.5370 &  0.2662 &  -0.032\\ %& 0.26834 & 0.5289 & 0.2667 & 0.5341  \\ 
Y Oph \& X Sgr excluded          &     &  0.0033  &  0.0020 & 0.0070 &  0.0020 &   0.057 &    0.0029  &  0.0016 & 0.0099 &  0.0035 &   0.069\\ %& 0.00069 & 0.0012 & 0.0017 & 0.0031  \\

$\sigma_{\theta} < 0.03$          &  79 &  0.5251  &  0.2705 & 0.5477 &  0.2723 &  -0.212 &    0.5263  &  0.2697 & 0.5319 &  0.2679 &  -0.030 \\ % 0.26905 & 0.5272 & 0.2705 & 0.5258  \\
Y Oph, X Sgr, \& FF Aql excluded &     &  0.0042  &  0.0026 & 0.0133 &  0.0028 &   0.119 &    0.0038  &  0.0020 & 0.0120 &  0.0030 &   0.096 \\ % & 0.00076 & 0.0013 & 0.0014 & 0.0023  \\

 & \\
Fouqu\'e \& Gieren (1997)        &     &  0.547   & 0.262  \\
                                 &     &  0.006   & 0.004  \\
Kervella et al. (2004a)          &     &  0.5354  & 0.2672 \\
                                 &     &  0.0012  & 0.0016 \\
G07                              &     &  0.5235  & 0.2752 \\
                                 &     &  0.0092  & 0.0045 \\
\hline 
\end{tabular} 

\tablefoot{Column 1 gives the conditions applied or the reference to the literature,
column 2 the number of data points, and 
columns 3-4 and 5-7 the coefficients of the SB relation. Columns 8-12 repeat these numbers based on the Monte Carlo simulation.
The second line gives the errors in the coefficients.
The adopted coefficients are indicated in boldface.
}
\label{TAB-SB}
\end{table*}

\section{Binary Cepheids}
\label{sect-bin}
 
A number of stars in the sample are known or suspected spectroscopic binaries. 
In order to apply the SB technique outlined in the previous section, the RV data have to be corrected for the binary motion.
The procedure is outlined in G08.
For a number of stars\footnote{ S Mus, S Nor, S Sge, SU Cas, SU Cyg, T Mon, U Vul, W Sgr, XX Cen, Z Lac}, the elements of G08 were used.
For FF Aql, V350 Sgr, V496 Aql, and VZ Cyg, new RV data allowed for an improved orbit compared to G08.
For X Sgr, the new RV data do not support the solution presented in G08. Although a period analysis continues to show a peak at 572 days, 
the binary solution is not satisfactory. For X Sgr no binary orbit is assumed here.
The orbital elements (three updated and five new orbits) are listed in Table~\ref{Tab-Bin}. 

For V496 Aql the period changed considerable. A period analysis showed that the period preferred in 
G08 (1331 days) only has the third highest peak in the current Fourier spectrum 
and that 1065 days clearly shows the better fit.
For V350 Sgr the orbit is updated and compares well to the recent orbit by Evans et al. (2011).

For the known spectroscopic binaries, the derived orbital parameters are close to literature values, cf. 
RX Cam  (Imbert 1996), DL Cas (Gieren et al. 1994),
MW Cyg (Imbert 1996, and Rastorgouev et al. 1997. The eccentricity derived here is in between the two values quoted),
AW Per (Evans et al. 2000), and
U Aql (Welch et al. 1987).
For Y Oph, the very non-eccentric orbit proposed by Abt \& Levy (1978) is not confirmed.

\begin{table*} 
\caption{Derived orbital parameters of binary Cepheids. Quantities without error bar have been fixed.
} 
\begin{tabular}{rrrcccc} \hline \hline 
Name      & $\gamma$  (\ks)  &    $K$  (\ks) &               $e$        &   $\omega$ ($\degr$) &  $T_0$  (JD-2400000)  &    Binary Period (d)    \\  
\hline 
%FF Aql   &                    4.91 $\pm$ 0.07 & 0.027 $^{+0.041}_{-0.016}$ & 319   $\pm$ 45   &  45437   $\pm$ 178   &  1432.4  $\pm$   1.1  \\ G08
FF Aql   & -15.71 $\pm$ 0.07 & 4.98 $\pm$ 0.06 & 0.011 $^{+0.033}_{-0.008}$ & 362   $\pm$ 103  &  45609   $\pm$ 415   &  1432.7  $\pm$   0.9  \\

%V496 Aql  &                     3.0             & 0                        & 0                &  45606   $\pm$  25   &  1331    $\pm$   6.5  \\ G08
V496 Aql  &  4.51 $\pm$ 0.14 & 3.63  $\pm$ 0.18  & 0                        & 0                &  45480   $\pm$  17   &  1066.2    $\pm$   1.9  \\

%V350 Sgr  &                  11.21 $\pm$ 0.12 & 0.405 $^{+0.015}_{-0.014}$ & 279   $\pm$  2   &  47594   $\pm$   9   &  1482    $\pm$   2.4  \\  G08
V350 Sgr  & 11.43 $\pm$ 0.05  & 10.50 $\pm$ 0.07 & 0.351 $^{+0.007}_{-0.007}$ & 283   $\pm$  1   &  52015   $\pm$   4   &  1468.9    $\pm$   0.9  \\ 

%VZ Cyg    &                    3.46 $\pm$ 0.19 & 0                        & 0                &  45002   $\pm$ 50    &  2092    $\pm$  21    \\ G08 
VZ Cyg    & -16.96 $\pm$ 0.10 &  3.02 $\pm$ 0.16 & 0                        & 0                &  44811   $\pm$ 36    &  2183    $\pm$  10    \\ 

%X Sgr    &  2.30 $\pm$ 0.27 & 0                        & 0                &  48208   $\pm$  19   &   573.6  $\pm$   0.6  \\ G08

RX Cam    & -37.54 $\pm$ 0.06 & 14.27 $\pm$ 0.11 & 0.459 $^{+0.007}_{-0.006}$ &  78.4 $\pm$ 1.0   &  45931.3  $\pm$   1.8  &  1113.8  $\pm$   0.5  \\ % nearly identical to Imbert 1996
DL Cas    & -36.23 $\pm$ 0.06 & 16.43 $\pm$ 0.11 & 0.350 $^{+0.006}_{-0.006}$ &  27.3 $\pm$ 1.0   &  47161.6  $\pm$   1.5  &   684.27 $\pm$   0.16 \\ % nearly identical to Gieren et al. 1994
MW Cyg    & -13.37 $\pm$ 0.12 &  6.43 $\pm$ 0.19 & 0.140 $^{+0.030}_{-0.025}$ &  78   $\pm$ 13    &  48862    $\pm$  15    &   439.61 $\pm$   0.18 \\ % ecc in between Imbert and Rast.
AW Per    &   6.60 $\pm$ 0.35 & 12.06 $\pm$ 0.32 & 0.499 $^{+0.032}_{-0.030}$ & 254   $\pm$ 3     &  38721    $\pm$ 178    & 14293.70 $\pm$ 283    \\ % EVW00
U Aql     &   1.14 $\pm$ 0.14 &  7.75 $\pm$ 0.22 & 0.134 $^{+0.025}_{-0.021}$ & 163   $\pm$ 9.5   &  42634    $\pm$  49    &  1853.6  $\pm$   3.0  \\ % WELHBSM87

\hline 
\end{tabular} 
\label{Tab-Bin} 
\end{table*}

\section{Results}

\subsection{The period dependence of the $p$-factor}
\label{sect-pp}

In their most recent work, Storm et al. (2011a) proposed a $p$-factor of $(1.55 \pm 0.04) −- (0.186 \pm 0.06) \log P$, 
confirming their earlier result of  $p = (1.58 \pm 0.02) −- (0.15 \pm 0.05) \log P$ by Gieren et al. (2005).
The slope is derived from the requirement that the distance to the barycenter of the LMC should not depend on period.

Independently, for the Galactic Cepheids with HST parallaxes, one can determine the $p$-factor which
makes the BW distance equal to the HST-based distance.
Storm et al. (2011a) find $p = (1.65 \pm 0.07) −- (0.28 \pm 0.08) \log P$.
The $p$-factors derived in this way are reported in Col.~10 of Table~\ref{TAB-PFAC} and a fit to the six stars with and error $<$0.15 gives 
$p= (1.33 \pm 0.16) - (0.07 \pm 0.16) \log P$. The best-fit constant value is $p$= 1.27 $\pm$ 0.03.
In a similar fashion, Ngeow et al. (2012) considered not only stars with an HST parallax (from van Leeuwen et al. 2007) 
but also Cepheids in clusters (from Turner 2010), determining that $p$-factor that makes the BW distance 
(which they took from Storm et al.) equal to the independently known distance. They find $p = (1.462 \pm 0.087) + (-0.172 \pm 0.086) \log P$,
%, with σ = 0.107
or, excluding the outlier FF Aql, $p = (1.447 \pm 0.070) + (-0.159 \pm 0.070) \log P$.
%with a dispersion of 0.064

\begin{figure} 

\begin{minipage}{0.49\textwidth}
\resizebox{\hsize}{!}{\includegraphics{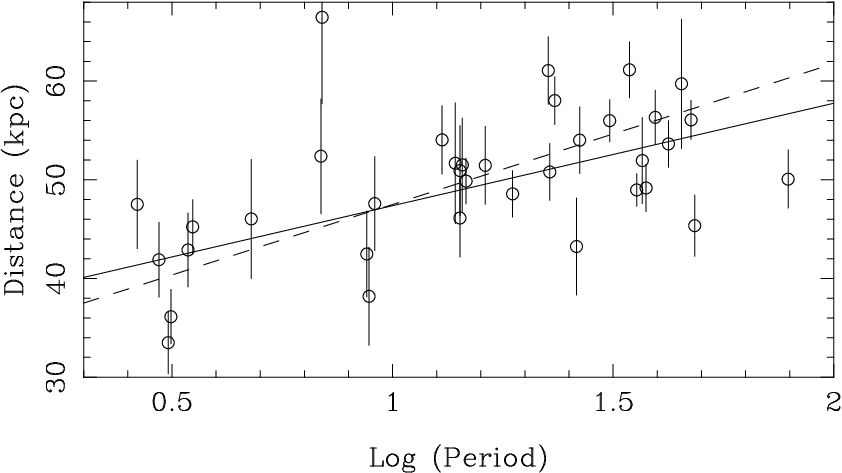}}
\end{minipage}
\begin{minipage}{0.49\textwidth}
\resizebox{\hsize}{!}{\includegraphics{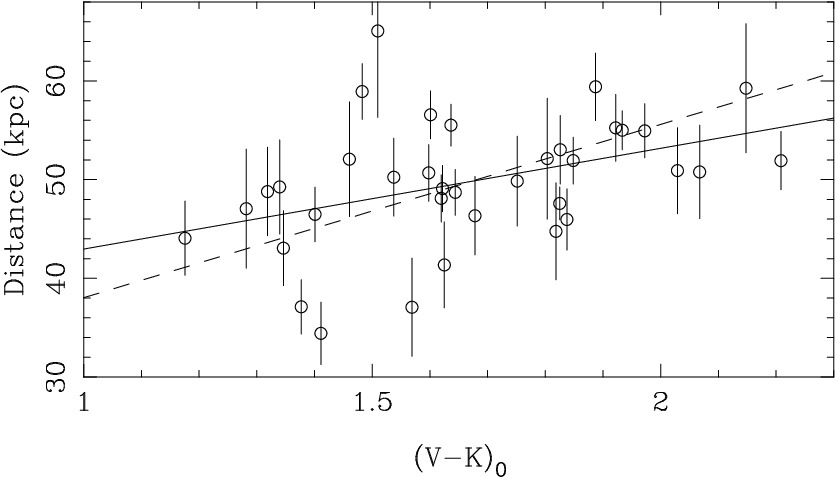}}
\end{minipage}

\caption[]{ 
Distance to the barycenter of the LMC against $\log P$ and against mean $(V-K)$ colour for a constant $p$-factor of 1.33.
The dashed line indicates the bisector fit, the solid line the least-squares solution.
} 
\label{Fig-LMCdist} 
\end{figure}

Figure~\ref{Fig-LMCdist} shows the distance  to the LMC Cepheids versus $\log P$ for $p$= 1.33.
As the distance is proportional to the $p$-factor and since the dependence on period is assumed to be linear in $\log P$, 
the slope in this plot indicates what the $p - \log P$ dependence should be to have no dependence of distance on period.
Depending on whether the slope is derived from the bisector (the dashed line), or a weighted least-squares fit (the solid line), 
the result is $-0.28 \pm 0.05$ or $-0.21 \pm 0.04$, respectively.

One can also make the consideration that the distance should be independent of (mean) $(V-K)$ colour. 
Figure~\ref{Fig-LMCdist} shows the result for $p$= 1.33. As period and colour are related, this is not independent of the earlier estimate.
Trying various coefficients, it is found that a slope of  $-0.21 \pm 0.05$ or $-0.24 \pm 0.05$ will give no 
dependence on colour, for the bisector and weighted least-squares fit, respectively.

The BW method is primarily of interest because it gives absolute distances.
For the Cepheids in the LMC, one can also expect and demand that the {\it slope} of the PL relation in the $V-$ and $K-$band is independent
of whether it is derived from $M_{\rm V}$ and $M_{\rm K}$ taking the BW distances or from the purely observed mean $V_0$ and $K_0$ magnitudes.

\begin{figure} 

\begin{minipage}{0.49\textwidth}
\resizebox{\hsize}{!}{\includegraphics{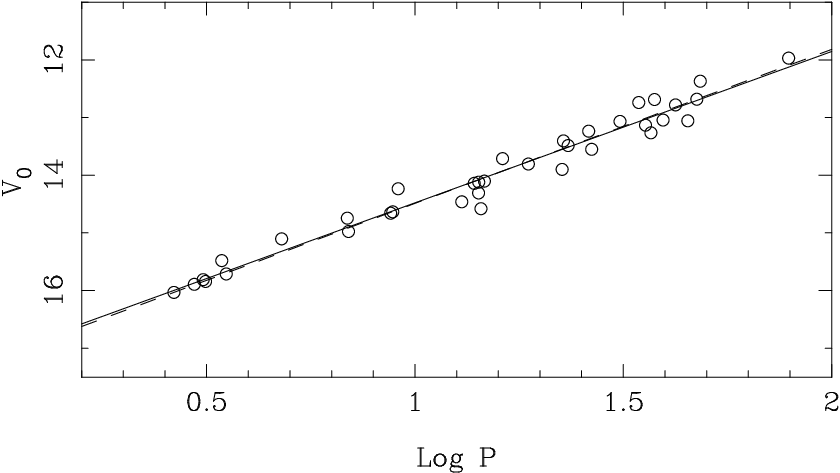}}
%\resizebox{\hsize}{!}{\includegraphics{PLobs_LMC_V_Pgt4lt170.ps}}
\end{minipage}
\begin{minipage}{0.49\textwidth}
\resizebox{\hsize}{!}{\includegraphics{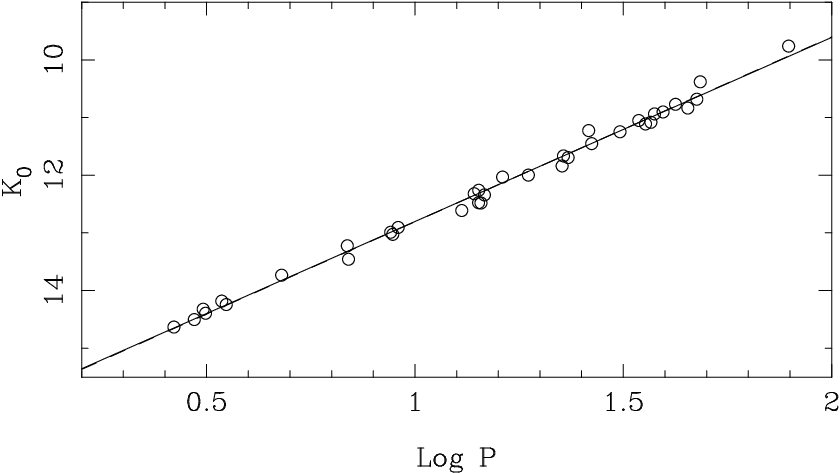}}
%\resizebox{\hsize}{!}{\includegraphics{PLobs_LMC_K_Pgt4lt170.ps}}
\end{minipage}
\begin{minipage}{0.49\textwidth}
\resizebox{\hsize}{!}{\includegraphics{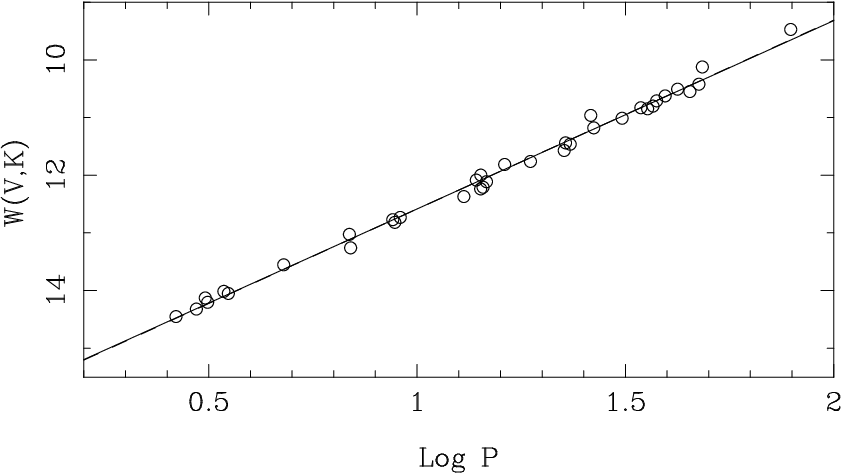}}
\end{minipage}

\caption[]{ 
Observed $V$-band (upper panel), $K$-band (middle) and $W(VK)$ (bottom) PL relations for the LMC Cepheids.
The dashed line indicates the bisector fit, the solid line the least-squares solution (indistinguishable except in the upper panel).
} 
\label{Fig-LMCPLobs} 
\end{figure}

The observed PL relations are listed in Table~\ref{Tab-lmcfits} and shown in Fig.~\ref{Fig-LMCPLobs}.
Apart from $V$ and $K$, the reddening free combination of these colours is also included, $W(VK)=  K - 0.13\; (V-K)$ (see Inno et al. 2012).
Three solutions are listed, depending on whether the LMC Cepheids are put at the barycenter or not. 
In the former case, one corrects for the tilt and orientation of the LMC disk. The model by van der Marel \& Cioni (2001) was used 
(as Storm et al. 2011b did). It is based mainly on AGB stars located between 2.5 and 6.7\degr\ from the LMC centre.
An alternative model used is that by Nikolaev et al. (2004), which is based on 2100 Cepheids within 4\degr\ from the centre.
The slope and zero point of the observed PL relations are very similar, independent of the type of correction. 
The smallest dispersion in the $K-$band is actually achieved when applying {\it no} correction.
The solution was calculated using a bisector and a weighted least-squares fit and is largely in agreement.
The slopes derived here from the observations are in excellent agreement with literature values, 
some of which are also listed in Table~\ref{Tab-lmcfits}.

\begin{figure} 

\begin{minipage}{0.49\textwidth}
\resizebox{\hsize}{!}{\includegraphics{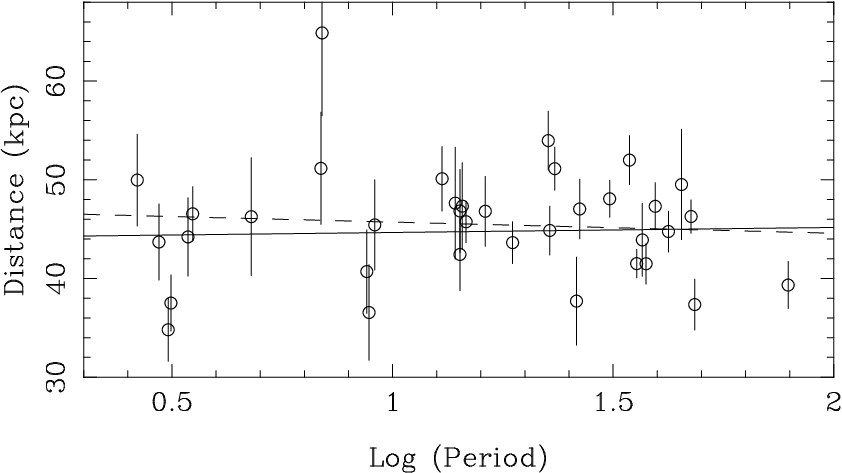}}
\end{minipage}
\begin{minipage}{0.49\textwidth}
\resizebox{\hsize}{!}{\includegraphics{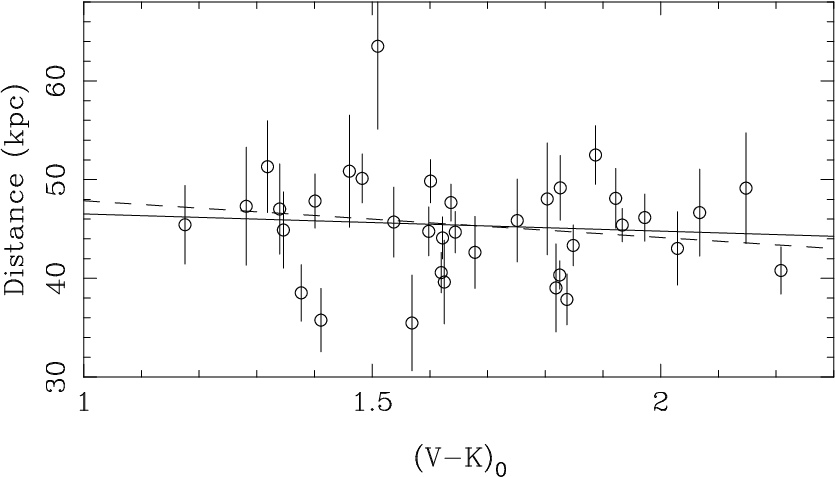}}
\end{minipage}

\caption[]{ 
Distance to the barycenter of the LMC against $\log P$ and against mean $(V-K)$ colour for a $p$-factor of $1.50 - 0.24 \log P$.
The dashed line indicates the bisector fit, the solid line the least-squares solution.
} 
\label{Fig-LMCdist1} 
\end{figure}

As the absolute magnitudes depend on the distance, which in turn depends on the $p$-factor, 
changing the slope in the $p - \log P$ relation will change the slope of the PL relation.
Demanding that the two are equal, it is found that the slope is $-0.25 \pm 0.05$ and $-0.25 \pm 0.05$, in the $K-$ and $V-$band, respectively.
Taking the average of these six determinations, the finally adopted relation is $p= p_o -0.24 \log P$ (with an estimated error of 0.03).
The dependence of distance on $\log P$ and $(V-K)$ colour for the LMC Cepheids is shown in Fig.~\ref{Fig-LMCdist1}.

\begin{table*} 

\caption{Observed, using a bisector and weighted least-squares fit, and literature PL relations in the LMC.
The zero point is listed in the first line, the slope in the second line.}

\begin{tabular}{lcclccccc} \hline \hline 

Relation                &      bisector    &       wLSQ        & remarks \\  
\hline

Observed $K$-band       & 16.01 $\pm$ 0.05 & 15.99 $\pm$ 0.04  &  Correction van der Marel \& Cioni (2001)   \\ %with offset vanM&C
                        & -3.20 $\pm$ 0.04 & -3.19 $\pm$ 0.04  &  \\ %with offset 

Observed $K$-band       & 16.03 $\pm$ 0.05 & 16.02 $\pm$ 0.04  & Correction Nikolaev et al. (2004)  \\ %with offset Nikolaev
                        & -3.22 $\pm$ 0.04 & -3.21 $\pm$ 0.04  & \\ 

Observed $K$-band       & 16.06 $\pm$ 0.04 & 16.05 $\pm$ 0.03  & No correction  \\ % NO offset 
                        & -3.25 $\pm$ 0.03 & -3.24 $\pm$ 0.03  &  \\ % NO offset 
 
Literature $K$-band     &                  & 15.996 $\pm$ 0.010 & Ngeow et al. (2009) \\
                        &                  & -3.194 $\pm$ 0.015 &  \\

Literature $K$-band     &                  & 16.070 $\pm$ 0.017 & Ripepi et al. (2012) \\
                        &                  & -3.295 $\pm$ 0.018 &  \\

Observed $V$-band       & 17.15 $\pm$ 0.08 & 17.09 $\pm$ 0.07  &   Correction van der Marel \& Cioni (2001) \\ %with offset vdM&C
                        & -2.67 $\pm$ 0.07 & -2.62 $\pm$ 0.07  &   \\ %with offset 

Observed $V$-band       & 17.18 $\pm$ 0.07 & 17.13 $\pm$ 0.07 & Correction Nikolaev et al. (2004) \\ % offset Nikolaev etal
                        & -2.69 $\pm$ 0.06 & -2.65 $\pm$ 0.06 &  \\ %  offset 

Observed $V$-band       & 17.21 $\pm$ 0.07 & 17.16 $\pm$ 0.07  &  No correction \\ % NO offset 
                        & -2.72 $\pm$ 0.06 & -2.68 $\pm$ 0.06  &   \\ % NO offset 

Literature $V$-band     &                  & 17.115 $\pm$ 0.015 & Ngeow et al. (2009) \\
                        &                  & -2.769 $\pm$ 0.023 &  \\

Literature $V$-band     &                  & -1.304 $\pm$ 0.065 & Turner et al. (2010) \\
                        &                  & -2.786 $\pm$ 0.075 & \\

Observed $W(VK)$       & 15.86 $\pm$ 0.05 & 15.85 $\pm$ 0.04  &   Correction van der Marel \& Cioni (2001) \\ %with offset vdM&C
                        & -3.27 $\pm$ 0.04 & -3.26 $\pm$ 0.04  &   \\ %with offset 

Observed $W(VK)$       & 15.88 $\pm$ 0.04 & 15.87 $\pm$ 0.04  &   Correction Nikolaev et al. (2004) \\ %with offset vdM&C
                        & -3.30 $\pm$ 0.04 & -3.29 $\pm$ 0.04  &   \\ %with offset 

Observed $W(VK)$       & 15.91 $\pm$ 0.03 & 15.91 $\pm$ 0.03  &   No Correction  \\ %with offset vdM&C
                        & -3.32 $\pm$ 0.03 & -3.32 $\pm$ 0.03  &   \\ %with offset 

Literature $W(VK)$     &                  & 15.901 $\pm$ 0.005 & Inno et al. (2012) \\
                        &                  & -3.326 $\pm$ 0.008 &  \\

Literature $W(VK)$     &                  & 15.870 $\pm$ 0.013 & Ripepi et al. (2012) \\
                        &                  & -3.325 $\pm$ 0.014 &  \\

\hline 

\end{tabular} 
\label{Tab-lmcfits}
\end{table*}

\subsection{The zero point of the $p - \log P$ relation}

As in Storm et al. (2011a) the zero point of the $p - \log P$ is based
on stars that have an independent distance, namely those with a
parallax (the ten stars listed in Table~\ref{TAB-PFAC}) and the Cepheids that are in clusters.
For the latter, Storm et al. (2011) used the associations and distances in Turner (2010).
However, this list is not complete (cf. Tammann et al. 2003), and some results have become available since 2010.
A discussion of the distances adopted for Cepheids in clusters in the present sample is presented in Appendix~A.

\medskip
\noindent
The relation finally adopted is  $p= 1.50 - 0.24 \log P$.
%% comparedist.dat Cluster.dat
%
For this relation, the weighted mean of the ratio of reference distance to BW distance (see later) is 1.000 $\pm$ 0.026 for eight stars which have a parallax 
(the two outliers that deviate by more than 3.4$\sigma$ and that are removed are W Sgr with a ratio of 2.20 $\pm$ 0.36, and FF Aql with 0.58 $\pm$ 0.11), 
and 0.999 $\pm$ 0.014 for the 18 stars in clusters. In the latter case, the three outliers that deviate by more than 4.5$\sigma$ and that are removed are 
SU Cyg with a ratio of 0.76 $\pm$ 0.05, VY Car with 1.57 $\pm$ 0.11, and X Cyg with 1.23 $\pm$ 0.04, 
while the recent determinations for $\delta$ Cep and $\zeta$ Gem were already considered when deriving the best distances 
for the stars with parallaxes (see Table~\ref{TAB-PFAC}).

In comparison, Storm et al. (2011a) base their zero point solely on nine stars with HST parallax 
(W Sgr is excluded as well, but its BW distance for FF Aql agrees with the HST-based one).
An aposteriori comparison to the Cepheids in clusters (with the distances from Turner 2010) using 16 stars in common with their Galactic Cepheid sample 
(of which SU Cas is excluded, as the result from Majaess et al. (2012c) was not available to them) 
shows an unweighted mean difference of 0.12 $\pm$ 0.06 in distance modulus.

Table~\ref{Tab-dist} lists the distances, radii, and absolute magnitudes obtained for the Galactic, LMC, and SMC Cepheids.
The table also lists the adopted $E(B-V)$ and error bar, the derived
period, and the $p$-factor following $p = 1.50 - 0.24 \log P$ adopted in the present paper. 
The distances, radii, and errors scale directly with $p$. 
The error in the period is a few units in the last decimal place. 
For the derived quantities two error bars are quoted. For the distance
and radius, the first error bar listed is the error in the fit, and for
the absolute magnitudes the error is due to the error in distance and $E(B-V)$. 
The second error quoted is based on a Monte Carlo simulation, where
(1) new datasets are generated based on the error bar in each individual $V$, $K$, and RV measurement, 
(2) the analysis takes into account an $E(B-V)$ value randomly
drawn from a Gaussian distribution based on the listed mean value and 1$\sigma$ error bar, 
(3) a random error in the $p$-factor of 0.02 units, and
(4) a variation in the number of harmonics used to describe the optical, infrared, and RV curves.
The second error quoted is the 1$\sigma$ dispersion in the derived quantities.

\begin{figure} 

\begin{minipage}{0.49\textwidth}
\resizebox{\hsize}{!}{\includegraphics{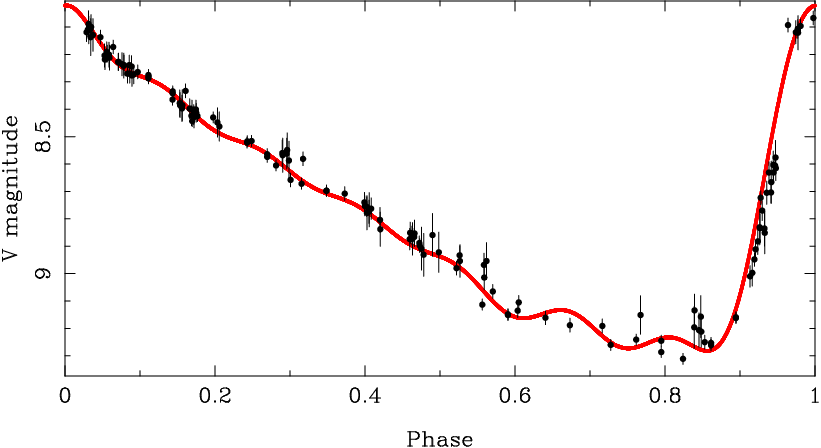}}
\end{minipage}
\begin{minipage}{0.49\textwidth}
\resizebox{\hsize}{!}{\includegraphics{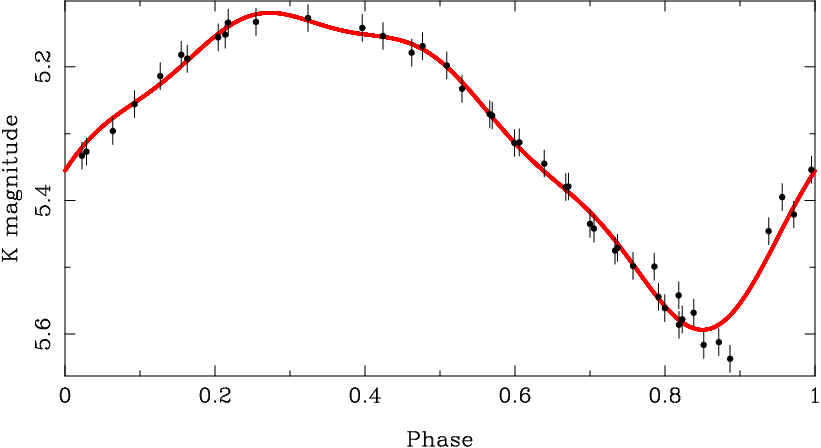}}
\end{minipage}
\begin{minipage}{0.49\textwidth}
\resizebox{\hsize}{!}{\includegraphics{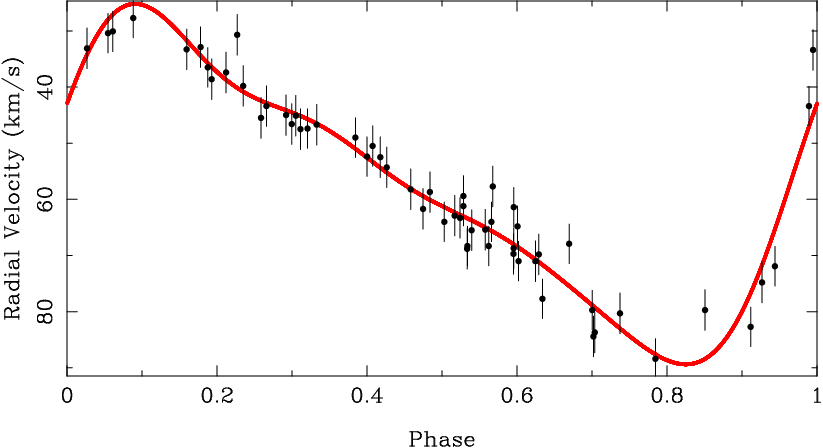}}
\end{minipage}

\caption[]{ 
Phased curves in $V$, $K$, and RV are shown for AQ Pup. Data
points are shown with errors bars and the line shows the harmonic fit. 
} 
\label{Fig-AQPUP1} 
\end{figure}

\begin{figure} 

\begin{minipage}{0.49\textwidth}
\resizebox{\hsize}{!}{\includegraphics{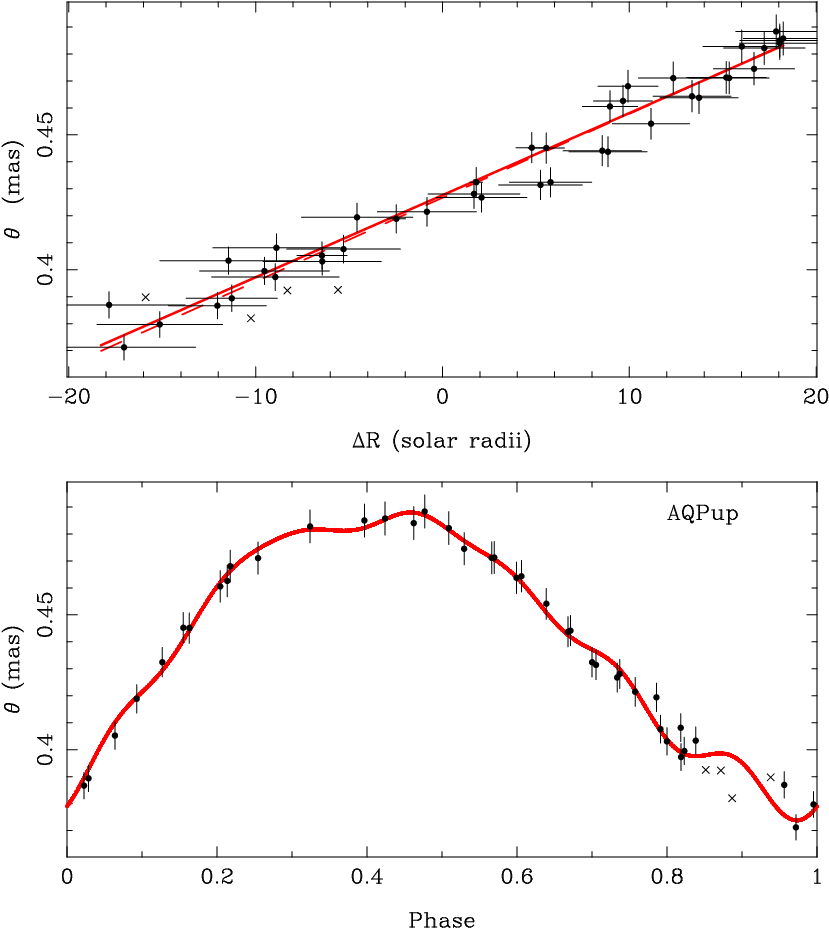}}
\end{minipage}

\caption[]{ 
For AQ Pup, the top panel shows the linear-bisector fit
to the angular diameter as a function of radial displacement.  The
bottom panel shows the angular diameter against phase. Crosses
represent data points not considered in the fit. 
%The fits to all stars in the sample are available from the author upon request.
% 
} 
\label{Fig-AQPUP2} 
\end{figure}

Figure~\ref{Fig-AQPUP1} illustrates the fit to the $V$, $K$, and RV
curve for AQ Pup, while Fig.~\ref{Fig-AQPUP2} shows the variation of
the angular diameter against phase and the change in angular diameter
derived from the SB relation against the change in radius from
integration of the RV curve from which the distance is derived (see
Eq.~\ref{eq-fit}). Figures similar to Figs.~\ref{Fig-AQPUP1} and
\ref{Fig-AQPUP2} for all stars in the sample are available from the author.

\begin{table*} 
\setlength{\tabcolsep}{1.3mm}

\caption{Distances, radii, and absolute magnitudes from the BW analysis. } 

%%%%% REFEREE VERSION %%%%%%%
%\scriptsize
%%%%% REFEREE VERSION %%%%%%%

\begin{tabular}{rcrcrrcc} \hline \hline 

Name       &     $E(B-V)$     & Period (d) & $p$   &                $d$ (pc)           &                $R$ (\rsol)    &              $M_{\rm V}$      &           $M_{\rm K}$         \\  
\hline
\multicolumn{8}{c}{Galactic Cepheids}  \\
AK Cep    & 0.635 $\pm$  0.049 &  7.233310 & 1.294 &  4035.8 $\pm$  378.2 $\pm$   628.0 &  59.5  $\pm$  5.6 $\pm$   9.7 & -3.90 $\pm$ 0.26 $\pm$  0.47 & -5.52 $\pm$ 0.20 $\pm$  0.43 \\ 
AN Aur    & 0.600 $\pm$  0.057 & 10.289226 & 1.257 &  5301.6 $\pm$  544.2 $\pm$   463.4 &  93.6  $\pm$  9.6 $\pm$   8.1 & -5.13 $\pm$ 0.29 $\pm$  0.31 & -6.56 $\pm$ 0.21 $\pm$  0.20 \\ 
AQ Pup    & 0.518 $\pm$  0.010 & 30.095462 & 1.145 &  2991.1 $\pm$   80.8 $\pm$   141.3 & 137.2  $\pm$  3.7 $\pm$   6.6 & -5.31 $\pm$ 0.07 $\pm$  0.12 & -7.27 $\pm$ 0.06 $\pm$  0.10 \\ 
AV Sgr    & 1.267 $\pm$  0.078 & 15.411587 & 1.215 &  2452.7 $\pm$  104.0 $\pm$   198.3 & 106.2  $\pm$  4.5 $\pm$   8.4 & -4.77 $\pm$ 0.28 $\pm$  0.36 & -6.66 $\pm$ 0.10 $\pm$  0.18 \\ 
AW Per    & 0.489 $\pm$  0.011 &  6.463585 & 1.305 &   971.5 $\pm$   79.4 $\pm$    93.5 &  53.9  $\pm$  4.4 $\pm$   5.3 & -4.06 $\pm$ 0.18 $\pm$  0.23 & -5.41 $\pm$ 0.17 $\pm$  0.22 \\ 
BB Sgr    & 0.281 $\pm$  0.009 &  6.637113 & 1.303 &   814.3 $\pm$   25.8 $\pm$    29.0 &  50.0  $\pm$  1.6 $\pm$   1.8 & -3.53 $\pm$ 0.07 $\pm$  0.09 & -5.15 $\pm$ 0.07 $\pm$  0.08 \\ 
BE Mon    & 0.565 $\pm$  0.038 &  2.705541 & 1.396 &  1884.0 $\pm$  283.2 $\pm$   233.4 &  25.9  $\pm$  3.9 $\pm$   3.2 & -2.66 $\pm$ 0.33 $\pm$  0.32 & -3.86 $\pm$ 0.30 $\pm$  0.28 \\ 
beta Dor  & 0.052 $\pm$  0.009 &  9.842554 & 1.262 &   329.5 $\pm$    6.0 $\pm$     8.8 &  63.9  $\pm$  1.2 $\pm$   1.7 & -4.00 $\pm$ 0.05 $\pm$  0.07 & -5.65 $\pm$ 0.04 $\pm$  0.06 \\ 
BF Oph    & 0.235 $\pm$  0.010 &  4.067677 & 1.354 &   760.2 $\pm$   32.2 $\pm$    34.9 &  33.5  $\pm$  1.4 $\pm$   1.5 & -2.82 $\pm$ 0.10 $\pm$  0.10 & -4.31 $\pm$ 0.09 $\pm$  0.10 \\ 
BG Lac    & 0.300 $\pm$  0.016 &  5.331921 & 1.326 &  1760.4 $\pm$   74.4 $\pm$    78.7 &  41.9  $\pm$  1.8 $\pm$   1.9 & -3.31 $\pm$ 0.10 $\pm$  0.12 & -4.80 $\pm$ 0.09 $\pm$  0.10 \\ 
BM Per    & 0.871 $\pm$  0.048 & 22.958209 & 1.173 &  2253.4 $\pm$   77.1 $\pm$   740.1 &  91.9  $\pm$  3.2 $\pm$  30.9 & -4.16 $\pm$ 0.18 $\pm$  0.56 & -6.33 $\pm$ 0.08 $\pm$  0.49 \\ 
BN Pup    & 0.416 $\pm$  0.018 & 13.672436 & 1.227 &  3570.8 $\pm$   72.8 $\pm$    95.2 &  76.3  $\pm$  1.6 $\pm$   1.9 & -4.18 $\pm$ 0.07 $\pm$  0.09 & -5.99 $\pm$ 0.04 $\pm$  0.06 \\ 
BZ Cyg    & 0.882 $\pm$  0.054 & 10.141721 & 1.259 &  2167.7 $\pm$  246.1 $\pm$   251.0 &  79.4  $\pm$  9.0 $\pm$   9.3 & -4.36 $\pm$ 0.30 $\pm$  0.34 & -6.10 $\pm$ 0.23 $\pm$  0.28 \\ 
CD Cyg    & 0.493 $\pm$  0.015 & 17.074070 & 1.204 &  2267.0 $\pm$   51.1 $\pm$    83.1 &  85.9  $\pm$  2.0 $\pm$   3.1 & -4.40 $\pm$ 0.07 $\pm$  0.10 & -6.26 $\pm$ 0.05 $\pm$  0.08 \\ 
CF Cas    & 0.553 $\pm$  0.011 &  4.875110 & 1.335 &  2934.9 $\pm$  187.0 $\pm$   221.1 &  38.6  $\pm$  2.5 $\pm$   2.9 & -3.01 $\pm$ 0.14 $\pm$  0.19 & -4.59 $\pm$ 0.13 $\pm$  0.18 \\ 
CK Sct    & 0.784 $\pm$  0.077 &  7.415939 & 1.291 &  2360.3 $\pm$  255.4 $\pm$   319.8 &  60.0  $\pm$  6.5 $\pm$   8.1 & -3.85 $\pm$ 0.35 $\pm$  0.43 & -5.52 $\pm$ 0.23 $\pm$  0.31 \\ 
CP Cep    & 0.702 $\pm$  0.050 & 17.863306 & 1.200 &  3073.5 $\pm$  106.3 $\pm$   120.1 &  85.2  $\pm$  3.0 $\pm$   3.5 & -4.14 $\pm$ 0.18 $\pm$  0.21 & -6.17 $\pm$ 0.08 $\pm$  0.10 \\ 
CR Cep    & 0.709 $\pm$  0.016 &  6.233239 & 1.309 &  1299.0 $\pm$   67.0 $\pm$   135.2 &  45.3  $\pm$  2.3 $\pm$   4.8 & -3.25 $\pm$ 0.12 $\pm$  0.25 & -4.91 $\pm$ 0.11 $\pm$  0.25 \\ 
CR Ser    & 0.961 $\pm$  0.087 &  5.301346 & 1.326 &  2610.9 $\pm$  413.2 $\pm$   519.6 &  70.4  $\pm$ 11.1 $\pm$  14.2 & -4.37 $\pm$ 0.44 $\pm$  0.54 & -5.91 $\pm$ 0.32 $\pm$  0.42 \\ 
CS Vel    & 0.737 $\pm$  0.029 &  5.904737 & 1.315 &  3297.7 $\pm$  136.6 $\pm$   182.7 &  43.0  $\pm$  1.8 $\pm$   2.4 & -3.30 $\pm$ 0.13 $\pm$  0.17 & -4.85 $\pm$ 0.09 $\pm$  0.13 \\ 
CV Mon    & 0.722 $\pm$  0.021 &  5.378664 & 1.325 &  1595.3 $\pm$   56.9 $\pm$    71.5 &  40.7  $\pm$  1.5 $\pm$   1.9 & -3.09 $\pm$ 0.11 $\pm$  0.13 & -4.71 $\pm$ 0.08 $\pm$  0.10 \\ 
DD Cas    & 0.450 $\pm$  0.016 &  9.811612 & 1.262 &  2406.5 $\pm$  119.2 $\pm$   118.3 &  50.5  $\pm$  2.5 $\pm$   2.5 & -3.49 $\pm$ 0.12 $\pm$  0.12 & -5.14 $\pm$ 0.11 $\pm$  0.11 \\ 
del Cep   & 0.075 $\pm$  0.009 &  5.366249 & 1.325 &   249.8 $\pm$   10.5 $\pm$    17.1 &  39.1  $\pm$  1.6 $\pm$   2.8 & -3.26 $\pm$ 0.10 $\pm$  0.15 & -4.69 $\pm$ 0.09 $\pm$  0.15 \\ 
DL Cas    & 0.488 $\pm$  0.010 &  8.000458 & 1.283 &  1845.1 $\pm$  168.9 $\pm$   182.6 &  60.7  $\pm$  5.6 $\pm$   6.0 & -3.95 $\pm$ 0.19 $\pm$  0.23 & -5.57 $\pm$ 0.19 $\pm$  0.23 \\ 
DT Cyg    & 0.042 $\pm$  0.011 &  2.499189 & 1.405 &   976.8 $\pm$  250.6 $\pm$   195.4 &  56.5  $\pm$ 14.5 $\pm$  11.3 & -4.31 $\pm$ 0.50 $\pm$  0.52 & -5.53 $\pm$ 0.50 $\pm$  0.52 \\ 
eta Aql   & 0.130 $\pm$  0.009 &  7.176813 & 1.295 &   270.6 $\pm$    6.8 $\pm$     9.3 &  51.1  $\pm$  1.3 $\pm$   1.8 & -3.67 $\pm$ 0.06 $\pm$  0.08 & -5.23 $\pm$ 0.05 $\pm$  0.08 \\ 
FF Aql    & 0.196 $\pm$  0.010 &  4.470838 & 1.344 &   666.2 $\pm$   80.2 $\pm$    89.5 &  62.0  $\pm$  7.5 $\pm$   8.3 & -4.39 $\pm$ 0.25 $\pm$  0.34 & -5.70 $\pm$ 0.25 $\pm$  0.33 \\ 
FM Aql    & 0.589 $\pm$  0.012 &  6.114278 & 1.311 &  1192.0 $\pm$   45.1 $\pm$    47.0 &  59.3  $\pm$  2.2 $\pm$   2.3 & -4.03 $\pm$ 0.09 $\pm$  0.10 & -5.55 $\pm$ 0.08 $\pm$  0.09 \\ 
FM Cas    & 0.325 $\pm$  0.017 &  5.809293 & 1.317 &  1694.1 $\pm$   79.5 $\pm$   135.0 &  39.0  $\pm$  1.8 $\pm$   3.1 & -3.07 $\pm$ 0.11 $\pm$  0.20 & -4.63 $\pm$ 0.10 $\pm$  0.18 \\ 
FN Aql    & 0.483 $\pm$  0.009 &  9.481631 & 1.266 &  1172.9 $\pm$   32.8 $\pm$    30.6 &  51.3  $\pm$  1.4 $\pm$   1.3 & -3.54 $\pm$ 0.07 $\pm$  0.07 & -5.18 $\pm$ 0.06 $\pm$  0.06 \\ 
GH Lup    & 0.335 $\pm$  0.018 &  9.277222 & 1.268 &  1003.4 $\pm$   67.4 $\pm$   138.8 &  56.5  $\pm$  3.8 $\pm$   7.8 & -3.48 $\pm$ 0.15 $\pm$  0.30 & -5.32 $\pm$ 0.14 $\pm$  0.30 \\ 
GY Sge    & 1.187 $\pm$  0.068 & 51.697277 & 1.089 &  2091.6 $\pm$  132.2 $\pm$   624.4 & 157.1  $\pm$ 10.0 $\pm$  44.8 & -5.35 $\pm$ 0.27 $\pm$  0.73 & -7.45 $\pm$ 0.14 $\pm$  0.68 \\ 
KN Cen    & 0.797 $\pm$  0.087 & 34.049845 & 1.132 &  3747.6 $\pm$   88.9 $\pm$   159.8 & 171.6  $\pm$  4.1 $\pm$   5.8 & -5.58 $\pm$ 0.30 $\pm$  0.35 & -7.67 $\pm$ 0.06 $\pm$  0.12 \\ 
KQ Sco    & 0.869 $\pm$  0.020 & 28.696478 & 1.150 &  2398.3 $\pm$   55.7 $\pm$    97.5 & 149.2  $\pm$  3.5 $\pm$   5.9 & -4.91 $\pm$ 0.09 $\pm$  0.12 & -7.28 $\pm$ 0.05 $\pm$  0.09 \\ 
l Car     & 0.147 $\pm$  0.013 & 35.557201 & 1.128 &   436.8 $\pm$    7.2 $\pm$    11.5 & 138.1  $\pm$  2.3 $\pm$   3.5 & -4.94 $\pm$ 0.06 $\pm$  0.08 & -7.16 $\pm$ 0.04 $\pm$  0.06 \\ 
LS Pup    & 0.461 $\pm$  0.015 & 14.147255 & 1.224 &  4470.1 $\pm$   80.1 $\pm$   120.8 &  79.3  $\pm$  1.4 $\pm$   2.1 & -4.28 $\pm$ 0.06 $\pm$  0.08 & -6.07 $\pm$ 0.04 $\pm$  0.06 \\ 
MW Cyg    & 0.635 $\pm$  0.017 &  5.954727 & 1.314 &  1185.8 $\pm$   64.5 $\pm$    95.1 &  38.3  $\pm$  2.1 $\pm$   3.1 & -2.96 $\pm$ 0.13 $\pm$  0.20 & -4.58 $\pm$ 0.12 $\pm$  0.19 \\ 
QZ Nor    & 0.253 $\pm$  0.016 &  3.786578 & 1.361 &  1394.0 $\pm$  107.8 $\pm$    91.2 &  32.0  $\pm$  2.5 $\pm$   2.1 & -2.68 $\pm$ 0.17 $\pm$  0.16 & -4.20 $\pm$ 0.16 $\pm$  0.15 \\ 
RR Lac    & 0.319 $\pm$  0.013 &  6.416324 & 1.306 &  1727.7 $\pm$   79.6 $\pm$   126.5 &  41.0  $\pm$  1.9 $\pm$   3.0 & -3.36 $\pm$ 0.11 $\pm$  0.17 & -4.79 $\pm$ 0.10 $\pm$  0.16 \\ 
RS Cas    & 0.784 $\pm$  0.071 &  6.295970 & 1.308 &  1207.3 $\pm$   49.2 $\pm$    79.8 &  40.5  $\pm$  1.7 $\pm$   2.7 & -3.01 $\pm$ 0.25 $\pm$  0.30 & -4.68 $\pm$ 0.09 $\pm$  0.15 \\ 
RS Ori    & 0.352 $\pm$  0.012 &  7.566893 & 1.289 &  1585.7 $\pm$   89.8 $\pm$   117.6 &  50.5  $\pm$  2.9 $\pm$   3.8 & -3.72 $\pm$ 0.13 $\pm$  0.18 & -5.22 $\pm$ 0.12 $\pm$  0.17 \\ 
RS Pup    & 0.457 $\pm$  0.009 & 41.457711 & 1.112 &  1567.6 $\pm$   30.3 $\pm$    72.4 & 155.9  $\pm$  3.0 $\pm$   7.5 & -5.41 $\pm$ 0.05 $\pm$  0.11 & -7.49 $\pm$ 0.04 $\pm$  0.10 \\ 
RT Aur    & 0.059 $\pm$  0.013 &  3.728309 & 1.363 &   473.5 $\pm$   15.0 $\pm$    29.9 &  35.1  $\pm$  1.1 $\pm$   2.2 & -3.09 $\pm$ 0.08 $\pm$  0.15 & -4.46 $\pm$ 0.07 $\pm$  0.15 \\ 
RU Sct    & 0.921 $\pm$  0.012 & 19.703219 & 1.189 &  1798.1 $\pm$   82.0 $\pm$    59.5 &  97.8  $\pm$  4.5 $\pm$   3.3 & -4.76 $\pm$ 0.10 $\pm$  0.09 & -6.55 $\pm$ 0.10 $\pm$  0.07 \\ 
RW Cam    & 0.633 $\pm$  0.016 & 16.415933 & 1.208 &  1967.8 $\pm$   66.5 $\pm$   139.7 &  97.5  $\pm$  3.3 $\pm$   7.0 & -4.87 $\pm$ 0.09 $\pm$  0.17 & -6.58 $\pm$ 0.07 $\pm$  0.16 \\ 
RW Cas    & 0.380 $\pm$  0.018 & 14.791859 & 1.219 &  3200.3 $\pm$  200.8 $\pm$   186.6 &  94.6  $\pm$  6.0 $\pm$   5.7 & -4.49 $\pm$ 0.15 $\pm$  0.15 & -6.44 $\pm$ 0.13 $\pm$  0.13 \\ 
RX Aur    & 0.263 $\pm$  0.012 & 11.624107 & 1.244 &  1546.7 $\pm$   98.1 $\pm$   118.7 &  68.1  $\pm$  4.3 $\pm$   5.2 & -4.12 $\pm$ 0.14 $\pm$  0.18 & -5.78 $\pm$ 0.13 $\pm$  0.17 \\ 
RX Cam    & 0.532 $\pm$  0.011 &  7.912153 & 1.284 &   818.1 $\pm$   45.1 $\pm$    66.8 &  49.9  $\pm$  2.8 $\pm$   4.1 & -3.62 $\pm$ 0.12 $\pm$  0.19 & -5.17 $\pm$ 0.12 $\pm$  0.18 \\ 
RY Cas    & 0.613 $\pm$  0.059 & 12.138139 & 1.240 &  2104.9 $\pm$  130.8 $\pm$   182.2 &  60.8  $\pm$  3.8 $\pm$   5.4 & -3.64 $\pm$ 0.24 $\pm$  0.30 & -5.48 $\pm$ 0.13 $\pm$  0.19 \\ 
RY CMa    & 0.239 $\pm$  0.010 &  4.678419 & 1.339 &  1248.2 $\pm$  128.7 $\pm$    99.5 &  38.7  $\pm$  4.0 $\pm$   3.1 & -3.14 $\pm$ 0.22 $\pm$  0.19 & -4.64 $\pm$ 0.21 $\pm$  0.19 \\ 
RY Sco    & 0.718 $\pm$  0.018 & 20.321239 & 1.186 &  1087.6 $\pm$   24.4 $\pm$    49.9 &  88.6  $\pm$  2.0 $\pm$   4.0 & -4.50 $\pm$ 0.08 $\pm$  0.12 & -6.31 $\pm$ 0.05 $\pm$  0.10 \\ 
RY Vel    & 0.547 $\pm$  0.010 & 28.124039 & 1.152 &  2002.2 $\pm$   26.0 $\pm$    96.1 & 110.8  $\pm$  1.4 $\pm$   5.2 & -4.90 $\pm$ 0.04 $\pm$  0.11 & -6.78 $\pm$ 0.03 $\pm$  0.10 \\ 
RZ CMa    & 0.443 $\pm$  0.016 &  4.254980 & 1.349 &  1633.3 $\pm$  121.6 $\pm$   133.3 &  31.3  $\pm$  2.3 $\pm$   2.6 & -2.81 $\pm$ 0.16 $\pm$  0.20 & -4.21 $\pm$ 0.16 $\pm$  0.19 \\ 
RZ Gem    & 0.563 $\pm$  0.025 &  5.529040 & 1.322 &  1990.3 $\pm$  182.5 $\pm$   193.9 &  39.2  $\pm$  3.6 $\pm$   3.8 & -3.28 $\pm$ 0.21 $\pm$  0.23 & -4.70 $\pm$ 0.19 $\pm$  0.22 \\ 
RZ Vel    & 0.299 $\pm$  0.009 & 20.399697 & 1.186 &  1410.6 $\pm$   17.3 $\pm$    38.2 &  99.8  $\pm$  1.2 $\pm$   2.6 & -4.58 $\pm$ 0.04 $\pm$  0.07 & -6.54 $\pm$ 0.03 $\pm$  0.06 \\ 

\hline 

\end{tabular} 
\label{Tab-dist}
\end{table*}

\setcounter{table}{9}
\begin{table*} 
\setlength{\tabcolsep}{1.3mm}

\caption{Continued. } 

%%%%% REFEREE VERSION %%%%%%%
%\scriptsize
%%%%% REFEREE VERSION %%%%%%%

\begin{tabular}{rcrcrrcc} \hline \hline 

Name       &     $E(B-V)$     & Period (d) & $p$   &                $d$  (pc)           &                $R$ (\rsol)    &              $M_{\rm V}$      &           $M_{\rm K}$         \\  
\hline

S Mus     & 0.212 $\pm$  0.017 &  9.659971 & 1.264 &   820.6 $\pm$   33.3 $\pm$    31.1 &  61.7  $\pm$  2.5 $\pm$   2.4 & -4.12 $\pm$ 0.10 $\pm$  0.10 & -5.64 $\pm$ 0.09 $\pm$  0.09 \\ 
S Nor     & 0.179 $\pm$  0.009 &  9.754255 & 1.263 &   814.1 $\pm$   23.0 $\pm$    28.4 &  59.5  $\pm$  1.7 $\pm$   2.1 & -3.70 $\pm$ 0.07 $\pm$  0.08 & -5.46 $\pm$ 0.06 $\pm$  0.07 \\ 
S Sge     & 0.100 $\pm$  0.010 &  8.382073 & 1.278 &   680.2 $\pm$   18.1 $\pm$    20.8 &  56.8  $\pm$  1.5 $\pm$   1.8 & -3.85 $\pm$ 0.07 $\pm$  0.08 & -5.44 $\pm$ 0.06 $\pm$  0.07 \\ 
SS Sct    & 0.325 $\pm$  0.009 &  3.671330 & 1.364 &  1394.4 $\pm$  130.8 $\pm$   111.2 &  45.6  $\pm$  4.3 $\pm$   3.6 & -3.58 $\pm$ 0.20 $\pm$  0.20 & -5.00 $\pm$ 0.19 $\pm$  0.20 \\ 
ST Tau    & 0.368 $\pm$  0.030 &  4.034249 & 1.355 &  1174.3 $\pm$  118.7 $\pm$   115.0 &  39.0  $\pm$  3.9 $\pm$   3.9 & -3.32 $\pm$ 0.23 $\pm$  0.25 & -4.69 $\pm$ 0.21 $\pm$  0.23 \\ 
SU Cas    & 0.259 $\pm$  0.010 &  1.949329 & 1.430 &   425.7 $\pm$   26.1 $\pm$    27.6 &  28.9  $\pm$  1.8 $\pm$   1.9 & -3.03 $\pm$ 0.13 $\pm$  0.15 & -4.12 $\pm$ 0.13 $\pm$  0.14 \\ 
SU Cyg    & 0.098 $\pm$  0.014 &  3.845553 & 1.360 &   958.5 $\pm$   58.8 $\pm$    53.1 &  37.4  $\pm$  2.3 $\pm$   2.0 & -3.34 $\pm$ 0.14 $\pm$  0.14 & -4.63 $\pm$ 0.13 $\pm$  0.13 \\ 
SV Mon    & 0.234 $\pm$  0.009 & 15.234488 & 1.216 &  1978.6 $\pm$   73.2 $\pm$    95.0 &  71.6  $\pm$  2.7 $\pm$   3.5 & -3.94 $\pm$ 0.09 $\pm$  0.12 & -5.84 $\pm$ 0.08 $\pm$  0.11 \\ 
SV Per    & 0.408 $\pm$  0.018 & 11.129260 & 1.249 &  2809.0 $\pm$  120.6 $\pm$   249.8 &  80.4  $\pm$  3.5 $\pm$   7.3 & -4.58 $\pm$ 0.11 $\pm$  0.22 & -6.16 $\pm$ 0.09 $\pm$  0.21 \\ 
S Vul     & 0.727 $\pm$  0.043 & 68.711152 & 1.059 &  3879.9 $\pm$  138.7 $\pm$   195.8 & 272.4  $\pm$  9.8 $\pm$  13.2 & -6.36 $\pm$ 0.16 $\pm$  0.19 & -8.62 $\pm$ 0.08 $\pm$  0.12 \\ 
SV Vul    & 0.461 $\pm$  0.021 & 45.027988 & 1.103 &  2173.5 $\pm$   53.1 $\pm$    72.2 & 189.2  $\pm$  4.6 $\pm$   8.2 & -5.95 $\pm$ 0.09 $\pm$  0.11 & -7.93 $\pm$ 0.05 $\pm$  0.08 \\ 
SW Cas    & 0.467 $\pm$  0.018 &  5.440909 & 1.323 &  2126.2 $\pm$  202.8 $\pm$   219.4 &  45.1  $\pm$  4.3 $\pm$   4.8 & -3.45 $\pm$ 0.21 $\pm$  0.25 & -4.96 $\pm$ 0.20 $\pm$  0.24 \\ 
SW Vel    & 0.344 $\pm$  0.009 & 23.439599 & 1.171 &  2088.8 $\pm$   23.9 $\pm$    62.3 &  95.6  $\pm$  1.1 $\pm$   2.9 & -4.54 $\pm$ 0.04 $\pm$  0.07 & -6.49 $\pm$ 0.03 $\pm$  0.06 \\ 
SX Vel    & 0.236 $\pm$  0.011 &  9.550418 & 1.265 &  1408.8 $\pm$  126.4 $\pm$   132.2 &  43.1  $\pm$  3.9 $\pm$   3.9 & -3.22 $\pm$ 0.19 $\pm$  0.20 & -4.83 $\pm$ 0.19 $\pm$  0.20 \\ 
SY Cas    & 0.430 $\pm$  0.039 &  4.071120 & 1.354 &  1870.5 $\pm$   99.6 $\pm$   190.8 &  30.6  $\pm$  1.6 $\pm$   3.2 & -2.85 $\pm$ 0.17 $\pm$  0.27 & -4.18 $\pm$ 0.11 $\pm$  0.22 \\ 
SZ Aql    & 0.537 $\pm$  0.016 & 17.139773 & 1.204 &  1863.5 $\pm$   21.7 $\pm$    48.3 &  93.2  $\pm$  1.1 $\pm$   2.4 & -4.43 $\pm$ 0.06 $\pm$  0.09 & -6.39 $\pm$ 0.03 $\pm$  0.06 \\ 
SZ Cyg    & 0.571 $\pm$  0.015 & 15.109910 & 1.217 &  2320.8 $\pm$  120.0 $\pm$   114.5 &  89.5  $\pm$  4.6 $\pm$   4.5 & -4.24 $\pm$ 0.12 $\pm$  0.12 & -6.27 $\pm$ 0.11 $\pm$  0.11 \\ 
SZ Tau    & 0.295 $\pm$  0.011 &  3.148925 & 1.380 &   599.5 $\pm$   15.8 $\pm$    28.2 &  39.1  $\pm$  1.0 $\pm$   1.8 & -3.34 $\pm$ 0.07 $\pm$  0.11 & -4.69 $\pm$ 0.06 $\pm$  0.10 \\ 
T Mon     & 0.181 $\pm$  0.011 & 27.032092 & 1.156 &  1125.9 $\pm$   27.5 $\pm$    33.0 & 114.0  $\pm$  2.8 $\pm$   3.4 & -4.67 $\pm$ 0.06 $\pm$  0.08 & -6.79 $\pm$ 0.05 $\pm$  0.07 \\ 
TT Aql    & 0.438 $\pm$  0.011 & 13.754828 & 1.227 &   987.1 $\pm$   30.9 $\pm$    31.1 &  79.9  $\pm$  2.5 $\pm$   2.7 & -4.24 $\pm$ 0.08 $\pm$  0.08 & -6.08 $\pm$ 0.07 $\pm$  0.07 \\ 
T Vel     & 0.289 $\pm$  0.009 &  4.639811 & 1.340 &   976.5 $\pm$    9.8 $\pm$    32.5 &  35.4  $\pm$  0.4 $\pm$   1.2 & -2.86 $\pm$ 0.04 $\pm$  0.08 & -4.41 $\pm$ 0.02 $\pm$  0.07 \\ 
T Vul     & 0.064 $\pm$  0.011 &  4.435422 & 1.345 &   512.8 $\pm$   12.8 $\pm$    14.3 &  33.7  $\pm$  0.8 $\pm$   0.9 & -2.99 $\pm$ 0.06 $\pm$  0.07 & -4.37 $\pm$ 0.05 $\pm$  0.06 \\ 
TW Nor    & 1.157 $\pm$  0.013 & 10.786356 & 1.252 &  2293.8 $\pm$  202.5 $\pm$   205.6 &  73.0  $\pm$  6.4 $\pm$   6.4 & -3.92 $\pm$ 0.19 $\pm$  0.21 & -5.84 $\pm$ 0.18 $\pm$  0.20 \\ 
TY Sct    & 0.937 $\pm$  0.059 & 11.053867 & 1.250 &  2004.4 $\pm$  124.6 $\pm$   151.0 &  58.7  $\pm$  3.7 $\pm$   4.2 & -3.74 $\pm$ 0.24 $\pm$  0.29 & -5.44 $\pm$ 0.13 $\pm$  0.17 \\ 
TZ Mon    & 0.431 $\pm$  0.029 &  7.428178 & 1.291 &  4325.6 $\pm$  327.9 $\pm$   385.2 &  59.2  $\pm$  4.5 $\pm$   5.3 & -3.81 $\pm$ 0.19 $\pm$  0.23 & -5.49 $\pm$ 0.16 $\pm$  0.20 \\ 
U Aql     & 0.360 $\pm$  0.010 &  7.024078 & 1.297 &   563.0 $\pm$   25.6 $\pm$    30.6 &  46.4  $\pm$  2.1 $\pm$   2.5 & -3.48 $\pm$ 0.10 $\pm$  0.13 & -5.02 $\pm$ 0.10 $\pm$  0.12 \\ 
U Car     & 0.265 $\pm$  0.010 & 38.819559 & 1.119 &  1401.4 $\pm$   49.2 $\pm$    42.2 & 140.7  $\pm$  5.0 $\pm$   4.2 & -5.27 $\pm$ 0.08 $\pm$  0.07 & -7.30 $\pm$ 0.07 $\pm$  0.07 \\ 
U Nor     & 0.862 $\pm$  0.024 & 12.644184 & 1.236 &  1364.5 $\pm$   31.9 $\pm$    43.5 &  76.7  $\pm$  1.8 $\pm$   2.4 & -4.26 $\pm$ 0.09 $\pm$  0.12 & -6.01 $\pm$ 0.05 $\pm$  0.07 \\ 
U Sgr     & 0.403 $\pm$  0.009 &  6.745308 & 1.301 &   584.2 $\pm$   21.8 $\pm$    21.2 &  47.1  $\pm$  1.8 $\pm$   1.8 & -3.45 $\pm$ 0.08 $\pm$  0.09 & -5.03 $\pm$ 0.08 $\pm$  0.08 \\ 
UU Mus    & 0.399 $\pm$  0.015 & 11.636138 & 1.244 &  2841.9 $\pm$   59.4 $\pm$   135.8 &  63.9  $\pm$  1.3 $\pm$   3.0 & -3.75 $\pm$ 0.07 $\pm$  0.12 & -5.59 $\pm$ 0.05 $\pm$  0.10 \\ 
U Vul     & 0.603 $\pm$  0.011 &  7.990749 & 1.283 &   651.3 $\pm$   28.0 $\pm$    30.2 &  53.4  $\pm$  2.3 $\pm$   2.5 & -3.91 $\pm$ 0.10 $\pm$  0.12 & -5.35 $\pm$ 0.09 $\pm$  0.10 \\ 
UZ Sct    & 1.071 $\pm$  0.066 & 14.747555 & 1.220 &  3087.7 $\pm$  147.8 $\pm$   140.9 &  85.6  $\pm$  4.1 $\pm$   3.8 & -4.69 $\pm$ 0.24 $\pm$  0.27 & -6.30 $\pm$ 0.10 $\pm$  0.11 \\ 
V1162 Aql & 0.205 $\pm$  0.021 &  5.376180 & 1.325 &  1249.2 $\pm$  123.7 $\pm$   305.1 &  42.7  $\pm$  4.2 $\pm$  10.9 & -3.34 $\pm$ 0.22 $\pm$  0.46 & -4.84 $\pm$ 0.21 $\pm$  0.46 \\ 
V340 Ara  & 0.546 $\pm$  0.048 & 20.810876 & 1.184 &  3711.8 $\pm$   72.7 $\pm$   242.3 &  99.1  $\pm$  2.0 $\pm$   8.2 & -4.38 $\pm$ 0.17 $\pm$  0.23 & -6.50 $\pm$ 0.05 $\pm$  0.14 \\ 
V340 Nor  & 0.321 $\pm$  0.018 & 11.288422 & 1.247 &  1833.7 $\pm$  120.8 $\pm$   154.0 &  72.8  $\pm$  4.8 $\pm$   6.1 & -3.96 $\pm$ 0.15 $\pm$  0.20 & -5.85 $\pm$ 0.14 $\pm$  0.19 \\ 
V350 Sgr  & 0.299 $\pm$  0.009 &  5.154258 & 1.329 &   916.5 $\pm$   24.6 $\pm$    51.7 &  39.8  $\pm$  1.1 $\pm$   2.3 & -3.30 $\pm$ 0.07 $\pm$  0.12 & -4.72 $\pm$ 0.06 $\pm$  0.13 \\ 
V386 Cyg  & 0.841 $\pm$  0.017 &  5.257679 & 1.327 &   978.0 $\pm$   54.6 $\pm$    90.2 &  40.4  $\pm$  2.3 $\pm$   3.8 & -3.08 $\pm$ 0.13 $\pm$  0.22 & -4.70 $\pm$ 0.12 $\pm$  0.21 \\ 
V402 Cyg  & 0.391 $\pm$  0.025 &  4.364891 & 1.346 &  2094.5 $\pm$  222.1 $\pm$   192.1 &  35.5  $\pm$  3.8 $\pm$   3.3 & -3.01 $\pm$ 0.23 $\pm$  0.23 & -4.46 $\pm$ 0.22 $\pm$  0.21 \\ 
V459 Cyg  & 0.730 $\pm$  0.018 &  7.251260 & 1.294 &  1573.4 $\pm$   90.5 $\pm$    99.2 &  34.5  $\pm$  2.0 $\pm$   2.0 & -2.79 $\pm$ 0.14 $\pm$  0.15 & -4.37 $\pm$ 0.12 $\pm$  0.14 \\ 
V495 Cyg  & 0.977 $\pm$  0.055 &  6.718375 & 1.301 &  3181.2 $\pm$  469.3 $\pm$   506.6 &  95.1  $\pm$ 14.0 $\pm$  15.4 & -5.12 $\pm$ 0.36 $\pm$  0.41 & -6.58 $\pm$ 0.30 $\pm$  0.36 \\ 
V496 Aql  & 0.397 $\pm$  0.010 &  6.807017 & 1.300 &  1157.4 $\pm$  123.8 $\pm$   123.2 &  57.2  $\pm$  6.1 $\pm$   6.2 & -3.86 $\pm$ 0.22 $\pm$  0.26 & -5.44 $\pm$ 0.22 $\pm$  0.25 \\ 
V538 Cyg  & 0.642 $\pm$  0.059 &  6.119215 & 1.311 &  2684.6 $\pm$  222.2 $\pm$   295.1 &  50.6  $\pm$  4.2 $\pm$   5.8 & -3.80 $\pm$ 0.26 $\pm$  0.34 & -5.24 $\pm$ 0.17 $\pm$  0.26 \\ 
V600 Aql  & 0.798 $\pm$  0.016 &  7.238847 & 1.294 &  1811.8 $\pm$   97.0 $\pm$   157.7 &  53.1  $\pm$  2.8 $\pm$   4.6 & -3.87 $\pm$ 0.12 $\pm$  0.21 & -5.33 $\pm$ 0.11 $\pm$  0.20 \\ 
V916 Aql  & 1.089 $\pm$  0.064 & 13.442663 & 1.229 &  3591.9 $\pm$   97.1 $\pm$   204.6 &  90.7  $\pm$  2.5 $\pm$   4.9 & -5.54 $\pm$ 0.22 $\pm$  0.28 & -6.62 $\pm$ 0.06 $\pm$  0.14 \\ 
V Car     & 0.169 $\pm$  0.010 &  6.696707 & 1.302 &   910.2 $\pm$   30.7 $\pm$    50.4 &  38.2  $\pm$  1.3 $\pm$   2.1 & -2.97 $\pm$ 0.08 $\pm$  0.13 & -4.57 $\pm$ 0.07 $\pm$  0.12 \\ 
V Cen     & 0.292 $\pm$  0.012 &  5.493980 & 1.322 &   694.1 $\pm$   18.1 $\pm$    25.2 &  41.8  $\pm$  1.1 $\pm$   1.5 & -3.31 $\pm$ 0.07 $\pm$  0.09 & -4.81 $\pm$ 0.06 $\pm$  0.08 \\ 
VW Cen    & 0.428 $\pm$  0.022 & 15.037338 & 1.217 &  3687.2 $\pm$   62.7 $\pm$   169.6 &  88.1  $\pm$  1.5 $\pm$   4.0 & -3.95 $\pm$ 0.09 $\pm$  0.13 & -6.17 $\pm$ 0.04 $\pm$  0.10 \\ 
VX Cyg    & 0.830 $\pm$  0.058 & 20.133213 & 1.187 &  2618.8 $\pm$  127.2 $\pm$   115.8 &  95.5  $\pm$  4.7 $\pm$   4.0 & -4.72 $\pm$ 0.22 $\pm$  0.24 & -6.51 $\pm$ 0.11 $\pm$  0.10 \\ 
VX Per    & 0.475 $\pm$  0.011 & 10.886268 & 1.251 &  4139.4 $\pm$  488.9 $\pm$   483.6 & 117.0  $\pm$ 13.8 $\pm$  13.8 & -5.33 $\pm$ 0.25 $\pm$  0.28 & -6.97 $\pm$ 0.24 $\pm$  0.29 \\ 
VY Car    & 0.237 $\pm$  0.009 & 18.904990 & 1.194 &  1363.4 $\pm$   33.1 $\pm$    41.0 &  76.4  $\pm$  1.9 $\pm$   2.3 & -3.97 $\pm$ 0.06 $\pm$  0.07 & -5.95 $\pm$ 0.05 $\pm$  0.07 \\ 
VY Cyg    & 0.606 $\pm$  0.018 &  7.857165 & 1.285 &  1091.9 $\pm$   76.5 $\pm$   145.7 &  29.3  $\pm$  2.1 $\pm$   3.8 & -2.57 $\pm$ 0.16 $\pm$  0.32 & -4.03 $\pm$ 0.15 $\pm$  0.31 \\ 
VY Sgr    & 1.283 $\pm$  0.077 & 13.558324 & 1.228 &  2702.0 $\pm$   71.2 $\pm$   143.9 &  98.0  $\pm$  2.6 $\pm$   5.2 & -4.88 $\pm$ 0.27 $\pm$  0.32 & -6.57 $\pm$ 0.06 $\pm$  0.13 \\ 
VZ Cyg    & 0.266 $\pm$  0.011 &  4.864379 & 1.335 &  1848.7 $\pm$   65.6 $\pm$    74.3 &  39.4  $\pm$  1.4 $\pm$   1.6 & -3.22 $\pm$ 0.08 $\pm$  0.10 & -4.68 $\pm$ 0.08 $\pm$  0.09 \\ 
VZ Pup    & 0.459 $\pm$  0.011 & 23.174946 & 1.172 &  4134.1 $\pm$   62.9 $\pm$    98.2 &  97.3  $\pm$  1.5 $\pm$   2.3 & -4.89 $\pm$ 0.05 $\pm$  0.07 & -6.58 $\pm$ 0.03 $\pm$  0.05 \\ 
W Gem     & 0.255 $\pm$  0.010 &  7.913488 & 1.284 &  1222.7 $\pm$  126.4 $\pm$   143.1 &  68.0  $\pm$  7.0 $\pm$   8.0 & -4.28 $\pm$ 0.22 $\pm$  0.27 & -5.84 $\pm$ 0.21 $\pm$  0.27 \\ 
W Sgr     & 0.108 $\pm$  0.011 &  7.594968 & 1.289 &   203.3 $\pm$   22.8 $\pm$    16.1 &  25.8  $\pm$  2.9 $\pm$   2.1 & -2.21 $\pm$ 0.23 $\pm$  0.18 & -3.77 $\pm$ 0.23 $\pm$  0.17 \\ 
WZ Car    & 0.370 $\pm$  0.011 & 23.015214 & 1.173 &  3217.4 $\pm$   46.1 $\pm$   164.5 &  91.6  $\pm$  1.3 $\pm$   5.2 & -4.42 $\pm$ 0.05 $\pm$  0.12 & -6.37 $\pm$ 0.03 $\pm$  0.12 \\ 
WZ Sgr    & 0.431 $\pm$  0.011 & 21.850159 & 1.179 &  1619.8 $\pm$   51.7 $\pm$    47.0 & 108.5  $\pm$  3.5 $\pm$   3.2 & -4.38 $\pm$ 0.08 $\pm$  0.07 & -6.63 $\pm$ 0.07 $\pm$  0.07 \\ 

\hline 

\end{tabular} 
\end{table*}

\setcounter{table}{9}
\begin{table*} 
\setlength{\tabcolsep}{1.3mm}

\caption{Continued. } 

%%%%% REFEREE VERSION %%%%%%%
%\scriptsize
%%%%% REFEREE VERSION %%%%%%%

\begin{tabular}{rcrcrrcc} \hline \hline 

Name       &     $E(B-V)$     & Period (d) & $p$   &                $d$ (pc)           &                $R$ (\rsol)    &              $M_{\rm V}$      &           $M_{\rm K}$         \\  
\hline

X Cyg     & 0.228 $\pm$  0.011 & 16.385764 & 1.209 &  1036.2 $\pm$   24.1 $\pm$    27.9 &  89.7  $\pm$  2.1 $\pm$   2.5 & -4.40 $\pm$ 0.06 $\pm$  0.07 & -6.32 $\pm$ 0.05 $\pm$  0.06 \\ 
X Lac     & 0.336 $\pm$  0.011 &  5.444534 & 1.323 &  1604.7 $\pm$   97.5 $\pm$    86.5 &  47.8  $\pm$  2.9 $\pm$   2.5 & -3.72 $\pm$ 0.13 $\pm$  0.13 & -5.11 $\pm$ 0.13 $\pm$  0.12 \\ 
X Pup     & 0.402 $\pm$  0.009 & 25.965762 & 1.161 &  2533.8 $\pm$   40.2 $\pm$   104.8 & 107.7  $\pm$  1.7 $\pm$   4.8 & -4.75 $\pm$ 0.05 $\pm$  0.10 & -6.73 $\pm$ 0.03 $\pm$  0.09 \\ 
X Sgr     & 0.237 $\pm$  0.015 &  7.012747 & 1.297 &   317.4 $\pm$    8.1 $\pm$    14.5 &  46.4  $\pm$  1.2 $\pm$   2.1 & -3.72 $\pm$ 0.07 $\pm$  0.11 & -5.08 $\pm$ 0.05 $\pm$  0.10 \\ 
X Vul     & 0.742 $\pm$  0.019 &  6.319553 & 1.308 &   870.1 $\pm$   54.2 $\pm$    64.2 &  40.1  $\pm$  2.5 $\pm$   3.0 & -3.29 $\pm$ 0.15 $\pm$  0.19 & -4.73 $\pm$ 0.13 $\pm$  0.17 \\ 
XX Cen    & 0.266 $\pm$  0.011 & 10.953515 & 1.251 &  1402.4 $\pm$   37.2 $\pm$    39.5 &  57.8  $\pm$  1.5 $\pm$   1.6 & -3.76 $\pm$ 0.07 $\pm$  0.07 & -5.42 $\pm$ 0.06 $\pm$  0.06 \\ 
XX Sgr    & 0.521 $\pm$  0.016 &  6.424310 & 1.306 &  1327.3 $\pm$   91.2 $\pm$    77.2 &  45.8  $\pm$  3.2 $\pm$   2.7 & -3.44 $\pm$ 0.15 $\pm$  0.14 & -4.99 $\pm$ 0.14 $\pm$  0.13 \\ 
Y Lac     & 0.207 $\pm$  0.016 &  4.323760 & 1.347 &  2683.6 $\pm$   81.7 $\pm$    81.1 &  44.8  $\pm$  1.4 $\pm$   1.4 & -3.66 $\pm$ 0.08 $\pm$  0.09 & -5.00 $\pm$ 0.07 $\pm$  0.07 \\ 
Y Oph     & 0.645 $\pm$  0.015 & 17.126144 & 1.204 &   558.7 $\pm$   17.6 $\pm$    19.8 &  85.4  $\pm$  2.7 $\pm$   3.0 & -4.70 $\pm$ 0.08 $\pm$  0.10 & -6.30 $\pm$ 0.07 $\pm$  0.08 \\ 
Y Sct     & 0.757 $\pm$  0.012 & 10.341254 & 1.257 &  1770.0 $\pm$  113.5 $\pm$   119.2 &  71.1  $\pm$  4.6 $\pm$   4.8 & -4.08 $\pm$ 0.14 $\pm$  0.15 & -5.84 $\pm$ 0.13 $\pm$  0.15 \\ 
Y Sgr     & 0.191 $\pm$  0.009 &  5.773372 & 1.317 &   415.2 $\pm$   57.8 $\pm$    42.2 &  37.5  $\pm$  5.2 $\pm$   3.9 & -2.96 $\pm$ 0.28 $\pm$  0.25 & -4.55 $\pm$ 0.28 $\pm$  0.25 \\ 
YZ Aur    & 0.601 $\pm$  0.058 & 18.192968 & 1.198 &  4245.2 $\pm$  219.8 $\pm$   428.5 & 103.7  $\pm$  5.4 $\pm$  10.7 & -4.74 $\pm$ 0.23 $\pm$  0.32 & -6.64 $\pm$ 0.11 $\pm$  0.23 \\ 
YZ Sgr    & 0.281 $\pm$  0.010 &  9.553741 & 1.265 &  1119.7 $\pm$   61.7 $\pm$    57.8 &  58.2  $\pm$  3.2 $\pm$   3.0 & -3.80 $\pm$ 0.12 $\pm$  0.12 & -5.45 $\pm$ 0.12 $\pm$  0.12 \\ 
zeta Gem  & 0.014 $\pm$  0.009 & 10.149922 & 1.258 &   359.3 $\pm$    9.2 $\pm$    10.2 &  64.7  $\pm$  1.6 $\pm$   1.8 & -3.93 $\pm$ 0.07 $\pm$  0.07 & -5.65 $\pm$ 0.05 $\pm$  0.06 \\ 
Z Lac     & 0.370 $\pm$  0.011 & 10.885697 & 1.251 &  1813.0 $\pm$   44.0 $\pm$    70.1 &  67.0  $\pm$  1.6 $\pm$   2.6 & -4.06 $\pm$ 0.06 $\pm$  0.10 & -5.74 $\pm$ 0.05 $\pm$  0.09 \\ 

\multicolumn{8}{c}{LMC Cepheids}  \\
HV 1005   & 0.100 $\pm$  0.005 & 18.714651 & 1.195 & 44096.6 $\pm$ 1787.4 $\pm$  1141.7 &  82.6  $\pm$  3.4 $\pm$   2.1 & -4.39 $\pm$ 0.09 $\pm$  0.06 & -6.20 $\pm$ 0.09 $\pm$  0.06 \\ 
HV 1006   & 0.100 $\pm$  0.005 & 14.216644 & 1.223 & 42626.4 $\pm$ 1557.0 $\pm$  3293.9 &  73.9  $\pm$  2.7 $\pm$   5.3 & -4.04 $\pm$ 0.08 $\pm$  0.17 & -5.90 $\pm$ 0.08 $\pm$  0.16 \\ 
HV 1023   & 0.070 $\pm$  0.005 & 26.554194 & 1.158 & 48109.6 $\pm$ 1770.2 $\pm$  2438.1 & 120.3  $\pm$  4.5 $\pm$   5.5 & -4.81 $\pm$ 0.08 $\pm$  0.12 & -6.92 $\pm$ 0.08 $\pm$  0.12 \\ 
HV 12197  & 0.060 $\pm$  0.005 &  3.143795 & 1.381 & 38519.9 $\pm$ 2388.9 $\pm$  1541.2 &  22.6  $\pm$  1.4 $\pm$   0.9 & -2.03 $\pm$ 0.13 $\pm$  0.09 & -3.48 $\pm$ 0.13 $\pm$  0.09 \\ 
HV 12198  & 0.060 $\pm$  0.005 &  3.522766 & 1.369 & 47828.2 $\pm$ 1786.8 $\pm$  2073.1 &  30.2  $\pm$  1.1 $\pm$   1.3 & -2.63 $\pm$ 0.08 $\pm$  0.10 & -4.10 $\pm$ 0.08 $\pm$  0.10 \\ 
HV 12199  & 0.060 $\pm$  0.005 &  2.639167 & 1.399 & 51317.8 $\pm$ 3516.8 $\pm$  3040.6 &  26.8  $\pm$  1.8 $\pm$   1.6 & -2.46 $\pm$ 0.14 $\pm$  0.14 & -3.87 $\pm$ 0.14 $\pm$  0.14 \\ 
HV 12202  & 0.060 $\pm$  0.005 &  3.101216 & 1.382 & 35765.2 $\pm$ 2292.9 $\pm$  2234.2 &  21.7  $\pm$  1.4 $\pm$   1.4 & -1.89 $\pm$ 0.14 $\pm$  0.14 & -3.38 $\pm$ 0.13 $\pm$  0.14 \\ 
HV 12203  & 0.060 $\pm$  0.004 &  2.954123 & 1.387 & 44902.6 $\pm$ 2957.3 $\pm$  2481.8 &  24.8  $\pm$  1.6 $\pm$   1.4 & -2.32 $\pm$ 0.14 $\pm$  0.14 & -3.70 $\pm$ 0.14 $\pm$  0.13 \\ 
HV 12204  & 0.060 $\pm$  0.005 &  3.438749 & 1.371 & 45433.9 $\pm$ 2154.3 $\pm$  3331.3 &  28.5  $\pm$  1.4 $\pm$   2.1 & -2.75 $\pm$ 0.10 $\pm$  0.17 & -4.04 $\pm$ 0.10 $\pm$  0.17 \\ 
HV 12452  & 0.058 $\pm$  0.005 &  8.738897 & 1.274 & 39620.7 $\pm$ 1448.4 $\pm$  3968.2 &  47.8  $\pm$  1.8 $\pm$   4.8 & -3.39 $\pm$ 0.08 $\pm$  0.22 & -5.06 $\pm$ 0.08 $\pm$  0.22 \\ 
HV 12505  & 0.100 $\pm$  0.005 & 14.393280 & 1.222 & 46661.2 $\pm$ 1577.9 $\pm$  4095.0 &  76.3  $\pm$  2.6 $\pm$   6.2 & -3.79 $\pm$ 0.07 $\pm$  0.19 & -5.90 $\pm$ 0.07 $\pm$  0.19 \\ 
HV 12717  & 0.058 $\pm$  0.005 &  8.843849 & 1.273 & 35477.4 $\pm$ 3154.6 $\pm$  3676.3 &  41.8  $\pm$  3.7 $\pm$   4.4 & -3.18 $\pm$ 0.19 $\pm$  0.23 & -4.78 $\pm$ 0.18 $\pm$  0.23 \\ 
HV 12815  & 0.070 $\pm$  0.005 & 26.115120 & 1.160 & 39034.2 $\pm$ 2303.9 $\pm$  3812.0 & 105.2  $\pm$  6.2 $\pm$  11.0 & -4.65 $\pm$ 0.13 $\pm$  0.21 & -6.66 $\pm$ 0.12 $\pm$  0.21 \\ 
HV 12816  & 0.070 $\pm$  0.005 &  9.108991 & 1.270 & 47026.6 $\pm$ 3305.1 $\pm$  3137.1 &  53.5  $\pm$  3.8 $\pm$   3.6 & -4.05 $\pm$ 0.15 $\pm$  0.16 & -5.38 $\pm$ 0.15 $\pm$  0.15 \\ 
HV 2257   & 0.060 $\pm$  0.005 & 39.388561 & 1.117 & 46153.9 $\pm$ 1085.6 $\pm$  2119.3 & 156.6  $\pm$  3.7 $\pm$   6.5 & -5.33 $\pm$ 0.05 $\pm$  0.10 & -7.47 $\pm$ 0.05 $\pm$  0.10 \\ 
HV 2282   & 0.100 $\pm$  0.005 & 14.677123 & 1.220 & 44678.3 $\pm$ 1082.1 $\pm$  1790.5 &  75.1  $\pm$  1.8 $\pm$   2.9 & -4.20 $\pm$ 0.05 $\pm$  0.09 & -5.96 $\pm$ 0.05 $\pm$  0.09 \\ 
HV 2338   & 0.040 $\pm$  0.005 & 42.194159 & 1.110 & 43337.5 $\pm$ 0707.4 $\pm$  1935.7 & 153.4  $\pm$  2.5 $\pm$   6.8 & -5.48 $\pm$ 0.04 $\pm$  0.10 & -7.48 $\pm$ 0.04 $\pm$  0.10 \\ 
HV 2369   & 0.095 $\pm$  0.005 & 48.392674 & 1.096 & 37861.3 $\pm$ 1268.6 $\pm$  2233.8 & 151.2  $\pm$  5.1 $\pm$   9.2 & -5.49 $\pm$ 0.07 $\pm$  0.13 & -7.48 $\pm$ 0.07 $\pm$  0.13 \\ 
HV 2405   & 0.070 $\pm$  0.005 &  6.923455 & 1.298 & 63519.3 $\pm$ 5494.6 $\pm$  6336.9 &  60.8  $\pm$  5.3 $\pm$   6.0 & -4.08 $\pm$ 0.18 $\pm$  0.25 & -5.60 $\pm$ 0.18 $\pm$  0.24 \\ 
HV 2527   & 0.070 $\pm$  0.005 & 12.949600 & 1.233 & 49165.1 $\pm$ 2326.8 $\pm$  2290.7 &  73.8  $\pm$  3.5 $\pm$   3.4 & -4.03 $\pm$ 0.10 $\pm$  0.11 & -5.89 $\pm$ 0.10 $\pm$  0.10 \\ 
HV 2538   & 0.100 $\pm$  0.005 & 13.871003 & 1.226 & 48045.6 $\pm$ 3441.4 $\pm$  4496.4 &  79.8  $\pm$  5.7 $\pm$   7.2 & -4.25 $\pm$ 0.15 $\pm$  0.20 & -6.07 $\pm$ 0.15 $\pm$  0.20 \\ 
HV 2549   & 0.058 $\pm$  0.005 & 16.218520 & 1.210 & 45696.6 $\pm$ 2762.6 $\pm$  2200.1 &  86.7  $\pm$  5.3 $\pm$   4.2 & -4.64 $\pm$ 0.13 $\pm$  0.11 & -6.32 $\pm$ 0.13 $\pm$  0.11 \\ 
HV 2827   & 0.080 $\pm$  0.005 & 78.824495 & 1.045 & 40791.1 $\pm$ 1652.4 $\pm$  1721.4 & 222.2  $\pm$  9.0 $\pm$   9.6 & -6.00 $\pm$ 0.09 $\pm$  0.10 & -8.21 $\pm$ 0.09 $\pm$  0.10 \\ 
HV 5655   & 0.100 $\pm$  0.005 & 14.212586 & 1.223 & 45859.2 $\pm$ 2057.8 $\pm$  3650.0 &  72.7  $\pm$  3.3 $\pm$   5.3 & -4.05 $\pm$ 0.10 $\pm$  0.18 & -5.88 $\pm$ 0.10 $\pm$  0.17 \\ 
HV 6093   & 0.058 $\pm$  0.005 &  4.784880 & 1.337 & 47300.7 $\pm$ 3548.0 $\pm$  4795.1 &  37.2  $\pm$  2.8 $\pm$   3.7 & -3.23 $\pm$ 0.16 $\pm$  0.22 & -4.59 $\pm$ 0.16 $\pm$  0.22 \\ 
HV 873    & 0.130 $\pm$  0.005 & 34.436191 & 1.131 & 50117.9 $\pm$ 1128.8 $\pm$  2200.7 & 148.5  $\pm$  3.4 $\pm$   6.7 & -5.84 $\pm$ 0.05 $\pm$  0.10 & -7.53 $\pm$ 0.05 $\pm$  0.10 \\ 
HV 876    & 0.100 $\pm$  0.005 & 22.715624 & 1.174 & 44756.6 $\pm$ 1588.5 $\pm$  1883.9 &  97.4  $\pm$  3.5 $\pm$   3.8 & -4.86 $\pm$ 0.08 $\pm$  0.09 & -6.60 $\pm$ 0.08 $\pm$  0.09 \\ 
HV 877    & 0.100 $\pm$  0.005 & 45.158119 & 1.103 & 49137.9 $\pm$ 2137.0 $\pm$  5164.0 & 171.7  $\pm$  7.5 $\pm$  18.4 & -5.41 $\pm$ 0.09 $\pm$  0.27 & -7.64 $\pm$ 0.09 $\pm$  0.27 \\ 
HV 878    & 0.058 $\pm$  0.005 & 23.306146 & 1.172 & 49845.2 $\pm$ 1349.1 $\pm$  1707.2 & 111.2  $\pm$  3.0 $\pm$   3.9 & -5.06 $\pm$ 0.06 $\pm$  0.08 & -6.85 $\pm$ 0.06 $\pm$  0.07 \\ 
HV 879    & 0.060 $\pm$  0.005 & 36.831567 & 1.124 & 43030.6 $\pm$ 2890.7 $\pm$  2305.0 & 131.8  $\pm$  8.9 $\pm$   7.3 & -4.94 $\pm$ 0.14 $\pm$  0.12 & -7.13 $\pm$ 0.14 $\pm$  0.12 \\ 
HV 881    & 0.030 $\pm$  0.005 & 35.743231 & 1.127 & 40336.0 $\pm$ 0949.4 $\pm$  1065.7 & 121.2  $\pm$  2.9 $\pm$   3.2 & -4.96 $\pm$ 0.05 $\pm$  0.06 & -6.97 $\pm$ 0.05 $\pm$  0.06 \\ 
HV 899    & 0.110 $\pm$  0.005 & 31.050706 & 1.142 & 47670.4 $\pm$ 1231.5 $\pm$  1397.6 & 128.0  $\pm$  3.3 $\pm$   3.8 & -5.34 $\pm$ 0.06 $\pm$  0.07 & -7.16 $\pm$ 0.06 $\pm$  0.07 \\ 
HV 900    & 0.058 $\pm$  0.005 & 47.481696 & 1.098 & 45391.7 $\pm$ 0988.8 $\pm$  1358.3 & 165.0  $\pm$  3.6 $\pm$   4.7 & -5.65 $\pm$ 0.05 $\pm$  0.07 & -7.65 $\pm$ 0.05 $\pm$  0.07 \\ 
HV 909    & 0.058 $\pm$  0.005 & 37.558988 & 1.122 & 40579.0 $\pm$ 1215.3 $\pm$  1637.7 & 128.5  $\pm$  3.9 $\pm$   5.0 & -5.40 $\pm$ 0.07 $\pm$  0.09 & -7.16 $\pm$ 0.06 $\pm$  0.09 \\ 
HV 914    & 0.070 $\pm$  0.005 &  6.878394 & 1.299 & 50856.1 $\pm$ 5016.2 $\pm$  2652.7 &  53.2  $\pm$  5.3 $\pm$   2.8 & -3.81 $\pm$ 0.20 $\pm$  0.12 & -5.32 $\pm$ 0.20 $\pm$  0.12 \\ 
U 1       & 0.100 $\pm$  0.005 & 22.542693 & 1.175 & 52497.2 $\pm$ 1546.8 $\pm$  2528.2 & 113.2  $\pm$  3.4 $\pm$   5.1 & -4.76 $\pm$ 0.07 $\pm$  0.11 & -6.82 $\pm$ 0.06 $\pm$  0.10 \\

\multicolumn{8}{c}{SMC Cepheids}  \\

HV 1345   & 0.030 $\pm$  0.005 & 13.478117 & 1.229 & 41067.2 $\pm$ 1705.7 $\pm$  1484.6 &  54.4  $\pm$  2.3 $\pm$   1.9 & -3.38 $\pm$ 0.09 $\pm$  0.08 & -5.22 $\pm$ 0.09 $\pm$  0.08 \\ 
HV 1335   & 0.090 $\pm$  0.005 & 14.380836 & 1.222 & 43362.0 $\pm$ 1891.0 $\pm$  2268.8 &  50.4  $\pm$  2.2 $\pm$   2.8 & -3.68 $\pm$ 0.09 $\pm$  0.12 & -5.23 $\pm$ 0.09 $\pm$  0.11 \\ 
HV 1328   & 0.004 $\pm$  0.003 & 15.835971 & 1.212 & 55005.4 $\pm$ 3950.6 $\pm$  4054.3 &  77.2  $\pm$  5.5 $\pm$   5.8 & -4.57 $\pm$ 0.15 $\pm$  0.16 & -6.11 $\pm$ 0.15 $\pm$  0.16 \\ 
HV 1333   & 0.070 $\pm$  0.005 & 16.295258 & 1.209 & 65988.1 $\pm$ 2898.9 $\pm$  4126.9 &  85.9  $\pm$  3.8 $\pm$   5.6 & -4.59 $\pm$ 0.09 $\pm$  0.14 & -6.29 $\pm$ 0.09 $\pm$  0.14 \\ 
HV 822    & 0.030 $\pm$  0.005 & 16.742306 & 1.206 & 67441.1 $\pm$ 2403.2 $\pm$  3533.0 &  97.2  $\pm$  3.5 $\pm$   5.4 & -4.66 $\pm$ 0.08 $\pm$  0.12 & -6.52 $\pm$ 0.08 $\pm$  0.12 \\ 
HV 837    & 0.042 $\pm$  0.005 & 42.705045 & 1.109 & 56363.4 $\pm$ 3097.9 $\pm$  2949.2 & 162.1  $\pm$  8.9 $\pm$   8.5 & -5.61 $\pm$ 0.12 $\pm$  0.12 & -7.59 $\pm$ 0.12 $\pm$  0.12 \\ 

\hline 

\end{tabular} 
\end{table*}

\subsection{PL(Z) relations}

Table~\ref{Tab-plfits} presents PL and PLZ relations of the form  $M = \alpha \log P + \beta + \gamma$ [Fe/H]
in the $V$, $W(VK)$, and $K-$band. The results are given for the SMC, LMC, Galaxy, and for all Cepheids. 
Because of the small numbers of SMC Cepheids, the results for that galaxy are not reliable and are given for completeness only.
For the Galactic Cepheids only and for the complete sample, a few clear outliers are removed (deviating by more than 0.8 mag 
from the PL relation).

\begin{figure} 

\begin{minipage}{0.49\textwidth}
\resizebox{\hsize}{!}{\includegraphics{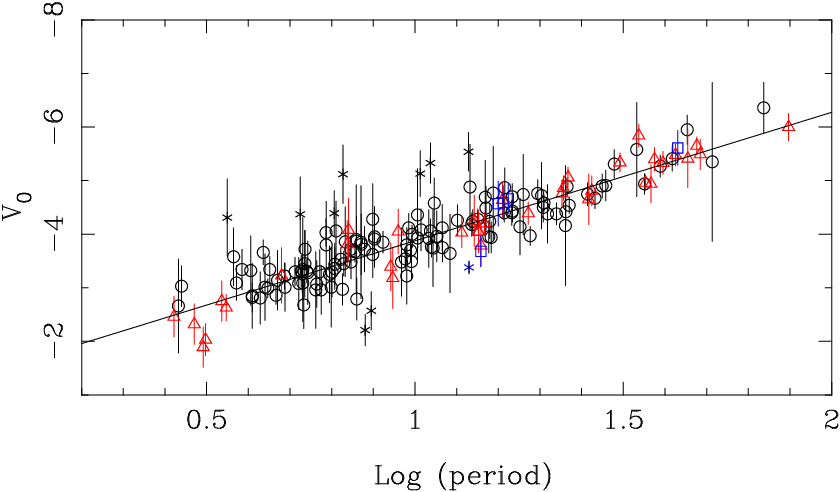}}
\end{minipage}
\begin{minipage}{0.49\textwidth}
\resizebox{\hsize}{!}{\includegraphics{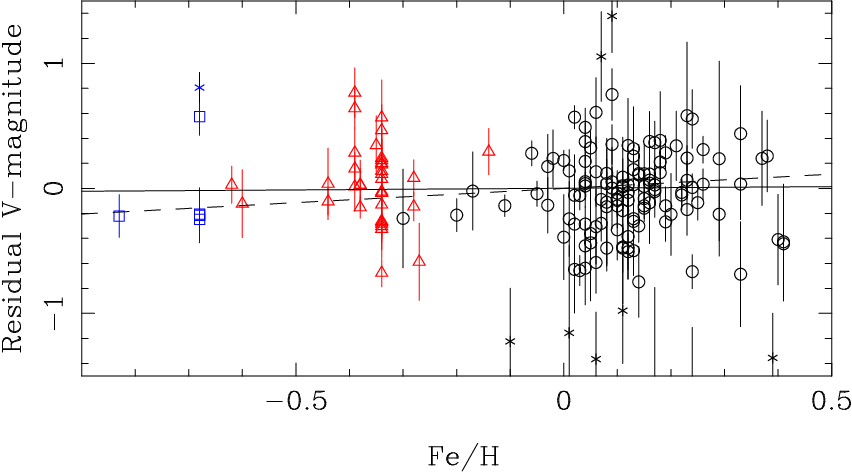}}
\end{minipage}

\caption[]{ 
The PL relation in the $V$-band. 
Galactic objects are plotted as (black) open circles, LMC Cepheids as  (red) triangles, 
and SMC Cepheids as (blue) squares.
Stars plotted with a cross symbol are excluded from the fit. 
The bottom panel shows the residual plotted versus metallicity.
The solid line shows the fit when the residual is fitted, 
the solid line the dependence from a two-parameter fit (as given in Table~\ref{Tab-plfits})
} 
\label{Fig-PLV} 
\end{figure}

\begin{figure} 

\begin{minipage}{0.49\textwidth}
\resizebox{\hsize}{!}{\includegraphics{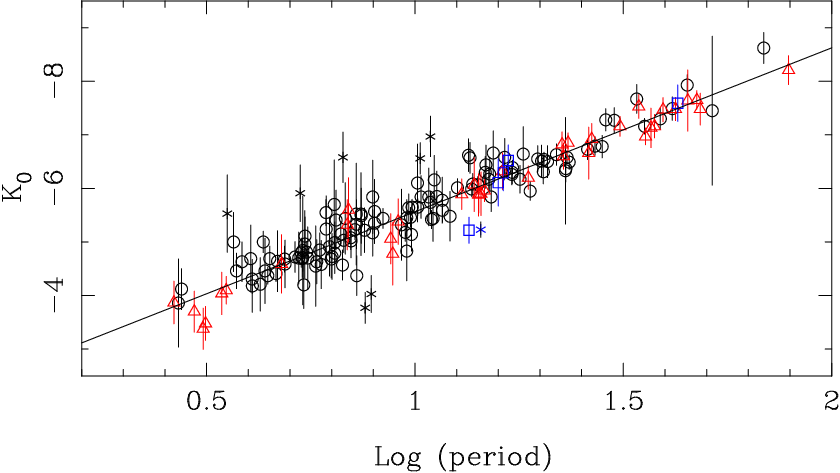}}
\end{minipage}
\begin{minipage}{0.49\textwidth}
\resizebox{\hsize}{!}{\includegraphics{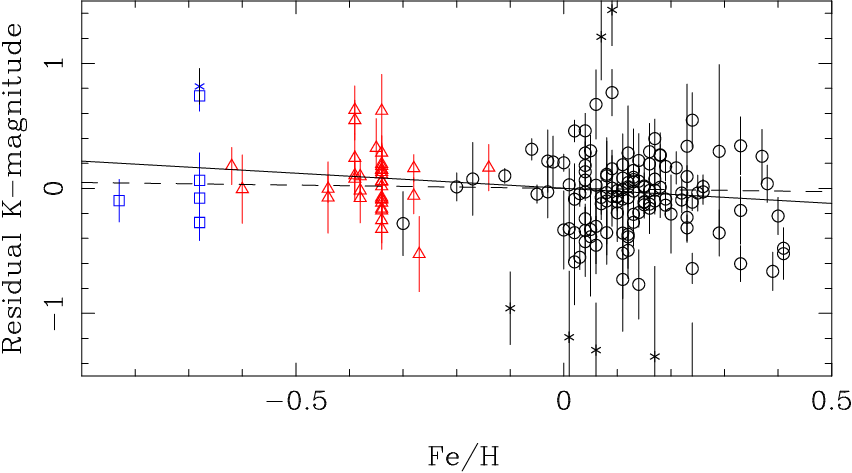}}
\end{minipage}

\caption[]{ 
As Fig.~\ref{Fig-PLV} in the $K$-band. 
} 
\label{Fig-PLK} 
\end{figure}

Figures~\ref{Fig-PLV} and \ref{Fig-PLK} show the PL relation in
the $V$-band and $K$-band for the complete sample; in the bottom panel, the residual plotted against metallicity is shown. 
The first and second error bar quoted for the magnitudes in Table~\ref{Tab-dist} have been added in quadrature.

The metallicity dependence was determined in two ways: 1) by
first fitting a linear PL relation and fitting the residual with a
linear relation against [Fe/H] (as shown in Figs.~\ref{Fig-PLV} and
\ref{Fig-PLK} by the solid line, done mainly for easy visualisation), 
and 2) by making a linear fit in the two variables $\log P$ and [Fe/H], as listed in Table~\ref{Tab-plfits} 
and shown in Figs.~\ref{Fig-PLV} and \ref{Fig-PLK} by the dashed line.

The slopes of the LMC ($-3.21 \pm 0.13$) and Galactic ($-3.03 \pm 0.08$) PL relation in the $K$-band are formally 
consistent at the 1$\sigma$ level; in the $V$-band they are consistent only at the 3$\sigma$ level 
($-2.69 \pm 0.12$) for the LMC, ($-2.21 \pm 0.09$) for the Galaxy).
The better agreement in $K$ and lesser agreement in $V$ between the two galaxies is in line with theoretically predicted relations 
(Bono et al. 2010), which give slopes at 
$Z$= 0.004, 0.008, 0.02 in the $K$-band of $-3.19 \pm 0.09$, $-3.28 \pm  0.09$, and $-3.22 \pm 0.15$, respectively,
(and  $\delta$slope/$\delta \log Z= 0.08 \pm 0.07$ over all metallicities they considered), 
and in the $V$-band of $-2.87 \pm 0.09$, $-2.80 \pm  0.15$, and $-2.51 \pm 0.24$, respectively
(and $\delta$slope/$\delta \log Z= 0.67 \pm 0.09$).
The slope found for the Galactic Cepheids in the $V$-band ($-2.21 \pm 0.09)$ agrees at the 1.5$\sigma$ level with 
the value of ($-2.43 \pm 0.12$) in Benedict et al. (2007) based on the ten stars with HST parallaxes.
In the $K$-band, the slope ($-3.03 \pm 0.08$) agrees at the 2$\sigma$ level with the ($-3.32 \pm 0.12$) found by Benedict et al.
The slopes found for the LMC Cepheids in the $V$-band ($-2.69 \pm 0.12$) and $K$-band ($-3.21 \pm 0.13$) 
are in excellent agreement with various determinations in the literature, see Sect.~\ref{sect-pp} and Tab.~\ref{Tab-lmcfits}.

The metallicity dependence quoted in G08 based on 68 Galactic Cepheids were 
$\gamma = (+0.27 \pm 0.30)$ mag/dex ($V$-band) and $(-0.11 \pm 0.24)$ ($K$-band).
Now, values of $(+0.17 \pm 0.25)$ and $(+0.07 \pm 0.20)$ from 121 Galactic Cepheids and  
$(+0.23 \pm 0.11)$ and $(-0.05 \pm 0.10)$ mag/dex from the complete sample are derived. 
Compared to G08, the larger sample of Galactic Cepheids has reduced the error bar, but the main reduction in error bar has 
come from adding the MC Cepheids.
The outcome is that the iron dependence of the PL relation in the $K$-band is not significant and 
only marginally significant (2$\sigma$) in the $V$ band.
A fit were the {\it slope} of the PL relation is also allowed to vary linearly with metallicity is also included in 
Table~\ref{Tab-plfits} but the error bars in the coefficients are large and the result is not significant.

\begin{table*} 

\caption{PL(Z) relations of the form $M = \alpha + \beta \log P + \gamma$ [Fe/H] + $\delta \, \log P \, \cdot$ [Fe/H], 
in the $V$, $W(VK)$, and $K-$band and for different galaxies. }

\begin{tabular}{lcrrrrr} \hline \hline 

Band    & Galaxy & N & $\alpha$ & $\beta$ & $\gamma$ & $\delta$   \\  
\hline 

K & ALL & 162 & -2.50 $\pm$ 0.08 & -3.06 $\pm$ 0.06 & - \\
K & GAL & 121 & -2.55 $\pm$ 0.09 & -3.03 $\pm$ 0.08 & - \\
K & LMC &  36 & -2.26 $\pm$ 0.17 & -3.21 $\pm$ 0.13 & - \\
K & SMC &   6 & -0.36 $\pm$ 0.98 & -4.56 $\pm$ 0.78 & - \\

K & ALL & 162 & -2.49 $\pm$ 0.08 & -3.07 $\pm$ 0.07 & -0.05 $\pm$ 0.10 \\ % -0.24 +- 0.07
K & GAL & 121 & -2.56 $\pm$ 0.09 & -3.03 $\pm$ 0.08 &  0.07 $\pm$ 0.20 \\
K & LMC &  36 & -2.27 $\pm$ 0.18 & -3.22 $\pm$ 0.13 &  0.19 $\pm$ 0.37 \\
K & SMC &   6 & -0.66 $\pm$ 0.99 & -4.23 $\pm$ 0.81 & -1.31 $\pm$ 0.86 \\

K & ALL & 162 & -2.47 $\pm$ 0.08 & -3.08 $\pm$ 0.07 & -0.59 $\pm$ 0.42 &  0.42 $\pm$ 0.31 \\

V & ALL & 160 & -1.48 $\pm$ 0.08 & -2.40 $\pm$ 0.07 & - \\
V & GAL & 119 & -1.68 $\pm$ 0.10 & -2.21 $\pm$ 0.09 & - \\
V & LMC &  36 & -1.10 $\pm$ 0.17 & -2.69 $\pm$ 0.12 & - \\
V & SMC &   6 &  0.73 $\pm$ 0.93 & -4.03 $\pm$ 0.74 & - \\

V & ALL & 160 & -1.55 $\pm$ 0.09 & -2.33 $\pm$ 0.07 &  0.23 $\pm$ 0.11 \\ % 0.03 +- 0.09
V & GAL & 121 & -1.69 $\pm$ 0.10 & -2.21 $\pm$ 0.09 &  0.17 $\pm$ 0.25 \\
V & LMC &  36 & -1.09 $\pm$ 0.17 & -2.68 $\pm$ 0.12 & -0.14 $\pm$ 0.35 \\
V & SMC &   6 &  0.48 $\pm$ 0.95 & -3.74 $\pm$ 0.77 & -1.19 $\pm$ 0.82 \\

V & ALL & 160 & -1.54 $\pm$ 0.09 & -2.34 $\pm$ 0.08 &  -0.04 $\pm$ 0.46 & 0.21 $\pm$ 0.34  \\ % 
 
WVK & ALL & 158 & -2.68 $\pm$ 0.08 & -3.12 $\pm$ 0.06 & - \\
WVK & GAL & 120 & -2.69 $\pm$ 0.09 & -3.12 $\pm$ 0.08 & - \\
WVK & LMC &  36 & -2.41 $\pm$ 0.18 & -3.27 $\pm$ 0.13 & - \\
WVK & SMC &   6 & -0.51 $\pm$ 0.98 & -4.63 $\pm$ 0.78 & - \\

WVK & ALL & 158 & -2.69 $\pm$ 0.08 & -3.11 $\pm$ 0.07 &  0.04 $\pm$ 0.10 \\ % 0.03 +- 0.09
WVK & GAL & 120 & -2.72 $\pm$ 0.09 & -3.13 $\pm$ 0.08 &  0.34 $\pm$ 0.20 \\
WVK & LMC &  36 & -2.42 $\pm$ 0.18 & -3.29 $\pm$ 0.13 & +0.23 $\pm$ 0.37 \\
WVK & SMC &   6 & -0.81 $\pm$ 0.99 & -4.29 $\pm$ 0.81 & -1.34 $\pm$ 0.85 \\

WVK & ALL & 158 & -2.70 $\pm$ 0.09 & -3.11 $\pm$ 0.07 &  +0.17 $\pm$ 0.44 & -0.10 $\pm$ 0.33  \\ % 

\hline 

\end{tabular} 
\label{Tab-plfits}
\end{table*}

\subsection{The PR relation}

\begin{figure} 

\begin{minipage}{0.49\textwidth}
\resizebox{\hsize}{!}{\includegraphics{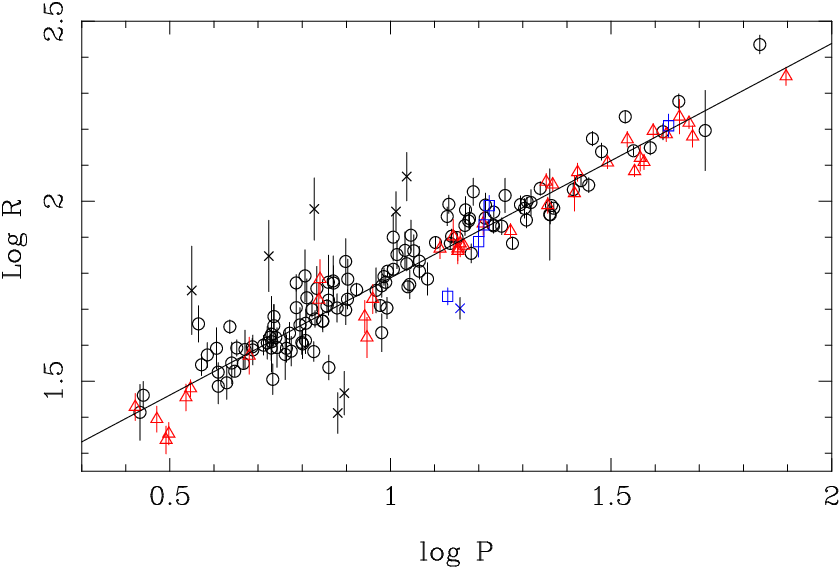}}
\end{minipage}

\caption[]{ 
The PR relation the all Cepheids.
Galactic objects are plotted as  (black) open circles, LMC Cepheids as (red) triangles, 
and SMC Cepheids as (blue) squares.
Stars plotted with a cross symbol are excluded from the fit. 
The line is the best fit listed in Eq.~\ref{Eq-PRall}
} 
\label{PRall} 
\end{figure}

Figure~\ref{PRall} shows the PR relation for all Galactic and MC Cepheids. The best fit is 
\begin{equation} 
\log R = (0.651 \pm  0.012) \log P + (1.136 \pm 0.014), \sigma= 0.055,
\label{Eq-PRall} 
\end{equation} 
where 8 outliers (deviating by 0.17 dex, or about 3$\sigma$) have been 
removed\footnote{These are AN Aur, CR Ser, DT Cyg, V495 Cyg, VX Per, VY Cyg, W Sgr, and HV 1335.}.

This result does depend on the adopted $p$-factor relation. For a constant $p= 1.33$, for example, 
the relation would become $\log R = 0.737 \log P + 1.074$. This is in agreement with Molinari et al. (2011), 
who find $\log R = (0.75 \pm 0.03) \log P + (1.10 \pm 0.03)$ for a constant $p= 1.27$.
A negative dependence of the $p$-factor on period leads to a shallower slope. 
The PR relation in Eq.~\ref{Eq-RAD} based on stars with known distance has a slope of $0.696 \pm 0.033$. 
This is in agreement with the present one, which depends on the $p$-factor.
The slopes in the PR relation that are found in the present work are shallower than often quoted in the literature   
(see Molinari et al. (2011), and Turner et al. (2010) for recent compilations), but these also depend on the $p$-factor.
For example,  Gieren et al. (1998) find $\log R = (0.750 \pm 0.024) \log P + (1.075 \pm 0.007)$ for $p= 1.39 -0.03 \log P$, 
while Turner \& Burke (2002) find      $\log R = (0.750 \pm 0.006) \log P + (1.071 \pm 0.006)$ for $p= 1.31$.

Theory predicts slopes shallower than this and more in agreement with the slopes found in the present paper.
Recent models by Petroni et al. (2003) lead to $\log R = (0.676 \pm 0.006) \log P + (1.173 \pm 0.008)$ for solar-metallicity.
Both theory and radii for stars with known distances lead to slopes in the PR relation shallower than found for 
BW-type analysis with a constant $p$-factor and thus support a (steep) period dependence of the $p$-factor.

\section{Summary and discussion}

The PL relation in the $V$ and $K$-band (and the
corresponding Wesenheit index) and the dependence on metallicity are
investigated for a sample of 128 Galactic, 6 SMC, and 36 LMC Cepheids
with an individual metallicity determination from high-resolution
spectroscopy.  Distances are derived using the Baade-Wesselink
technique implementing the most recent surface-brightness relation and
estimates for the projection factor.

The slope of the $p$-factor relation is found to be $-0.24$ from
demanding that the distance to the LMC does not depend on period and
$(V-K)$ colour and that the slope of the PL relation in $V$ and $K$
be the same for the observed relations based on apparent and absolute magnitudes.
This result agrees within the errors with the slope found by Storm et al. (2011a).
The slope found by Storm et al. and in the present work is much steeper than the one predicted by theory, 
$p = (-−0.08 \pm 0.05) \log P + (1.31 \pm 0.06)$ (Nardetto et al. 2009).

The zero point of the relation is tight to the ten Cepheids that have HST and improved Hipparcos 
parallaxes (and a cluster distance for two Cepheids) and to the 18 Cepheids that have only a cluster distance.
The finally adopted relation is $p= 1.50 -0.24 \log P$.

The metallicity dependence of the PL relation is investigated. 
No significant dependence is found in $K$, and $W(VK)$, and a 2$\sigma$ result in $V$, $\gamma = +0.23 \pm 0.11$.

The distance scale found here is shorter than that in Storm et al. (2011a, b) because of the smaller $p$-factor found here.
At a typical period of ten days, $(1.55 -0.186 \log P)/(1.50 -0.24 \log P)= 1.364 / 1.26 = 1.08$.
This is reflected in the distance to the MCs. The median distance to the 36 LMC and 6 SMC Cepheids is 
45.5 and 55.7 kpc (Distance Modulus (DM) of 18.29 and  18.73). The error in the mean is 0.47 and 1.4 kpc (0.02 and 0.06 in DM).

Although the absolute distances are short for the adopted $p$-factor, the difference in DM between SMC and LMC is 0.44 $\pm$ 0.06, 
which is in agreement with other independent determinations 
(Type {\sc ii} Cepheids, red clump, tip of the RGB), see the discussion in Matsunaga et al. (2011). 
Storm et al. (2011b) found 0.47 $\pm$ 0.14.

The difference with respect to Storm et al. in the distance scale is unlikely to be related to reddening.
Their values for $E(B-V)$ have been adopted for the MC Cepheids. For the 74 Galactic Cepheids in common, 
the average difference in adopted $E(B-V)$ is only 0.002.
%absolute difference in $E(B-V)$ is only 0.005.

In general, however, changes in reddening do influence the derived distances.
For example, the present Galactic sample has 50 stars in common with the sample in Pejcha \& Kochanek (2012). 
The average difference in $E(B-V)$ is $-0.023$ (theirs minus the present work), while for 21 LMC and 
3 SMC stars in common with Pejcha \& Kochanek, the difference is +0.034 and +0.030.

As a test the distances were re-computed, changing all $E(B-V)$ by +0.05 for the MC Cepheids 
and by $-0.03$ for the Galactic ones.  The average LMC (SMC) distance is increased by +640 (+900) pc, 
while the distance to the Cepheids with parallaxes and in clusters is decreased by on average $-0.85\%$.
In other words, systematic effects in $E(B-V)$ of this magnitude would increase the DM to LMC and SMC to 18.34 and 18.78, respectively.

The exact procedure of comparing BW to HST-based distances also plays a role. Here, the weighted mean of the ratio has been used.
Storm et al.  use the unweighted mean of the difference in distances relative to the average of the two distances.
If this procedure is adopted then a distance scale longer by 3.7\% is found (corresponding to DM of 18.37 and 18.81, respectively).

The distance to the LMC is shorter than most of the recent determinations, which are in the range 18.42-18.55 
(Walker 2012, who quotes 49.7 kpc with a range of $\pm$3\%).
Within the current framework, this difference could be reconciled by suggesting that the $p$-factor is larger at lower metallicity 
by about $\sim 8-9$\% at the metallicity of both the LMC and SMC. This is not predicted by theory (Nardetto et al. 2009), 
but then theory currently also does not predict the steep dependence of the $p$-factor on period either.

\medskip

Further improvements could still be made on the observational side.
Metallicity determination from high-resolution spectroscopy for the 5 SMC and (a significant subset of the) 22 LMC Cepheids in the sample 
without such determination would likely further improve the determination of the metallicity term in the PL relation.
Interferometry for more stars could improve on better constraining the
surface-brightness relation, possibly further investigating the hint
of a metallicity dependence noted in Sect.~4.  The observations of T Vul 
(Gallenne et al. 2012) demonstrate that angular diameters as small
as 0.60\arcsec\ can be determined reliably in the $K$-band on 300m baselines.

For reference, Table~\ref{TAB-INTER} lists the stars with minimum
angular diameters $\ge$0.59\arcsec\ and angular diameter amplitude
$\ge$0.06\arcsec, sorted by the latter quantity. It includes objects
for which interferometric observations already exist (including objects that would benefit from
additional observations, like X Sgr and W Sgr).

\begin{table} 

\caption{Galactic Cepheids with the largest angular diameters, sorted by amplitude.} 

\begin{tabular}{rccrr} \hline \hline 

Name    & $\theta_{\rm min}$ & $\theta_{\rm max}$-$\theta_{\rm min}$ & [Fe/H]  \\  
        &  (\arcsec)        &  (\arcsec) &  \\
\hline 

T Vul\tablefootmark{*} &  0.59 & 0.06 & 0.08 \\
RT Aur   &  0.65 & 0.07 & 0.13 \\
AW Per   &  0.62 & 0.08 & 0.04 \\
S Mus    &  0.67 & 0.08 & 0.07 \\
S Nor    &  0.64 & 0.08 & 0.13 \\
W Gem    &  0.62 & 0.08 & 0.02 \\
RX Cam   &  0.68 & 0.09 & 0.11 \\
S Sge    &  0.73 & 0.09 & 0.15 \\
U Sgr    &  0.70 & 0.09 & 0.15 \\
U Vul    &  0.71 & 0.09 & 0.19 \\
GY Sge   &  0.61 & 0.11 & 0.29 \\
S Vul    &  0.60 & 0.11 & 0.12 \\
U Aql    &  0.70 & 0.11 & 0.17 \\
Y Sgr    &  0.79 & 0.11 & 0.12 \\
Y Oph\tablefootmark{*} &  1.36 & 0.12 & 0.13 \\
RY Sco   &  0.68 & 0.13 & 0.16 \\
X Sgr\tablefootmark{*} &  1.41 & 0.14 & -0.20 \\
TT Aql   &  0.65 & 0.16 & 0.22 \\
zeta Gem\tablefootmark{*} &  1.60 & 0.16 & 0.10 \\
del Cep\tablefootmark{*}  &  1.38 & 0.17 & 0.12 \\
RZ Vel   &  0.59 & 0.17 & 0.04 \\
X Cyg    &  0.70 & 0.18 & 0.17 \\
SV Vul   &  0.71 & 0.20 & 0.12 \\
beta Dor\tablefootmark{*} &  1.71 & 0.21 & 0.10 \\
eta Aql\tablefootmark{*}  &  1.65 & 0.21 & 0.15 \\
U Car    &  0.86 & 0.22 & 0.04 \\
RS Pup   &  0.78 & 0.24 & 0.22 \\
T Mon    &  0.80 & 0.24 & 0.23 \\
W Sgr\tablefootmark{*} &  1.11 & 0.25 & 0.09 \\
l Car\tablefootmark{*} &  2.64 & 0.59 & 0.13 \\

\hline 
\end{tabular} 
\tablefoot{
\tablefoottext{*}{A star that has been monitored interferometrically, see Table~\ref{TAB-PFAC}.}
}

\label{TAB-INTER} 
\end{table}

\acknowledgements{  
This research has made use of the SIMBAD database, operated at CDS, Strasbourg, France. 
MG would like to thank Alexandre Gallenne for providing the individual angular-diameter determinations for T Vul and FF Aql,
Giuseppe Bono and Laura Inno for providing their latest results on MC PL relations before publication, 
and Giuseppe, Jesper Storm and Chow Choong Ngeow for commenting on a earlier version of the paper.
}

{}

\begin{appendix}

\section{Cluster distances}

It is not our intention to give a review on the distances to Galactic clusters containing Cepheids.
Recent compilations of distances can be found in Turner (2010) and Tammann et al. (2003). 
The latter is based on Feast (1999), which is an update of Feast \& Walker (1987), with detailed remarks in Walker (1987b).

However, when comparing cluster distances quoted in the
literature to the BW distances obtained here and checking the
literature in more detail, it was obvious that not all distances were
given on the same distance scale. In addition, some results obtained since Turner (2010) can be included in the analysis.
 
Traditionally, the distances to clusters are based on Zero Age Main Sequence (ZAMS) fitting using $BV$ data and a reference ZAMS, 
which is very often that of Turner (1976, 1979a). It is tied to a DM to the Pleiades of 5.56.

Recently, Turner, Majaess and co-workers used ZAMS fitting with 2MASS $JHK$ data to derive distances to Cepheids containing clusters.
The procedure is outlined in Majaess et al. (2011), and the distances
to nine benchmark open clusters thathave HST and revised Hipparcos-based distances (van Leeuwen 2009) determined. There is the well-known
disagreement for the Pleiades, but the infrared ZAMS fitting distances
to the other clusters, and the comparison to the Hyades and Pleiades
that have HST-based parallaxes is excellent.
The 2MASS-based ZAMS fitting is therefore tied to a DM of 5.65 for the Pleiades, which is thus different from that implied when using 
the Turner ZAMS in the optical.

In Table~\ref{ClusterD}, the adopted cluster-based DM are listed for the Cepheids in the sample.
Infrared ZAMS fitting is preferred over earlier work in the optical.
Where appropriate, the older work is scaled to the adopted Pleiades distance.
A few Cepheids in clusters that are in our sample, but where the association is uncertain 
or the DM in the literature are very discrepant, have not been considered: KQ Sco, GY Sge, T Mon, SV Vul (see Hoyle et al. 2003)

\begin{table*} 

\caption{Distances to cluster containing Cepheids.}

\begin{tabular}{rrrcccccccccc} \hline \hline 

Name Cepheid &  Name Cluster & adopted DM   & Method & Reference & Remarks\\
             &               &              &        &           &         \\
\hline 

BB Sgr    & Collinder 394 &  9.38 $\pm$ 0.10 & JHK & Turner (2010) \\
V Cen     & NGC 5662      &  9.28 $\pm$ 0.05 & JHK & Turner (2010) \\
RU Sct    & Trumpler 35   & 11.11 $\pm$ 0.10 & JHK & Turner (2010) \\
SU Cyg    & Turner 9      &  9.33 $\pm$ 0.05 & JHK & Turner (2010) \\
S Vul     & Anon Vul OB   & 12.47 $\pm$ 0.29 & JHK & Turner (2011) \\
delta Cep & Cep OB6       &  7.21 $\pm$ 0.13 & JHK & Majaess et al. (2012a) \\
zeta Gem  & ADS 5742      &  7.75 $\pm$ 0.09 & JHK & Majaess et al. (2012b) \\
SU Cas    & Alessi 95     &  8.04 $\pm$ 0.08 & JHK & Majaess et al. (2012c) \\

VY Car    & Car OB2       & 11.66 $\pm$ 0.15\tablefootmark{*} & BV  & Turner (1977) & includes a +0.09 correction in DM \\
RZ Vel    & Vel OB1       & 11.32 $\pm$ 0.15\tablefootmark{*} & BV  & Turner (1979b) & includes a +0.09 correction in DM \\
CS Vel    & Ruprecht 79   & 12.55 $\pm$ 0.16 & BV  & Walker (1987c) & includes a +0.08 correction in DM \\
SZ Tau    & NGC 1647      &  8.76 $\pm$ 0.02 & BV  & Turner (1992)  & includes a +0.09 correction in DM \\
SW Vel    & Vel OB 5      & 12.08 $\pm$ 0.15 & BV  & Turner et al. (1993)  & includes a +0.09 correction in DM \\
X Cyg     & Ruprecht 175  & 10.52 $\pm$ 0.04 & BV  & Turner (1998)  & includes a +0.09 correction in DM \\

U Sgr     & IC 4725       &  9.05 $\pm$ 0.09 & BV  & \tablefootmark{a} & \tablefootmark{a} \\
DL Cas    & NGC 129       & 11.10 $\pm$ 0.07 & BV  & \tablefootmark{b} & \tablefootmark{b} \\
S Nor     & NGC 6087      &  9.82 $\pm$ 0.18 & BV  & \tablefootmark{c} & \tablefootmark{c} \\
TW Nor    & Lynga 6       & 11.40 $\pm$ 0.12 & BV  & \tablefootmark{d} & \tablefootmark{d} \\
QZ Nor, 
V340 Nor  & NGC 6067      & 11.15 $\pm$ 0.09 & BV  & \tablefootmark{e} & \tablefootmark{e}  \\
CV Mon    & vandenBergh 1 & 11.12 $\pm$ 0.15 & BV  & \tablefootmark{f} & \tablefootmark{f}  \\
WZ Sgr    & Turner 2      & 11.31 $\pm$ 0.10 & BV  & \tablefootmark{g} & \tablefootmark{g}  \\
CF Cas    & NGC 7790      & 12.63 $\pm$ 0.11 & BV  & \tablefootmark{h} & \tablefootmark{h}  \\

\hline 
\end{tabular} 
%\tablebib{
%(1) van Leeuwen et al. (2007); (2) Majaess et al. (2012a); (3) Majaess et al. (2012b);
%}
\tablefoot{
\tablefoottext{a}{The average of the distances quoted in 
An et al. (2007; 8.93 $\pm$ 0.08 plus a +0.02 correction), 
Hoyle et al. (2003; 9.08 $\pm$ 0.18 plus a +0.09 correction), and 
Pel et al. (1985; 8.95 $\pm$ 0.10 plus a +0.08 correction).}
\tablefoottext{b}{The average of the distances quoted in 
Turner et al. (1992; 11.11 $\pm$ 0.02 plus a +0.09 correction),
An et al. (2007; 11.04 $\pm$ 0.05 plus a +0.02 correction), and 
Hoyle et al. (2003; 10.94 $\pm$ 0.14 plus a +0.09 correction).}
\tablefoottext{c}{The average of the distances quoted in 
Turner (1986; 9.78 $\pm$ 0.03 plus a +0.09 correction), 
An et al. (2007; 9.65 $\pm$ 0.06 plus a +0.02 correction), and 
Pel et al. (1985; 9.84 $\pm$ 0.10 plus a +0.08 correction).}
\tablefoottext{d}{The average of the distances quoted in 
An et al. (2007; 11.51 $\pm$ 0.13 plus a +0.02 correction), 
Hoyle et al. (2003; 11.33 $\pm$ 0.18 plus a +0.09 correction), and
Walker et al. (1985a;  11.15 $\pm$ 0.3 plus a +0.09 correction).}
\tablefoottext{e}{The average of the distances quoted in 
An et al. (2007; 11.03 $\pm$ 0.08 plus a +0.02 correction), 
Hoyle et al. (2003; 11.18 $\pm$ 0.12 plus a +0.09 correction), and
Walker et al. (1985b;  11.05 $\pm$ 0.10 plus a +0.09 correction).}
\tablefoottext{f}{Three distance determinations have been considered:
Turner et al. (1998; 11.08 $\pm$ 0.03 plus a +0.09 correction, adopting $E(B-V)= 0.75$), 
An et al. (2007; 10.74 $\pm$ 0.21 plus a +0.02 correction, adopting $E(B-V)= 0.57$), and
Hoyle et al. (2003; 11.34 $\pm$ 0.21 plus a +0.09 correction, adopting $E(B-V)= 0.90$).
The adopted distance is the average of the three, but the dispersion is large. 
This is likely due to the very different reddenings adopted. If a correction is made to a reddening of 0.75, 
adopting $\Delta$ DM/$\Delta$ E(B-V) $\sim 2$ (An et al. 2007), then the average becomes 11.14 with a very small dispersion.}
\tablefoottext{g}{The average of the distances quoted in 
Turner et al. (1993; 11.26 $\pm$ 0.10 plus a +0.09 correction), and 
Hoyle et al. (2003; 11.18 $\pm$ 0.16 plus a +0.09 correction).}
\tablefoottext{h}{The average of the distances quoted in 
An et al. (2007; 12.46 $\pm$ 0.11 plus a +0.02 correction), 
Hoyle et al. (2003; 12.58 $\pm$ 0.14 plus a +0.09 correction), and 
Romeo et al. (1989; 12.65 $\pm$ 0.15 plus a +0.08 correction).}
\tablefoottext{*}{No error quoted, conservative error adopted.}
}

\label{ClusterD}
\end{table*}

\end{appendix}

\end{document}